\documentclass{aa}
\usepackage[varg]{txfonts}
\usepackage{chemformula}
\usepackage{graphicx}
\usepackage[colorlinks=true,     linkcolor=blue, citecolor=blue, filecolor=blue, urlcolor=blue]{hyperref}
                        
\begin{document}

\title{The GAPS Programme at TNG\thanks{Based on observations made with the Italian Telescopio Nazionale Galileo (TNG) operated by the Fundaci\'{o}n Galileo Galilei (FGG) of the Istituto Nazionale di Astrofisica (INAF) at the Observatorio del Roque de los Muchachos (La Palma, Canary Islands, Spain).}}
\subtitle{LV. Multiple molecular species in the atmosphere of HAT-P-11\,b and review of the HAT-P-11 planetary system}
\author{
M. Basilicata\inst{1,2}
\and 
P. Giacobbe\inst{2}
\and
A.~S.~Bonomo\inst{2}
\and
G. Scandariato\inst{3}
\and
M. Brogi\inst{4, 2}
\and
V. Singh\inst{3}
\and
A. Di Paola\inst{4}
\and
L. Mancini\inst{1,2,5}
\and
A. Sozzetti\inst{2}
\and
A. F. Lanza\inst{3}
\and
P. E. Cubillos\inst{2, 14}
\and
M. Damasso\inst{2}
\and
S. Desidera\inst{13}
\and
K. Biazzo\inst{3}
\and
A. Bignamini\inst{10}
\and
F. Borsa\inst{6}
\and
L. Cabona\inst{13}
\and
I. Carleo\inst{11,2}
\and
A. Ghedina\inst{7}
\and
G. Guilluy\inst{2}
\and
A. Maggio\inst{8}
\and
G. Mainella\inst{7}
\and
G. Micela\inst{8}
\and
E. Molinari\inst{6}
\and
M. Molinaro\inst{10}
\and
D. Nardiello\inst{13}
\and
M. Pedani\inst{7}
\and
L. Pino\inst{12}
\and
E. Poretti\inst{6,7}
\and
J. Southworth\inst{9}
\and
M. Stangret\inst{13}
\and
D. Turrini\inst{2}
}
\institute{
Department of Physics, University of Rome ``Tor Vergata'', Via della Ricerca Scientifica 1, I-00133 Rome, Italy \\
\email{mario.basilicata@inaf.it}
\and 
INAF -- Osservatorio Astrofisico di Torino, Via Osservatorio 20, I-10025, Pino Torinese, Italy
\and
INAF -- Osservatorio Astrofisico di Catania, Via S. Sofia 78, I-95123, Catania, Italy
\and
Department of Physics, University of Turin, Via Pietro Giuria
1, I-10125 Torino, Italy
\and
Max Planck Institute for Astronomy, K\"{o}nigstuhl 17, 69117 -- Heidelberg, Germany
\and
INAF -- Osservatorio Astronomico di Brera, Via E. Bianchi 46, 23807 Merate (LC), Italy
\and
Fundaci\'{o}n Galileo Galilei -- INAF, Rambla Jos\'{e} Ana Fernandez P\'{e}rez 7, 38712 -- Bre\~{n}a Baja, Spain
\and
INAF -- Osservatorio Astronomico di Palermo, Piazza del Parlamento, 1, I-90134 Palermo, Italy
\and
Astrophysics Group, Keele University, Keele ST5 5BG, UK
\and
INAF -- Osservatorio Astronomico di Trieste, via Tiepolo 11, 34143 Trieste, Italy
\and
Instituto de Astrof\'{i}sica de Canarias (IAC), 38205 La Laguna, Tenerife, Spain
\and
INAF -- Osservatorio Astroﬁsico di Arcetri, Largo Enrico Fermi 5, I-50125 Firenze, Italy
\and
INAF -- Osservatorio Astronomico di Padova, Vicolo dell'Osservatorio 5, 35122 Padova, Italy
\and
Space Research Institute, Austrian Academy of Sciences, Schmiedl{\ss}trasse 6, 8042 Graz, Austria
}

\abstract{The atmospheric characterisation of hot and warm Neptune-size exoplanets is challenging mainly due to their relatively small radius and atmospheric scale height, which reduce the amplitude of atmospheric spectral features. The warm-Neptune HAT-P-11\,b is a remarkable target for atmospheric characterisation because of the large brightness of its host star ($V=9.46$\,mag; $H=7.13$\,mag).}
{The aims of this work are to review the main physical and architectural properties of the HAT-P-11 planetary system, and to probe the presence of eight molecular species in the atmosphere of HAT-P-11\,b through near-infrared (NIR) high-resolution transmission spectroscopy.}
{We reviewed the physical and architectural properties of the HAT-P-11 planetary system by analysing transits and occultations of HAT-P-11\,b from the \textit{Kepler} data set as well as HIRES at Keck archival radial-velocity data. We modelled the latter with Gaussian-process regression and a combined quasi-periodic and squared-exponential kernel to account for stellar variations on both (short-term) rotation and (long-term) activity-cycle timescales. In order to probe the atmospheric composition of HAT-P-11\,b, we observed four transits of this target with the NIR GIANO-B at TNG spectrograph and cross-correlated the data with template atmospheric transmission spectra.}
{We find that the long-period radial-velocity signal previously attributed to the HAT-P-11\,c planet ($P \sim 9.3$\,years; $M_{\rm p}\sin{i} \sim 1.6\,M_{\rm J}$; $e \sim 0.6$) is more likely due to the stellar magnetic activity cycle. Nonetheless, the Hipparcos-Gaia difference in the proper-motion anomaly suggests that an outer-bound companion might still exist. For HAT-P-11\,b, we measure a radius of $R_{\rm p}=0.4466\pm0.0059\, R_{\rm J}$, a mass of $M_{\rm p}=0.0787\pm0.0048\, M_{\rm J}$, a bulk density of $\rho_{\rm p} = 1.172\pm0.085$\,g\,cm$^{-3}$, and an orbital eccentricity of $e=0.2577^{+0.0033}_{-0.0025}$. These values are compatible with those from the literature. Probing its atmosphere, we detect the presence of two molecular species, \ch{H2O} and \ch{NH3}, with a S/N of $5.1$ and $5.3$, and a significance of $3.4\,\sigma$ and $5.0\,\sigma$, respectively. We also tentatively detect the presence of \ch{CO2} and \ch{CH4}, with a S/N of $3.0$ and $4.8$, and a significance of $3.2\ \sigma$ and $2.6\ \sigma$, respectively.}
{We revisit the HAT-P-11 planetary system,  confirm the presence of water vapour, and report the detection of \ch{NH3} in the atmosphere of HAT-P-11\,b, also finding hints for the presence of \ch{CO2} and \ch{CH4} that need to be confirmed by further observations.}
\maketitle

\section{Introduction}
\label{introduction}
The majority of known exoplanets have orbital and physical characteristics that are different from those of the Solar System planets. This is the case, for example, for hot and warm Neptunes, which are planets with a mass similar to that of Neptune and with equilibrium temperatures of $T_{\rm eq} \gtrsim 1000$\,K and $T_{\rm eq} \lesssim 1000 $\,K, respectively. The existence of such planets so close to their host stars (orbital periods $P_{\rm orb}\lesssim10$\,days) provides a unique opportunity to study atmospheric physical and chemical conditions that cannot be studied in the Solar System.

The study of exoplanetary atmospheres makes a crucial contribution to the exoplanet characterisation process. For example, by knowing the chemical composition of the primary atmosphere of an exoplanet, it is possible to constrain its formation and evolution path based on the study of relative elemental abundances (e.g. \citealt{oberg2011,madhusudhan2017,madhusudhan2019,Banzatti2020,bitsch2022,pacetti2022}).

By studying exoplanetary atmospheres in the near-infrared (NIR), it is possible to probe deep layers (down to pressures $P\approx0.1$\,bar) where molecular species dominate the atmospheric composition and absorb the IR light through thousands of rotational-vibrational transitions.
Over the past few years, the large number of transiting exoplanet discoveries and the possibility to use space instruments have made low-resolution spectroscopy (LRS) the most used technique for exoplanetary atmospheric characterisation ~\citep{madhusudhan2019}; in particular, for probing the terminator region of planetary atmospheres via the transmission-spectroscopy technique. This method is based on measuring how the planetary effective radius varies with wavelengths during transit because of the absorption of the stellar light by the chemical species present in the atmosphere of the target.

An alternative technique for probing exoplanetary atmospheres using high spectral-resolution instruments is high-resolution spectroscopy (HRS) from ground-based observatories (see \citealt{birkby2018} for a review). For what concerns the NIR studies, different molecular species have been identified in the atmosphere of hot Jupiters with the HRS technique, such as \ch{CO} \citep{snellen2010}, \ch{H2O} \citep{birkby2013}, \ch{TiO} \citep{nugroho2017}, \ch{HCN} \citep{hawker2018}, \ch{CH4} \citep{guilluy2019}, \ch{NH3} and \ch{C2H2} \citep{giacobbe2021}, and \ch{OH} \citep{nugroho2021}. Having improved the data analysis approach (e.g. employing the principal component analysis to remove the telluric and stellar contaminations and performing an optimal selection of the spectral orders), it is now possible to simultaneously detect multiple molecular species in the atmospheres of both hot and warm giant planets (e.g. \citealt{giacobbe2021,guilluy2022,carleo2022}).

With respect to LRS, HRS is more sensitive to line position than depth, allowing a self-calibration of data, and has the advantage of combining the signal of thousands of spectrally resolved absorption or emission lines rather than bands from a chemical species. On the other hand, LRS can take advantage of the two space telescopes currently in operation (HST and JWST), which allow observers to avoid telluric contamination.
A future combination of these two complementary techniques will improve the amount of information that can be inferred about the physics and chemistry of exoplanetary atmospheres~\citep{brogiline}.

The ideal targets for atmospheric studies are close-in giant exoplanets given the higher planet--star radius and flux contrast. In the case of Neptune-size planets, the relatively small radius and the smaller atmospheric scale height combined with the possible presence of clouds or hazes reduce the expected amplitude of the atmospheric absorption, making the atmospheric characterisation of these targets more difficult. Indeed, there are only a few works in the literature reporting the detection of molecular species in the atmosphere of warm and hot-Neptunes (i.e. \citealt{fraine2014, benneke2019, bezard2022, kreidberg2020,brande2022,athano2023,mikalevans2023}), and most of them report the detection of water vapour obtained with LRS using data from \textit{HST/WFC3} (\citealt{athano2023} also report the presence of titanium oxide on HAT-P-26\,b, while \citealt{bezard2022} only report the presence of methane on K2-18\,b).

A remarkable target for atmospheric studies is HAT-P-11\,b, a warm Neptune-size exoplanet (the first of this class of planets discovered with transit searches) orbiting a K4\,V-class star \citep{bakos2010}.
\citet{fraine2014}, \citet{tsiaras2018}, \citet{chachan2019}, and \citet{cubillos2022} detected the presence of water vapour in its atmosphere at low resolution by analysing transmission spectra from HST. With the same data, ~\citet{welbanks2019} estimated an abundance of log($X_{\ch{H2O}}$)$=-3.66^{+0.83}_{-0.57}$. The analyses presented by \citet{chachan2019} (\textit{HST/WFC3+STIS}) and by \citet{cubillos2022} (\textit{HST/WFC3}) also suggest the presence of methane; however, when also considering the \textit{Spitzer} observations, both these analyses found no evidence for the presence of this molecule due to an offset between the \textit{Spitzer} and HST transit depths. The detection of atmospheric absorption in the \ch{He} metastable $1083$ nm triplet during transit \citep{allart2018,mansfield2018} also provided complementary constraints on the size of the planet’s upper atmosphere (extending beyond two planetary radii) and the corresponding mass-loss rate (the planet has only lost up to a few percent of its mass over its history, leaving its bulk composition largely unaffected). Due to the large
brightness of its host star  ($H = 7.131\pm0.021$\,mag, ~\citealt{cutri2003}) and the presence of already detected chemical species (i.e. \ch{He} and \ch{H2O}) in its atmosphere, this target provides a great opportunity to characterise the atmosphere of warm Neptunes. In addition, the orbit of HAT-P-11\,b is eccentric and \citet{sanchis2011} estimated a high obliquity angle ($\psi = 106^{+15}_{-12}$\,deg) between the orbital plane of the  planet and the equatorial plane of its
host star, indicating a quasi-polar orbit. This planet therefore offers a rare occasion to explore complex planetary evolution paths, such as the one that led to its current orbital configuration.

To study the atmosphere of the planet HAT-P-11\,b at high spectral resolution, precise and accurate knowledge of its orbit (in particular the eccentricity $e$ and the planetary argument of periastron $\omega_{\rm p}$) is mandatory. Multiple analyses of the orbital parameters of HAT-P-11\,b can be found in the literature (i.e. \citealt{bakos2010,southworth2011,knutson2014,huber2017,allart2018,yee2018}). The most recent is that of \citet{yee2018}, which is based on the analysis of the radial-velocity (RV) data of HAT-P-11. In particular, these authors obtained the following values for the eccentricity and the planetary argument of periastron: $e = 0.218^{+0.034}_{-0.031}$; $\omega_{\rm p}=199^{+14}_{-16}$\,deg. The most precise estimation of the orbital parameters was reported by \citet{huber2017}. These latter authors simultaneously modelled the planetary transits and secondary eclipses in the {\it Kepler} data, obtaining the following values: $e=0.26459^{+0.00069}_{-0.00048}$; $\omega_{\rm p}=197.774^{+0.203}_{-0.094}$\,deg.
Although these two sets of estimations are compatible with each other at the $2\,\sigma$ level, mainly due to the large uncertainty on the values by \citet{yee2018}, it is important to note that the two values of $e$ differ by about $0.05$, with the analysis of \citet{huber2017} pointing towards a higher value of the eccentricity. Even such a small inaccuracy on the value of $e$ still produces an RV shift of roughly $\approx10$\,km\,s$^{-1}$ (larger than the GIANO-B spectral resolution of $6$\,km\,s$^{-1}$) during the transit, with severe impacts on the atmospheric characterisation using the HRS method. 
To improve the accuracy of the orbital solution, we decided to determine the HAT-P-11\,b orbital parameters through independent analysis of both the \textit{Kepler} photometric data and the RV measurements.
Moreover, by doing so, we reviewed the physical parameters of HAT-P-11\,b and the architecture of the HAT-P-11 planetary system. All the details of this preliminary analysis are described in the following subsections. The main parameters of the HAT-P-11 planetary system are summarised in Table~\ref{tab1}.

In this work, we report a review of the physical and architectural properties of the HAT-P-11 planetary system and the results of the study of four transit events of HAT-P-11\,b recorded with GIANO-B, the high-resolution NIR \'echelle spectrograph mounted at the 3.58\,m Telescopio Nazionale Galileo (TNG), in order to probe the presence of eight molecular species in its atmosphere with the transmission HRS as part of the Global Architecture of Planetary Systems (GAPS) Project\footnote{\url{https://theglobalarchitectureofplanetarysystems.wordpress.com/}} and in particular as part of the exoplanetary atmospheres characterisation subprogramme, described in \citet{guilluy2022}.

\begin{table}
\caption{Main physical and orbital parameters of the HAT-P-11 system.}
\label{tab1}
\resizebox{\hsize}{!}{  
\centering
\begin{tabular}{l l l}
\hline
\hline \\[-8pt]
Parameter\tablefootmark{a} & Value & Reference\tablefootmark{b}\\
\hline \\[-6pt]
\multicolumn{1}{l}{\textbf{Stellar Parameters}} \\ [2pt] %
   Spectral\ Class\dotfill & K4\,V & Ref. 1\\
    $M_{\star}\ $[M$_\odot$]\dotfill& $0.86\pm0.06$ & Ref. 2 \\
    $R_{\star}\ $[R$_\odot$]\dotfill& $0.76\pm0.01$ & Ref. 2\\
    $T_{\textrm{eff}}$ [K]\dotfill& $4780\pm50$ & Ref. 1\\
    $ $[Fe/H]\dotfill& $0.31\pm0.05$ & Ref. 1\\
    $\log{g}$ [log$_{10}$, cgs]\dotfill& $4.563^{+0.092}_{-0.080}$ & Ref. 3\\
    $V_{\textrm{sys}}$ [km\,s$^{-1}$]\dotfill& $-63.24\pm0.26$ & Ref. 4\\
    $ $Distance [ly]\dotfill & $123.17\pm0.11$ & Ref. 4\\
    $ $Age [Gyr]\dotfill& $6.5^{+5.9}_{-4.1}$& Ref. 2\\
    $H$\dotfill& $7.131\pm0.021$& Ref. 5\\ [4pt]
%\hline \\[-6pt]%
\multicolumn{1}{l}{\textbf{Planetary Parameters}} \\ [2pt] %
    $M_\textrm{p}\ $[M$_{\rm J}$]\dotfill& $0.0787\pm0.0048$ & This work\\
    $R_\textrm{p}\ $[R$_{\rm J}$]\dotfill& $0.4466\pm0.0059$ & This work \\
    $M_\textrm{p}\ $[M$_{\oplus}$]\dotfill& $25.0\pm1.5$ & This work\\
    $R_\textrm{p}\ $[R$_{\oplus}$]\dotfill& $4.901\pm0.065$ & This work \\
    $\rho_{\rm p}\ $[g\,cm$^{-3}$]\dotfill& $1.172\pm0.085$ & This work \\
    $T_\textrm{eq}$ [K]\dotfill& $847^{+46}_{-54}\ (663^{+36}_{-42})$ & This work \\ [2pt]
    $P_{\rm orb}$ [days]\dotfill& $4.887802443\pm0.000000034$ & Ref. 6 \\ [2pt]
    $T_\textrm{0}$ [BJD$_{\textrm{TDB}}$] \dotfill & $2\,454\,957.8132067\pm0.0000053$ & Ref. 6 \\ [2pt]
    $i$ [deg]\dotfill& $89.027\pm0.068$ & This work \\ [2pt]
    $e$ \dotfill& $0.2577^{+0.0033}_{-0.0025}$ & This work \\  [2pt]
    $\omega_\textrm{p}$ [deg]\dotfill& $192.0^{+2.9}_{-3.0}$ & This work \\  [2pt]
    $a$ [au]\dotfill& $0.0532\pm0.0010$ & This work\\
    $K_\textrm{p}$ [km\,s$^{-1}$]\dotfill& $123.4\pm9.9$ & This work \\  [2pt]
\hline                                            
\end{tabular}
}
\tablefoot{
        \tablefoottext{a}{The symbols of the parameters listed in the table have the following meanings: $M_\star$ - stellar mass; $R_\star$ - stellar radius; $T_{\rm eff}$ - stellar effective temperature; [Fe/H] - stellar metallicity; $\log{g}$ - logarithm of stellar surface gravity; $V_{\rm sys}$ - systemic radial velocity; $H$ - apparent magnitude in the photometric $H$ band; $M_{\rm p}$ - planetary mass; $R_{\rm p}$ - planetary radius; $\rho_{\rm p}$ - planetary mean density; $T_{\rm eq}$ - planetary equilibrium temperature computed at the planet-to-star distance during the secondary eclipse, assuming an inefficient (full) heat re-distribution for the derived geometrical albedo ($A_{\rm g}=0.440^{+0.044}_{-0.049}$); $P_{\rm orb}$ - orbital period; $T_0$ - transit epoch; $i$ - orbital inclination; $e$ - orbital eccentricity; $\omega_{\rm p}$ - planetary argument of periastron; $a$ - orbital semi-major axis; $K_{\rm p}$ - planetary radial-velocity semi-amplitude.}\\
        \tablefoottext{b}{The references of the values in the table are: 1.\,\citet{bakos2010}; 2.\,\citet{lundkvist2016}; 3.\,\citet{tic2019}; 4.\,\citet{gaia2018}; 5.\,\citet{cutri2003}; 6.\,\citet{huber2017}.}
}
\end{table}

The paper is organised as follows: in Sect.~\ref{revisitation} we review the HAT-P-11 planetary system; in Sect.~\ref{atmchar} we describe the high-resolution transit observations and data analysis process  in detail, and discuss the results of the atmospheric characterisation. Finally, our conclusions and future perspectives are reported in Sect.~\ref{conclusion}.

\section{Revisitation of the HAT-P-11 planetary system}
\label{revisitation}
\subsection{\textit{Kepler} light-curve data analysis}
\label{Kepler_Data_Analysis}
We downloaded the \textit{Kepler} light curves of HAT-P-11\,b from the Mikulski Archive for Space Telescopes (MAST\footnote{\url{https://archive.stsci.edu/missions-and-data/kepler}}). These consist of short-cadence (60~s) light curves from 14 quarters out of the total 17 observed from 2 May 2009 to 11 May 2013. The short-cadence light curves of each quarter are subdivided into tranches of three except for quarters 0 and 1 with a single and quarter 17 with two light curves. This results in a total of 37 separate light curves to be analysed independently. 

\subsubsection{Transits and occultations}
\label{trocc}
We trimmed the \textit{Kepler} light curves around the transits and occultations, whose central times are predicted using the ephemeris of \citet{huber2017}. The width of each time interval is equal to three times the transit duration $T_{14}$ and is large enough to include the time of the eclipses for both the orbital solutions discussed above. We discard all the transits and occultations that, due to gaps in the \textit{Kepler} light curve, do not cover the whole time interval of $3\,T_{14}$.

Many transits of HAT-P-11\,b show clear signs of starspot crossings (see, e.g. \citealt{sanchis2011,beky2014,morris2017,scandariato2017} for a detailed analysis of the transit anomalies of HAT-P-11\,b), which may bias the fit of the light curve. For this reason, we adopted an iterative approach aimed at selecting the transits with minimum evidence of anomalies. Using the orbital period derived by \citet{huber2017}, we phase-folded the planetary transits and we processed the phase-folded data using a running median average. The width of the running window was 15~s, which is less than the cadence of the light curves. This guarantees that the averaged transit profile is negligibly time-smoothed by our approach.

Once the average transit profile was obtained, we computed the Median Absolute Deviation (MAD) of the data with respect to it. We then rejected all the transits with at least one data point located more than 6\,MAD above the averaged transit profile. We iteratively repeated this process until no additional transit was rejected. At the end of this process, we noticed that a few transits passed our selection despite showing correlated noise due to either bad data detrending or stellar activity. We therefore refined the selection of the transits in the following way. For each transit, we first computed the residuals with respect to the averaged profile. Then, we associated each transit with the standard deviation of the corresponding residuals. Finally, we rejected the 10\% transits with the largest standard deviations. This produced a final list of 64 bona fide transits free from anomalies above the noise level. Of course, the possible presence of non-crossed starspots can influence the stellar flux level and therefore the measured transit depth; however, this effect mainly affects the value of the planetary radius (with an over-estimation of $\approx1\,\%$\footnote{This value is computed using Eq.~(12) in ~\citet{ballerini2012} with a maximum out-of-transit HAT-P-11 flux modulation of 2\% (see the top panel of Fig.~1 in ~\citealt{beky2014}).}) rather than the orbital parameters and is therefore negligible for the main scope of this work.

\begin{figure}
    \centering
    \includegraphics[width=\linewidth]{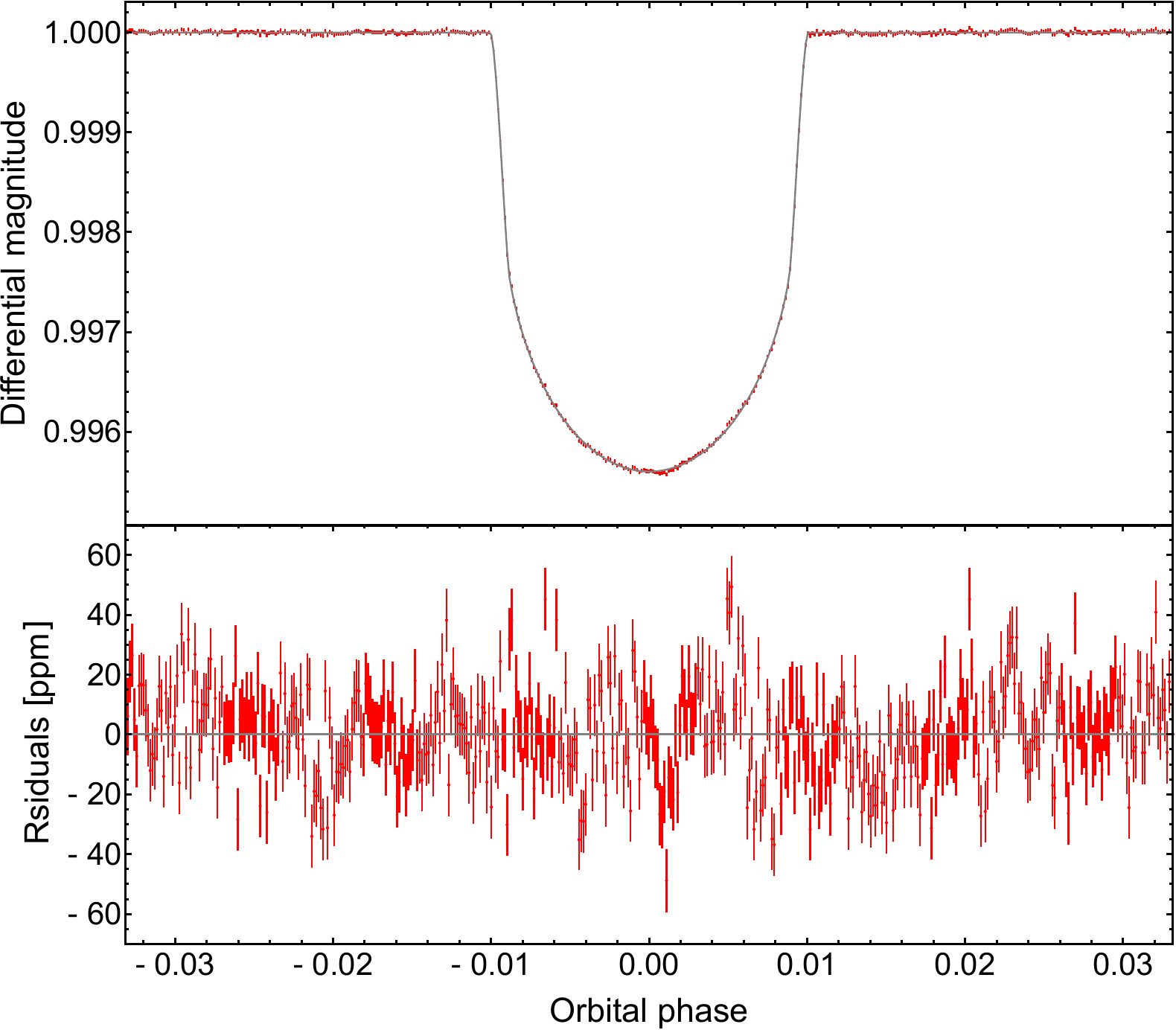}
    \caption{Transit light curve analysis. Top panel: Phase-folded {\it Kepler} transit light curve of HAT-P-11\,b binned by 1 minute, based on 64 bona fide transits (see text). Our best-fitting model of the transit in the Kepler bandpass is overplotted in grey. Bottom panel: Residuals of the fit.}
    \label{fig:trans}
\end{figure}
\begin{figure}
    \centering
    \includegraphics[width=\linewidth]{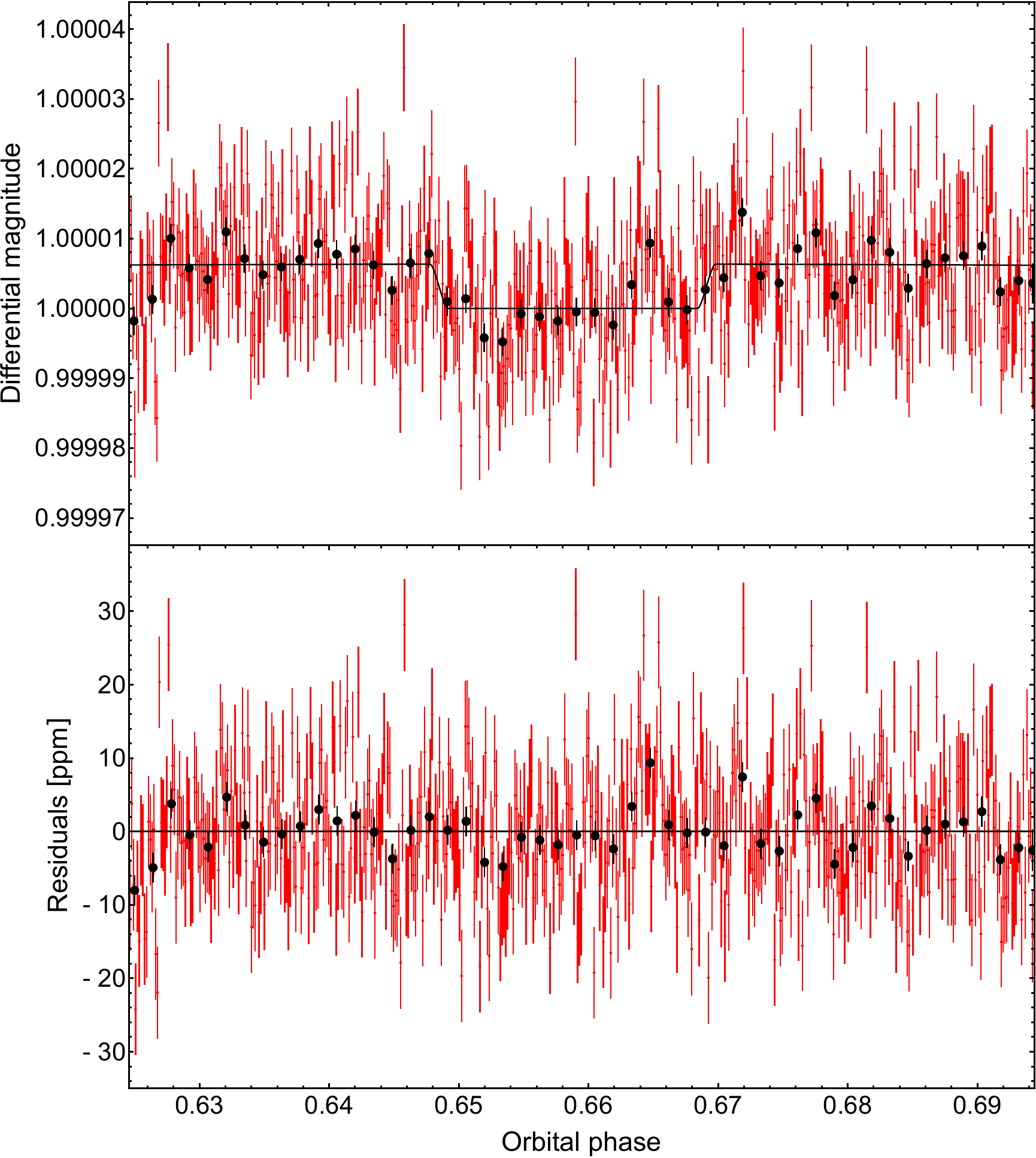}
    \caption{Occultation light curve analysis. Top panel: Phase-folded {\it Kepler} occultation light curve of HAT-P-11\,b binned by 1 minute, based on 196 occultations (see text). Our best-fitting model of the occultation in the Kepler bandpass is overplotted with a black solid line. The data have been binned for clarity (black dots with error bars). Bottom panel: Residuals of the fit.}
    \label{fig:ecl}
\end{figure}

For the occultations, we adopted a similar approach. The only difference is that we do not expect anomalies during the eclipses, and therefore the first iterative selection of ``unspotted'' light curves is skipped. The final selection is made up of 196 occultations.

In summary, the dataset that we used to fit the orbit of the planet is composed of two subsets of data, centred on the transits and occultations respectively. To save computation time, we fit the phase-folded and rebinned data (1\,min). The timestamps of the rebinned light curve were defined so that the time of transit provided by \citet{huber2017} corresponds to the origin of the time axis.

The transit profile was modelled with the quadratic limb darkening (LD) law provided by \citet{Mandel2002}, with the reparametrisation of the coefficients proposed by \citet{Kipping2013}. Similarly, the occultation profile was modelled following \citet{Mandel2002}, but assuming that the planetary dayside is uniformly bright \citep{singh2022,scandariato2022}. Since we have operated a different data rejection with respect to \citet{huber2017}, we re-derived all the orbital parameters except for the orbital period, which we fixed to the best estimate of \citet{huber2017} to phase-fold the data. The free parameters of the model are the stellar density $\rho_\star$, the time of transit $T_0$, the planet-to-star radius ratio $R_{\rm p}/R_\star$, the impact parameter $b$, the LD coefficients $q_1$ and $q_2$, the $e_{\rm c}=e\cos{\omega_\star}$ and $e_{\rm s}=e\sin{\omega_\star}$ parameters (where $e$ and $\omega_\star$ are the orbital eccentricity and the stellar argument of periastron, respectively) and the occultation depth $\delta_{\rm ecl}$.

In the model, we also included a jitter term and a renormalisation coefficient independently for the transit and eclipse subcurves. The two jitter terms take into account the fact that the two subcurves have different noise properties, being the combination of a different number of transits/eclipses. The renormalisation coefficients fix the preliminary normalisation of the transit and eclipse subcurves computed by the extraction pipeline.

We adopted a maximum-likelihood Bayesian approach where the data were fitted by running a Monte Carlo sampling of the parameter space. The parameter space was defined by the priors listed in Table~\ref{tab:parameters_kepler}. We remark that we have used uniform priors for all the parameters. The priors on $e_{\rm c}$ and $e_{\rm s}$ have been conveniently set to span a large range around the expected values and include both the orbital solutions of \citet{huber2017} and \citet{yee2018}, while saving computation time.

For the log-likelihood maximisation, we first searched the parameter space for the global maximum position using the python package \texttt{PyDE}\footnote{\url{https://github.com/hpparvi/PyDE}}. Afterwards, we sampled the posterior probability distribution of the model parameters using the \texttt{emcee} package version 3.1.3 \citep{Foreman2013}. Given the demand for resources for the model fitting, we ran the code in the HOTCAT computing infrastructure \citep{Bertocco2020,Taffoni2020}. We let the chains run for 250\,000 steps, long enough to ensure formal convergence. The best-fitting model of the light curve, together with the corresponding residuals, is shown separately in Fig.~\ref{fig:trans} and Fig.~\ref{fig:ecl}, for the transit and the occultation, respectively. The list of the free parameters and their corresponding priors and best-fitting $1\, \sigma$ confidence interval is given in Table~\ref{tab:parameters_kepler}. Our estimates are consistent with previous analyses within 2$\sigma$.

\begin{table*}[!h]
\begin{center}
\renewcommand{\arraystretch}{1.5} % Default value: 1
\caption{Model parameters for the fit of the \textit{Kepler} data. \textit{U} stands for a uniform prior.}\label{tab:parameters_kepler}
\begin{tabular}{lllcl}
\hline\hline
Parameters & Symbol & Units & C.I. & Prior\\
\hline
Stellar density & $\rho_\star$ & $\rho_\sun$  & 1.915$\pm0.081$ & U(1.5,2.5)\\
Time of transit & $T_0$ & days & 0.0000967$^{+0.0000084}_{-0.0000082}$ & U(-0.05,0.05)\\
Planet-to-star radius ratio & $R_{\rm p}/R_\star$ & - & 0.058993$^{+0.000065}_{-0.000070}$ & U(0.05,0.07)\\
Impact parameter & $b$ & - & 0.227$^{+0.013}_{-0.015}$ & U(0,0.6)\\
First LD coefficient & $q_1$ & - & 0.4644$^{+0.0062}_{-0.0063}$ & U(0,1)\\
Second LD coefficient & $q_2$ & - & 0.4813$^{+0.0064}_{-0.0062}$ & U(0,1)\\
 & $\sqrt{2e}\cos\omega_\star$ & - & 0.7021$^{+0.0035}_{-0.0041}$ & U(0.5,0.9)\\
 & $\sqrt{2e}\sin\omega_\star$ & - & 0.150$\pm0.037$ & U(0,0.5)\\
Secondary eclipse depth & $\delta_{\rm ecl}$ & ppm & 6.95$^{+0.66}_{-0.64}$ & U(0,20)\\
\hline
\hline
Derived parameters & Symbol & Units & C.I. & \\
\hline
Planetary radius\tablefootmark{a} & $R_{\rm p}$ & R$_{\rm J}$ & $0.4466\pm0.0059$\\
Scaled semi-major axis & $a/R_\star$ & - & 15.05$^{+0.21}_{-0.22}$\\
Orbital inclination & $i$ & deg & 89.027$\pm0.068$\\
Transit duration & $T_{14}$ & hr & 2.35562$^{+0.00093}_{-0.00094}$\\
Orbital eccentricity & $e$ & - & 0.2577$^{+0.0033}_{-0.0025}$\\
Stellar argument of periastron & $\omega_\star$ & degree & 12.0$^{+2.9}_{-3.0}$\\
\hline
\end{tabular}
\tablefoot{
        \tablefoottext{a}{The uncertainty includes the uncertainty on the stellar radius.}
}
\end{center}
\end{table*}

~\\
~\\
\subsubsection{Albedo and equilibrium temperature}
\label{albedo}
The day-side flux of an exoplanet is a combination of reflected starlight off the planet's illuminated hemisphere and its thermal irradiation. The former is parameterised by the geometric albedo $A_{\rm g}$ while the latter is parameterised by the brightness temperature $T_{\rm d}(\Delta\lambda)$, which is a measure of the day-side temperature in a given wavelength interval $\Delta\lambda$ \citep{Santerne2011A&A,singh2022}. Consequently, the occultation depth can be expressed as the following:
\begin{equation}
    \delta_{\rm ecl} = A_{\rm g}\left(\frac{R_{\rm p}}{d_{\rm sec}}\right)^{2} + 
    \pi\left(\frac{R_{\rm p}}{R_{\star}}\right)^{2}  \frac{ \int_{\Delta\lambda} \frac{2hc^{2}}{\lambda^{5}} \left[ \exp \left( \frac{hc}{k_{\rm B}\lambda T_{\rm d}}\right) - 1 \right]^{-1}   \Omega_{\lambda} d\lambda}{\int_{\Delta\lambda} S_{\lambda}^{\rm CK} \Omega_{\lambda} d\lambda}\,,
\end{equation}
where $h$ is the Planck constant, $k_{\rm B}$ the Boltzmann constant, $c$ the speed of light, $d_{\rm sec}$ the distance of the planet from the star during the secondary eclipse and $S_{\lambda}^{\rm CK}$ is the stellar Kurucz flux \citep{Castelli&Kurucz2003IAUS} (computed for $T_{\rm eff}=4750$\,K, $\log{g}=4.5$, and [Fe/H]$=0.2$). Both the planetary and the stellar fluxes are integrated over the \textit{Kepler} passband $\Delta\lambda$ with the corresponding response function  $\Omega_{\lambda}$.

\begin{figure}[h!]
\centering
\includegraphics[width=\linewidth]{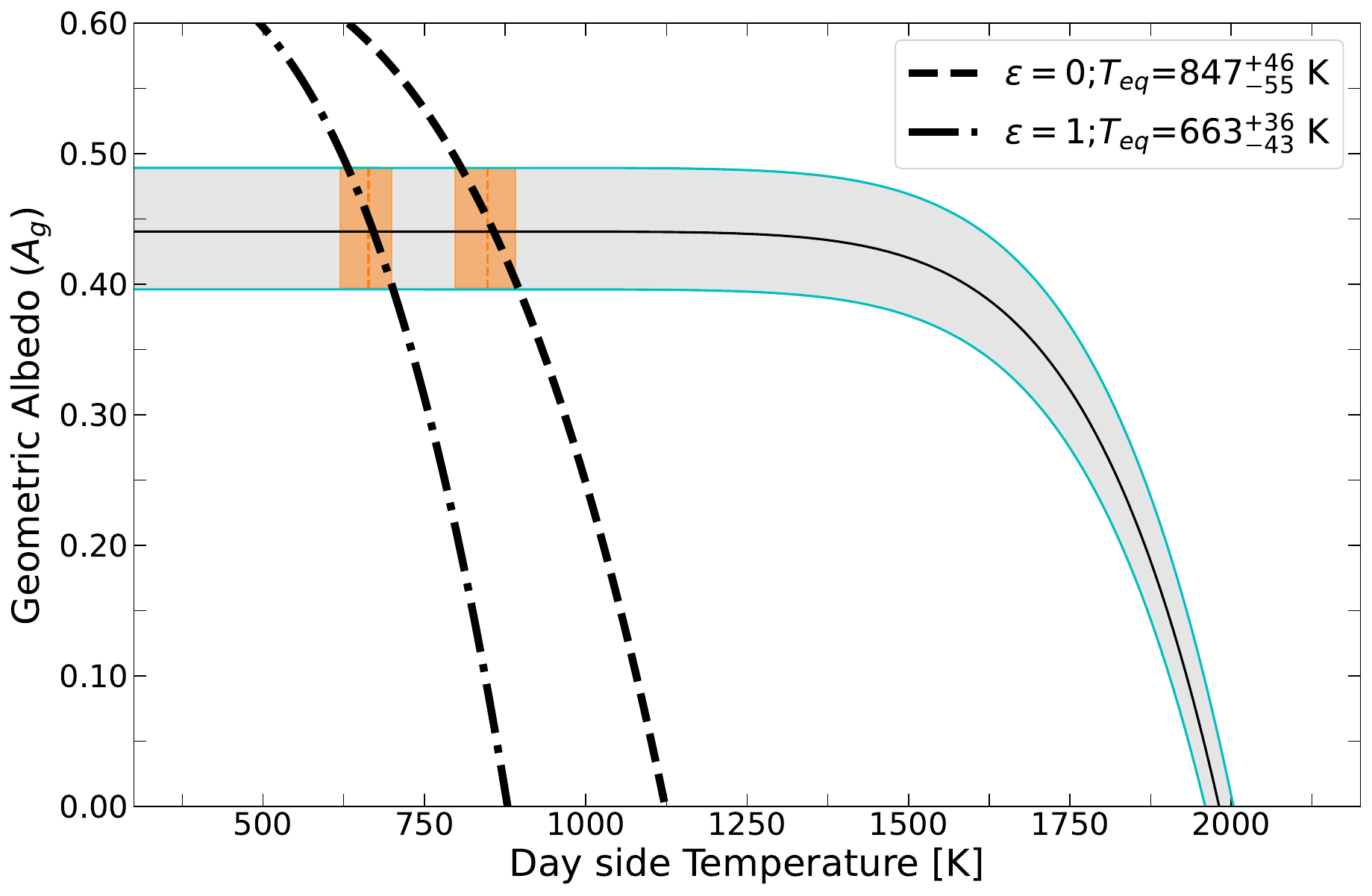}
\caption{Geometric albedo ($A_{\rm g}$) estimated as a function of the day-side temperature for the measured occultation depth ($\delta_{\rm ecl}=6.95^{+0.66}_{-0.64}$\,ppm). The cyan lines represent the $1\,\sigma$ uncertainty curves. The dash-dotted and dashed lines represent the variation of $A_{\rm g}$ with varying $T_{\rm d}$ for the 2 heat re-circulation cases ($\epsilon$) considered, computed at $d = d_{\rm sec}=14.83\pm0.30\,R_\star$ (that is, during the occultation). The 2 orange shaded regions correspond to the intersection of the curves, identifying the average HAT-P-11\,b day-side temperatures for the 2 extreme scenarios of $\epsilon$, computed at $d=d_{\rm sec}$. These 2 values of the equilibrium temperature with the associated uncertainties are reported in the legend.}
\label{fig: Ag_Tday}
\end{figure}

For the derived occultation depth, the relationship between the geometric albedo and the brightness temperature is shown in Fig.~\ref{fig: Ag_Tday}. For HAT-P-11\,b, the thermal contribution to the observed depth in optical passbands is practically negligible given the low temperature of its atmosphere. Therefore, the occultation depth is the result of a highly reflective atmosphere. We derive a geometric albedo of $A_{\rm g}=0.440^{+0.044}_{-0.049}$ corresponding to the occultation depth of $\delta_{\rm ecl}=6.95^{+0.66}_{-0.64}$\,ppm reported in Table~\ref{tab:parameters_kepler}. We therefore confirm and improve the results obtained by \citet{huber2017}.  Following \citet{han2014}, we assume that the geometric albedo and Bond albedo are related via $A_{\rm b} = \frac{3}{2} A_{\rm g}$. We use this $A_{\rm b}$ to estimate the planet's day-side equilibrium temperature\footnote{We equate the optical brightness temperature with the planet's equilibrium temperature, which is reasonable if the planetary emission spectrum closely resembles a Planck distribution at temperature $T_{\rm d}$.} as a function of time following \citep{CowanAgol2011ApJ}:
\begin{equation}\label{eqn:Dayside}
T_{\rm d}(t) = T_{\rm eff} \sqrt{\frac{R_{\star}}{d(t)}} \left(1 - A_{\rm b}\right)^{\frac{1}{4}} \left(\frac{2}{3} - \frac{5}{12}\epsilon \right)^\frac{1}{4}\,,
\end{equation}
where $\epsilon$ is the heat re-circulation efficiency and $d(t)$ the distance of the planet from the star as a function of the time (to take into account the eccentricity of the orbit). 

We considered two extreme scenarios: one with inefficient ($\epsilon=0$) and another with complete ($\epsilon = 1$) heat re-circulation. We report in the plot the equilibrium temperature estimates at the occultation position. As a result of the varying stellar irradiation, the temperature estimates at the periastron are: $970^{+52}_{-62}$~K and $759^{+41}_{-49}$~K for $\epsilon=0$ and $\epsilon=1$, respectively, and at the apoastron are $750^{+41}_{-47}$~K and $587^{+32}_{-37}$~K for $\epsilon=0$ and $\epsilon=1$, respectively. During the transit, the planet-to-star separation distance is $d_{\rm tr}=13.33\pm0.26$\,$R_{\star}$ and therefore, the corresponding day-side temperatures are $894^{+48}_{-57}$~K and $699^{+38}_{-44}$~K assuming no changes in planetary albedo throughout the eccentric orbit. The pairs of $T_{\rm d}$ values we report represent the extremes of the range in which the planet's equilibrium temperature is expected to be at different planet positions along its orbit. At the transit position, we consider the scenario $\epsilon=1$, that is, a uniform temperature distribution throughout the planet, so that we can use $T_{\rm eq}=699^{+38}_{-44}$~K as the equilibrium temperature around the day-night terminator to build our atmospheric transmission spectrum models described in Sect.~\ref{pl_extr}.

\subsection{Radial-velocity data analysis}
\label{RV_Data_Analysis}
We also analysed 180 publicly available radial velocities of HAT-P-11, which were obtained with the HIRES at Keck spectrograph by \citet{bakos2010} and \citet{yee2018}, after discarding the in-transit measurements to avoid the Rossiter-McLaughlin effect ~\citep{ross1924,mcl1924} and three outliers at the observing epochs  4334.9662, 4957.0433 and 7933.0122
$\rm BJD_{TDB}-2450000$, which were identified in the residuals of our RV model (Sect.~\ref{RV_modeling}) through the Chauvenet's criterion
(e.g. \citealt{Bonomoetal2023}).

\subsubsection{Planet c or magnetic activity cycle?}
\begin{figure}
\centering
\includegraphics[width=\linewidth]{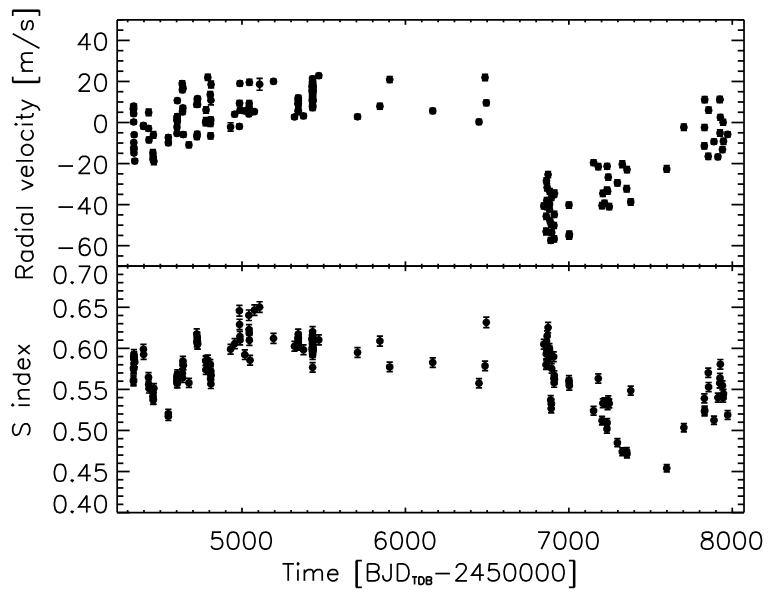}
\caption{HIRES RV (top panel) and CaII S-index (bottom panel) measurements of HAT-P-11. The two-time series show almost identical long-term variations with a shift of $\sim 500$~days in the minimum.
}
\label{figure_RV_Sindex_HAT-P-11}
\end{figure}

The HIRES RVs show a clear long-term trend (Fig.~\ref{figure_RV_HAT-P-11}, left panel), which was attributed by \citet{yee2018} to 
a second planet companion, HAT-P-11\,c, with $P_{\rm orb}\sim 3410$~d (9.3~yr), $M_{\rm p}\sin{i} \approx 1.6 \, M_{\rm J}$, and $e\approx0.6$,  
and, to a lesser extent, to the stellar magnetic activity cycle. 
Despite the very similar behaviour of the trend in both the S-index and the RVs (see Fig.~\ref{figure_RV_Sindex_HAT-P-11}), 
the stellar activity cycle was not deemed sufficient by \citet{yee2018} to account for the RV variations in the long term for three main reasons (see their Sect.~3 for more details): 
$(i$) the large RV semi-amplitude ($\sim 30$~\,m\,s$^{-1}$) of the long-term signal compared to semi-amplitudes of $\lesssim 10$\,m\,s$^{-1}$ observed by \citet{2011arXiv1107.5325L} for magnetic activity cycles in $\sim 300$ solar-type stars; 
$(ii$) the presence of a shift of $\sim 500$~days between the minimum of the S-index and that of the RVs (Fig.~\ref{figure_RV_Sindex_HAT-P-11}); 
and 
$(iii$) the relatively weak correlation between the S-index and RV measurements with a Pearson's coefficient of $\sim 0.34$. 

In our view, these three motivations do not provide strong evidence that the long-term signal is planetary in origin. 
Indeed, concerning $(i$), HAT-P-11 is considerably more active than the stars in the sample studied by \citet{2011arXiv1107.5325L},  
with a $\log{R^{'}_{\rm HK}}$ of $-4.35$ \citep{2017ApJ...848...58M} higher than the typical $\log{R^{'}_{\rm HK}}$ of $-4.8$~--~$-5.0$ of that stellar sample. Since one of those stars, namely HD\,21693, shows a semi-amplitude of $\sim 10$~\,m\,s$^{-1}$ for $\log{R^{'}_{\rm HK}}=-4.89$ (see Fig.~16 in \citealt{2011arXiv1107.5325L}), a semi-amplitude of $\sim 30$~\,m\,s$^{-1}$ for the RV variation associated with the activity cycle is certainly possible for an unusually active star such as HAT-P-11 \citep{2017ApJ...848...58M}.

Regarding $(ii$), detailed studies of the correlation between RV and S-index measurements by \citet{2019A&A...632A..81M} (therefore subsequent to \citealt{yee2018}) showed that a combination of geometrical effects (stellar inclinations and butterfly diagrams) and variations of magnetic activity level over time may easily produce hysteresis patterns, and hence temporal shifts of a few hundreds of days in the minima of the RV and S-index variations (see Fig.~8 in \citealt{2019A&A...632A..81M}). For example, the minimum of the long-term RV variations of the above-mentioned star HD\,21693, which are caused by the magnetic activity cycle, also leads the minimum of the $\log{R^{'}_{\rm HK}}$ variations by $\sim500$~d, similarly to HAT-P-11 \citep{2011arXiv1107.5325L, 2019A&A...632A..81M}. Given the high inclination $i_\star=100\pm2$\,deg of the host star HAT-P-11 and a starspot latitudinal distribution similar to the solar butterfly diagrams, as unveiled from the occultations of starspots by HAT-P-11\,b during transits \citep{morris2017}, the temporal difference in the minima of the RV and S-index variations could be due to the hysteresis patterns described by \citet{2019A&A...632A..81M}. After all, the fact that the magnetic activity cycle from the S-index was found to have the same periodicity as the hypothetical planet c \citep{2017ApJ...848...58M} remains suspicious.

\begin{figure}[t!]
\vspace{-1.5cm}
\hspace{-1.0cm}
\includegraphics[width=1.1\columnwidth]{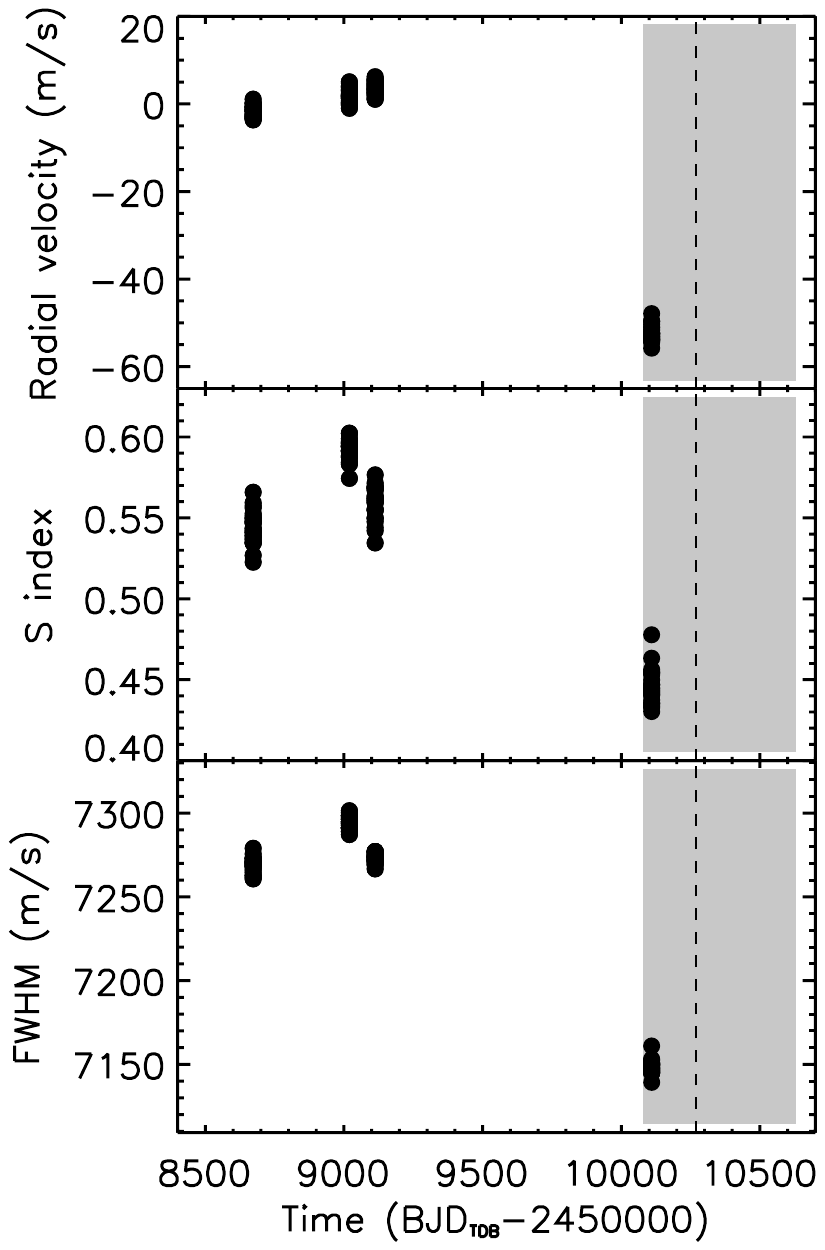}
\vspace{-1.75cm}
\caption{HARPS-N RV (top panel), CaII S-index (middle panel), and FWHM (bottom panel) measurements of HAT-P-11 during the four HAT-P-11\,b transit nights for atmospheric characterisation. The three time series are highly correlated showing the same long-term trend, with no apparent shifts in the minima of variations. The vertical dashed line indicates the predicted periastron time of the hypothetical planet c, and the grey area shows its $1\sigma$ error bar accounting for the uncertainty on the orbital period from \citet{yee2018}.
Note: the HARPS-N radial velocities were divided by their median of -63420.4~\,m\,s$^{-1}$, to make a straightforward comparison with the HIRES radial velocities in Fig.~\ref{figure_RV_Sindex_HAT-P-11}.}
\label{figure_RV_Sindex_FWHM_HARPS-N_HAT-P-11}
\end{figure}

Last but not least, even the absence of a strong correlation between the S-index and RV measurements does not necessarily lean towards the planetary origin of the RV long-term signal; this is because a relatively low correlation, at least partly, ensues from the temporal shift between the S-index and RV variations, while it is much higher in the first $\sim1000$~d of observations \citep{2014ApJ...785..126K}. Since we observed HAT-P-11 in GIARPS mode, we extracted the HAT-P-11 RVs and activity indicators from the HARPS-N spectra of the four transits of HAT-P-11\,b to look at their behaviour. For that purpose, we used the online v3.7 data reduction software and cross-correlated the HARPS-N spectra with a K5\,V synthetic stellar template \citep{2002A&A...388..632P}. The variations of the HARPS-N RVs, S-index and full width at half maximum (FWHM) of the cross-correlation function show an almost identical behaviour (Fig.~\ref{figure_RV_Sindex_FWHM_HARPS-N_HAT-P-11}): the Pearson's correlation coefficient is 0.94 between RVs and S-index, and 0.98 between RVs and FWHM. This suggests that the $\sim 9-10$~yr long-term RV signal is more likely due to the stellar activity cycle than to the long-period eccentric companion HAT-P-11\,c. Moreover, if the RV measurements of the last transit night actually caught the minimum of the activity cycle, given that the HARPS-N RV peak-to-peak variation of $\sim 60$~\,m\,s$^{-1}$ is the same as that observed by \citet{yee2018}, temporal shifts between the RV and S-index minima as caused by hysteresis phenomena may not have occurred in the current activity cycle. 

\begin{figure}[t!]
\centering
\includegraphics[width=0.95\columnwidth]{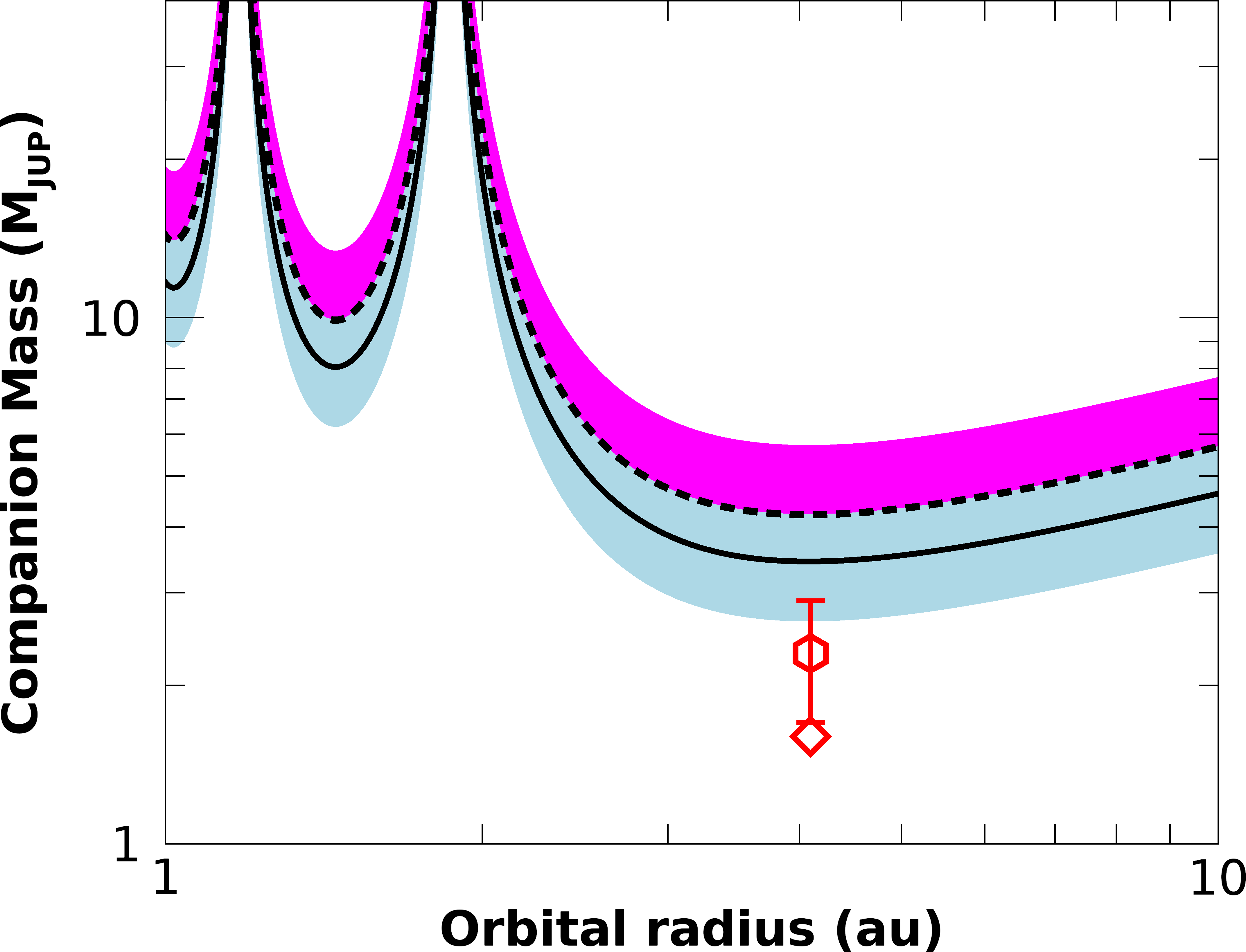}
\caption{Sensitivity of the PMA technique to companions of given mass and orbital separation orbiting HAT-P-11. The black long-dashed and solid curves correspond to the combinations of mass and orbital radius explaining the PMA values at the mean Gaia DR2 and DR3 epochs, respectively. The shaded light blue region corresponds to the $1\sigma$ uncertainty domain of the DR3 PMA, while the shaded magenta region encompasses the $1\ \sigma$ uncertainty of the DR2 PMA. The red diamond indicates the separation and minimum mass of the HAT-P-11\,c companion proposed by \citet{yee2018}, while the red hexagon corresponds to the true mass value determined by the \citet{Xuan2020} analysis.}
\label{fig_PMA}
\end{figure}

Even though our data suggest that the magnetic activity cycle is the most plausible origin of the long-term RV signal, Hipparcos-Gaia absolute astrometry still provides hints that a long-period companion may actually exist. The catalogues of astrometric accelerations produced by \citet{Brandt2018,Brandt2021} and \citet{Kervella2019,Kervella2022} indicate the presence of a proper motion anomaly (PMA) at the mean Gaia epoch, whose signal-to-noise ratio (S/N) grows from S/N$\simeq2.1$ to S/N$\simeq4.7$ and from S/N$\simeq2.8$ to S/N$\simeq4.3$ between the Gaia DR2 and Gaia EDR3 editions of the former and latter catalogue, respectively. Indeed, \citet{Xuan2020} used the Hipparcos-Gaia DR2 PMA values in combination with the \citet{bakos2010} RVs of HAT-P-11 to constrain the true mass and inclination of the putative companion HAT-P-11\,c. Fig. \ref{fig_PMA} shows the Hipparcos-Gaia DR2 and DR3 PMA sensitivity curves based on Eq.~(15) of \citet{Kervella2019}, along with the minimum-mass value of HAT-P-11\,c derived by \citet{yee2018} and the best-fit true mass obtained by \citet{Xuan2020}.
The Hipparcos-Gaia DR3 PMA sensitivity curve indicates that, at the orbital separation of HAT-P-11\,c, a companion inducing a statistically significant PMA should have a mass of $\sim3.5$ $M_\mathrm{Jup}$. In the \citet{Xuan2020} analysis, the true mass value of HAT-P-11\,c falls below the PMA sensitivity curve, with a companion having true mass equal to the minimum mass from \citet{yee2018} compatible at $\sim1.4\,\sigma$. This is somewhat surprising as such a companion is not expected to produce a PMA with S/N$\gtrsim3$. As the PMA technique heavily relies on the constraints from the RVs in order to successfully provide inferences on the mass and inclination of a companion, it is therefore possible that, if the long-term modulation in the RVs is actually dominated by the activity cycle, then HAT-P-11\,c exists at larger separation and with a different mass than those inferred by \citet{Xuan2020}. 

\subsubsection{Radial-velocity modelling and improved mass determination for HAT-P-11\,b}
\label{RV_modeling}
In the lack of strong evidence that the RV long-term trend is caused by the planet c with the orbital parameters given in \citet{yee2018} for the reasons explained above, we modelled the HIRES RVs with a Keplerian orbit for the transiting planet HAT-P-11\,b only, which has six free parameters: $T_{0}$, $P$, $e_{c}$, $e_{s}$, the RV semi-amplitude, $K_\star$, and the RV zero point, $\gamma$. 

To account for non-stationary stellar variations produced by magnetic activity phenomena, we used Gaussian-process (GP) regression (e.g. \citealt{2014MNRAS.443.2517H, 2015PhDT.......193H, 2015ApJ...808..127G}) 
with three different kernels, namely the squared-exponential (SE) kernel:
\begin{equation} %\notag
\label{equation_kernel_SE}
k(t, t^{\prime}) = h^{2} \cdot \exp{\left[ - \frac{(t-t^{\prime})^2}{2\lambda^2}\right]} 
+ \left[\sigma_{\rm RV}^2(t) + \sigma_{\rm jit}^2 \right] \cdot \delta_{t, t^{\prime}}, 
\end{equation}

\noindent
the quasi-periodic (QP) kernel
\begin{equation} 
\label{equation_kernel_QP}
k(t, t') = h^{2} \cdot \exp{\left[ - \frac{(t-t')^2}{2\lambda^2} - \frac{2 \sin^2{\left(\frac{\pi (t-t')}{P_{\rm rot}} \right)}}{w^2} \right]} 
+ \left[\sigma_{\rm RV}^2(t) + \sigma_{\rm jit}^2 \right] \cdot \delta_{\rm t, t'}
,\end{equation}

\noindent
and a third (QP-SE) kernel that is the sum of the QP and SE kernels in such a way as to simultaneously model 
the stellar activity variations on both short-term rotation timescales (Eq.~\ref{equation_kernel_QP}) 
and long-term activity cycle timescales (Eq.~\ref{equation_kernel_SE}), namely

\begin{equation} 
\label{equation_kernel_QPSE}\begin{split}
k(t, t') = h_{\rm rot}^{2} \cdot \exp{\left[ - \frac{(t-t')^2}{2\lambda_{\rm rot}^2} - \frac{2 \sin^2{\left(\frac{\pi (t-t')}{P_{\rm rot}} \right)}}{w_{\rm rot}^2} \right]}\\
+ h_{\rm cycle}^{2} \cdot \exp{\left[ - \frac{(t-t^{\prime})^2}{2\lambda_{\rm cycle}^2}\right]} 
+ \left[\sigma_{\rm RV}^2(t) + \sigma_{\rm jit}^2 \right] \cdot \delta_{\rm t, t'}
,\end{split}\end{equation}

\noindent
where $t$ and $t^{\prime}$ are the epochs at two different RV observations, $h$ is the semi-amplitude of the correlated noise, $\lambda$ is the correlation decay timescale, $P_{\rm rot}$ is the period of the quasi-periodic variations, $w$ is the inverse complexity harmonic parameter, $\sigma_{\rm RV}(t)$ is the formal uncertainty of the RV point at time $t$, and $\sigma_{\rm jit}$ is the uncorrelated jitter term, which would absorb any extra white noise not modelled by either the Keplerian or the GP.  

We employed Bayesian differential evolution Markov chain Monte Carlo (DE-MCMC; \citealt{TerBraak2006, 2013PASP..125...83E, 2015A&A...575A..85B}) techniques to derive the posterior distributions of the model parameters, by using the same prescriptions for the number and convergence of the DE-MCMC chains given
in \citet{2013PASP..125...83E} and \citet{2006ApJ...642..505F}. We imposed Gaussian priors on $T_{0}$ and $P$ from the \emph{Kepler} transit ephemeris and uniform priors on $K_\star$ and $\gamma$ as well as on the GP hyper-parameters $h$, $\lambda$, $P_{\rm rot}$, and $w$, with the boundaries specified in Table~\ref{TableRVparam} for each GP kernel. As for the eccentricity and stellar argument of periastron, we ran two analyses per kernel: in the first one, we used uninformative priors on $e$ and $\omega_{\star}$, while in the second one we adopted Gaussian priors from the results of the modelling of the optical secondary eclipse (Sect.~\ref{Kepler_Data_Analysis}). We took the medians and the 15.87\%-84.14\% quantiles of the posterior distributions as the values and $1\, \sigma$ uncertainties of the fitted and derived parameters.

In the first analysis with uniform priors on $e$ and $\omega$, we determined 
$e=0.277\pm0.025$ and $\omega_\star=26.8\pm 8.6$ with the SE kernel, 
$e=0.288\pm0.022$ and $\omega_\star=27.9\pm 7.0$ with the QP kernel, and 
$e=0.290\pm0.021$ and $\omega_\star=29.2\pm 6.7$ with the QP-SE kernel.
These are consistent with the $e$ and $\omega$ values derived in Sect.~\ref{Kepler_Data_Analysis} at 
$\lesssim1.5\, \sigma$ and $\lesssim2.3\, \sigma$, respectively.
In the second analysis, we found 
$e=0.2608\pm0.0086$ and $\omega_\star=13.5 \pm 2.7$\,deg with the SE kernel, 
$e=0.2638\pm0.0090$ and $\omega_\star=14.2 \pm 2.6$\,deg with the QP kernel, and 
$e=0.2654\pm0.0091$ and $\omega_\star=14.7 \pm 2.6$\,deg with the QP-SE kernel. 
These $e$ and $\omega$ determinations are closer to the values from the secondary eclipse ($\sim 0.6\, \sigma$ and $\sim 0.8\, \sigma$, respectively) 
as expected from the use of the Gaussian priors on them. The radial-velocity semi-amplitude, $K_{\star}$, does not vary from the first to the second analysis for a given kernel, but was found to be slightly higher for the SE kernel, that is, $K_\star=11.20 \pm 0.50$\,m\,s$^{-1}$, to be compared to 
$K_\star= 10.75 \pm 0.41 $\,m\,s$^{-1}$ and $K_\star= 10.78 \pm 0.42 $\,m\,s$^{-1}$ for the QP and QP-SE kernels, respectively
(see Table~\ref{TableRVparam}). 

\begin{figure*}
\centering
\includegraphics[width=6.5cm]{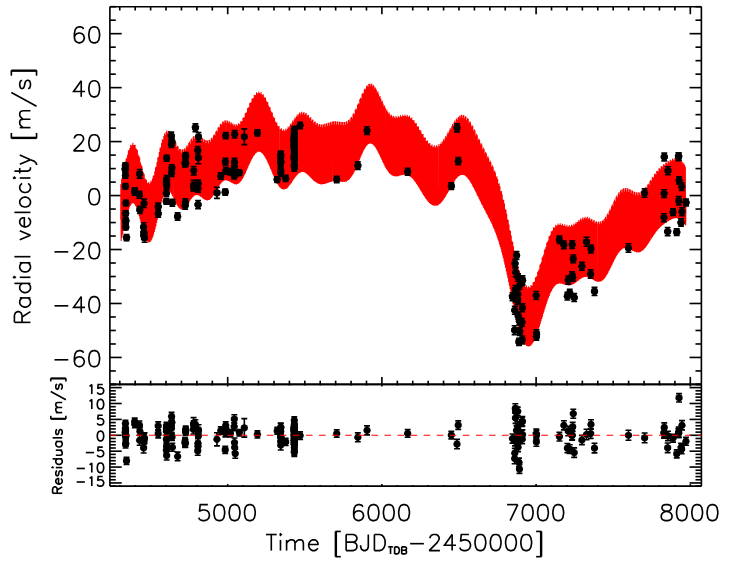}
\includegraphics[width=6.5cm]{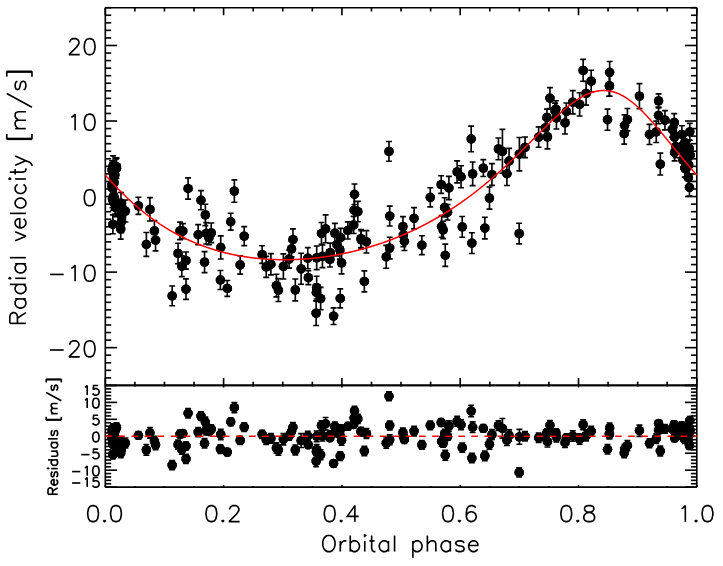}
\vspace{0.20cm}
\includegraphics[width=6.5cm]{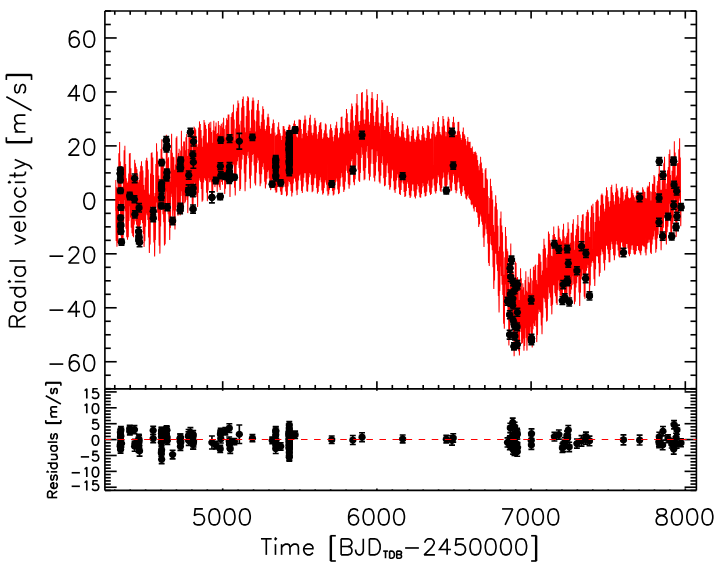}
\includegraphics[width=6.5cm]{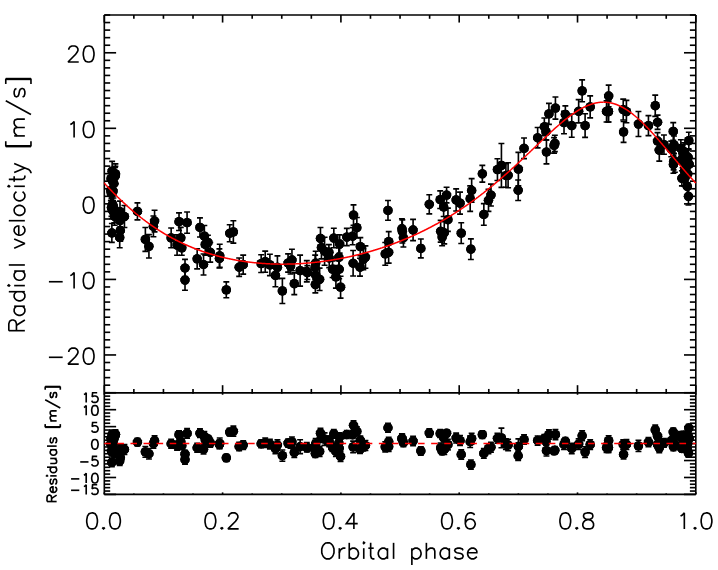}
\vspace{0.20cm}
\includegraphics[width=6.5cm]{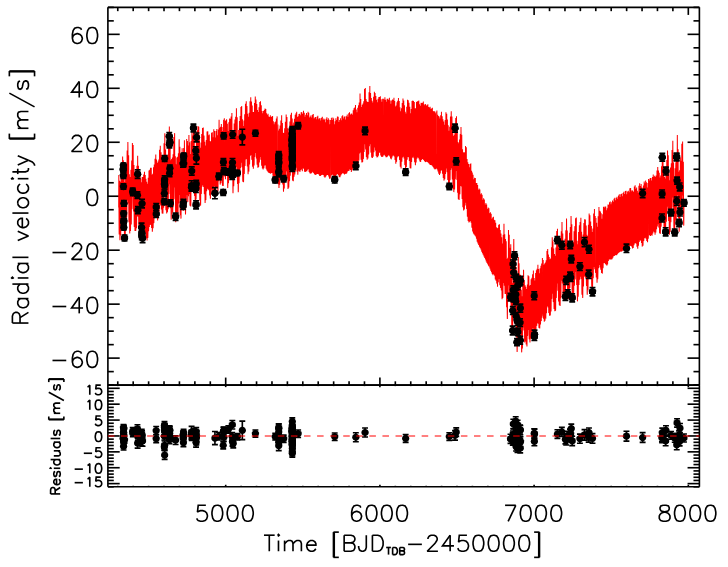}
\includegraphics[width=6.5cm]{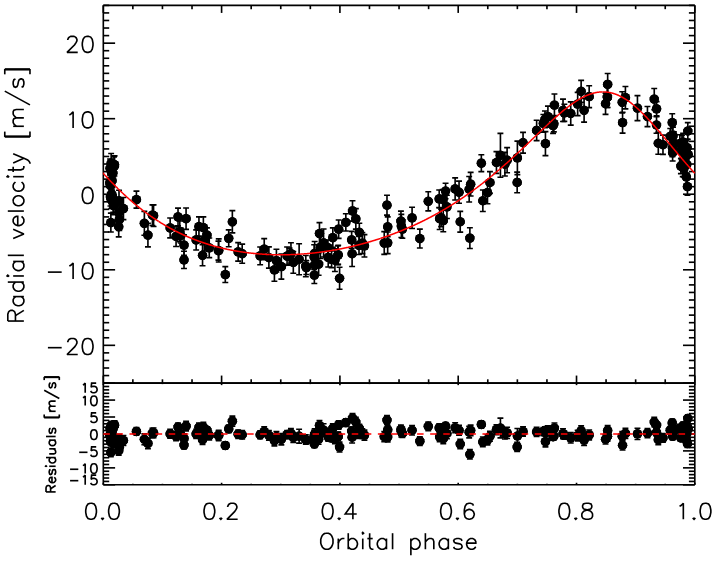}
\caption{
Radial-velocity data analysis. Left panels: HIRES radial velocities of HAT-P-11 (black circles) showing a clear long-term trend likely due to the stellar magnetic activity cycle. 
The best-fit models including both the Keplerian signal of HAT-P-11\,b and Gaussian-process regression with squared-exponential (top; Eq.~\ref{equation_kernel_SE}), quasi-periodic (middle; Eq.~\ref{equation_kernel_QP}), and quasi-periodic+squared-exponential (bottom; Eq.~\ref{equation_kernel_QPSE}) kernels are indicated with red solid lines.
Right panels: HAT-P-11 RVs (black circles) phase-folded with the transit ephemeris of HAT-P-11\,b, after removing the long-term trends modelled with Gaussian processes and the same kernels as in the corresponding left panels. The Keplerian eccentric model is displayed with a red solid line. 
We note the smaller and smaller scatter in the residuals as we move from top (SE kernel) to bottom (QP+SE kernel).}
\label{figure_RV_HAT-P-11}
\end{figure*}

By using the Bayesian information criterion (BIC) as a proxy for the Bayesian evidence, we found that the model with the QP-SE kernel is the most favoured, while that with the SE kernel is highly disfavoured. We therefore adopted the orbital parameters of the former (QP-SE) model (Table~\ref{TableRVparam}), which has also a more physical rationale because the rotation and activity cycle signals were modelled with two different (QP and SE) kernels. On the other hand, the QP kernel had to adapt to fit the activity cycle long-term variation in addition to the rotational signal, with its hyper-parameters $h$ and $\lambda$ taking intermediate values between $h_{\rm rot}$ and $h_{\rm cycle}$, and $\lambda_{\rm rot}$ and $\lambda_{\rm cycle}$ in the third kernel (Eq.~\ref{equation_kernel_QPSE}). We note that the GP models with both the QP and QP-SE kernels properly retrieved a stellar rotation period of $P_{\rm rot}\sim 32-33$~d, close to $P_{\rm rot}\sim 29-30$~d as estimated from the \emph{Kepler} photometry, 
despite the large uniform prior adopted (see Table~\ref{TableRVparam}). We show the best-fit GP+Keplerian models as a function of time in the left panel of Fig.~\ref{figure_RV_HAT-P-11}, and the Keplerian orbit due to HAT-P-11\,b as a function of the orbital phase, after the removal of the GP activity model, in the right panel of the same figure. 

For the rest of our analysis, we decided to use the orbital solution that we obtained from the analysis of transits and occultations due to the higher precision/accuracy in the determination of $e$ and $\omega$. We combined the stellar parameters, the transit parameters from the \emph{Kepler} light curve (Sect.~\ref{trocc}), and the RV parameters to derive a mass of $M_{\rm p}=0.0787\pm0.0048~M_{\rm J}$ ($M_{\rm p}=25.0\pm1.5\, M_{\oplus}$) and a mean density of $\rho_{\rm p}=1.172\pm0.085$\,g\,cm$^{-3}$, for HAT-P-11\,b. Finally, by knowing both the mass of the star and the planet, we computed the value of the planetary RV semi-amplitude $K_{\rm p}$, which we used for the atmospheric characterisation in Sect.~\ref{atmchar}. All the derived parameters of the HAT-P-11 planetary system are reported in Table~\ref{tab1}.

\begin{table*}[h!]
\begin{center}
\caption{Parameters and adopted priors of the RV models with Gaussian processes and three different kernels ($U$ and $N$ stand for uniform and Gaussian priors, respectively).}
\begin{tabular}{lllcl}
\hline\hline \\ [-8pt]
Parameter & Symbol & Units & C.I. & ~~~~Prior\\
\hline\hline \\ [-8pt]
~\\
\multicolumn{5}{c}{\emph{Squared-exponential Gaussian process model} [$\Delta \rm BIC=56.8$]} \\
~\\
%\hline
RV zero point & $\gamma$ & $\rm m\,s^{-1}$ & $-3.2\pm4.6$ & $U]-\infty, +\infty[$ \\ [2pt]
RV jitter & $\sigma_{\rm jit}$ & $\rm m\,s^{-1}$ & $3.29\pm0.24$ & $U[0, +\infty[$ \\ [2pt]
GP-SE amplitude & $h$ & $\rm m\,s^{-1}$ & $17.8^{+3.0}_{-2.4}$ & $U[0, +\infty[$ \\ [2pt]
GP-SE decay timescale  & $\lambda$ & days & $93.0^{+10.9}_{-9.9}$ & $U[0, 500]$ \\ [2pt]
Eccentricity  & $e$ & - & $0.2608\pm0.0086$ & $N(0.258, 0.004)$ \\ [2pt]
Stellar argument of periastron  & $\omega_\star$ & deg & $13.5\pm2.7$ & $N(12, 3)$ \\ [2pt]
Radial-velocity semi-amplitude & $K_\star$ & $\rm m\,s^{-1}$ & $11.20\pm0.50$ & $U[0, +\infty[$ \\ [2pt]
~\\
\hline
~\\
\multicolumn{5}{c}{\emph{Quasi-periodic Gaussian process model} [$\Delta \rm BIC=18.6$]} \\
~\\
RV zero point & $\gamma$ & $\rm m\,s^{-1}$ & $-3.11^{+4.38}_{-4.51}$ & $U]-\infty, +\infty[$ \\ [2pt]
RV jitter & $\sigma_{\rm jit}$ & $\rm m\,s^{-1}$ & $2.16 \pm 0.24$ & $U[0, +\infty[$ \\ [2pt]
GP amplitude & $h$ & $\rm m\,s^{-1}$ & $16.4^{+2.5}_{-2.0}$ & $U[0, +\infty[$ \\ [2pt]
GP decay timescale  & $\lambda$ & days & $146^{+34}_{-45}$ & $U[0, 500]$ \\ [2pt]
GP rotational period  & $P_{\rm rot}$ & days & $32.73^{+0.27}_{-0.45}$ & $U[25, 40]$ \\ [2pt]
GP harmonic inverse complexity & $w$ & & $1.87^{+0.40}_{-0.29}$ & $U[0.1, 5]$ \\ [2pt]
Eccentricity  & $e$ & - & $0.2638\pm0.0090$ & $N(0.258, 0.004)$ \\ [2pt]
Stellar argument of periastron  & $\omega_\star$ & deg & $14.2\pm2.5$ & $N(12, 3)$ \\ [2pt]
Radial-velocity semi-amplitude & $K_\star$ & $\rm m\,s^{-1}$ & $10.72\pm0.43$ & $U[0, +\infty[$ \\ [2pt]
~\\
\hline
~\\
\multicolumn{5}{c}{\emph{Quasi-periodic+squared-exponential Gaussian process model} [$\Delta \rm BIC=0$]} \\
~\\
RV zero point & $\gamma$ & $\rm m\,s^{-1}$ & $-3.3^{+11.7}_{-12.1}$ & $U]-\infty, +\infty[$ \\ [2pt]
RV jitter & $\sigma_{\rm jit}$ & $\rm m\,s^{-1}$ & $1.94^{+0.21}_{-0.19}$ & $U[0, +\infty[$ \\ [2pt]
GP rotational amplitude & $h_{\rm rot}$ & $\rm m\,s^{-1}$ & $6.84^{+1.34}_{-1.12}$ & $U[0, +\infty[$ \\ [2pt]
GP rotational decay timescale  & $\lambda_{\rm rot}$ & days & $58.1^{+13.4}_{-11.9}$ & $U[0, 500]$ \\ [2pt]
GP rotational period  & $P_{\rm rot}$ & days & $31.81^{+0.76}_{-0.82}$ & $U[25, 40]$ \\ [2pt]
GP rotational harmonic inverse complexity & $w_{\rm rot}$ & & $1.09^{+0.26}_{-0.21}$ & $U[0.1, 5]$ \\ [2pt]
GP cycle amplitude & $h_{\rm cycle}$ & $\rm m\,s^{-1}$ & $23.4^{+12.9}_{-6.6}$ & $U[0, +\infty[$ \\ [2pt]
GP cycle decay timescale  & $\lambda_{\rm cycle}$ & days & $435^{+156}_{-119}$ & $U[0, 1000]$ \\ [2pt]
Eccentricity  & $e$ & - & $0.2654\pm0.0091$ & $N(0.258, 0.004)$ \\ [2pt]
Stellar argument of periastron  & $\omega_\star$ & deg & $14.7\pm2.6$ & $N(12, 3)$ \\ [2pt]
Radial-velocity semi-amplitude & $K_\star$ & $\rm m\,s^{-1}$ & $10.78\pm0.42$ & $U[0, +\infty[$ \\ [2pt]
~\\
\hline\hline\\
\label{TableRVparam}
\end{tabular}
\end{center}
\end{table*}

\section{Atmospheric characterisation of HAT-P-11\,b at high spectral resolution}
\label{atmchar}
\subsection{Observations and data reduction}
Four transits of HAT-P-11\,b were simultaneously observed with the GIANO-B (wavelength range: $950-2450$\,nm, spectral resolving power $R\approx 50\,000$) and the HARPS-N (wavelength range: $383-693$\,nm, spectral resolving power $R\approx 115\,000$) high-resolution spectrographs in the GIARPS at TNG conﬁguration \citep{claudi2017} during the following nights: 7 July 2019; 18 June 2020; 19 September 2020; 13 June 2023. We only used the NIR (GIANO-B)
observations for the present work. A total of 240 spectra were collected during the four observing nights (60 during the first one, 60 during the second one, 58 during the third one, and 62 during the fourth one), each with an exposure time of 200\,s. The observations were performed with the nodding acquisition mode ABAB, where target and sky spectra were taken in pairs while alternating between two nodding positions along the slit (A and B) separated by $5^{\prime\prime}$, allowing an optimal subtraction of the detector noise and background. All the observations were scheduled in order to obtain spectra before, during, and after the transit with airmass between 1 and 2. The measured mean signal-to-noise ratio (S/N) per spectrum, averaged across the entire spectral range and dataset, is between 48 and 59. In Table~\ref{logobs} a schematic log of the observations is reported.

GIANO-B spectra cover the $Y, J, H, K$ spectral bands in 50 spectral orders. The raw spectra were dark-subtracted and extracted using the GOFIO pipeline Python 3 version \citep{rainer2018}. Although GOFIO also performs a preliminary wavelength calibration using U-Ne lamp spectra as a template, the mechanical instability of the instrument causes the wavelength solution to change during the observations. Since the U-Ne lamp spectrum is only acquired at the end of the observations to avoid persistence on the camera, the wavelength solution of the spectra determined by GOFIO is not sufficiently accurate and is expected to shift and jitter between consecutive exposures. In order to correct this shift, the spectra have been aligned to a common reference frame via cross-correlation with a time-averaged observed spectrum of the target used as a template. Thanks to this correction, we achieved a residual scatter in the measured peak position of the cross-correlation function (i.e. a residual shift of the spectra) well below $0.3$\,km\,s$^{-1}$ (approximately $1/10^{\rm th}$ of a pixel) for most of the spectral orders.

As these observations were performed from the ground, the spectra are contaminated by the presence of telluric lines (i.e. absorption lines due to the chemical species present in the Earth's atmosphere). However, the telluric spectrum provides a good wavelength-calibration source, since the lines' position does not change with time and the lines' wavelength is well known. Refined wavelength calibration is made by matching a set of telluric lines in the time-averaged observed spectrum with a high-resolution model of the Earth transmission spectrum generated via the ESO Sky Model Calculator \citep{noll2012}, and solving for the pixel-wavelength relation with a fourth-order polynomial fit. The spectral orders that showed either heavily saturated telluric lines or a high residual drift ($>0.4$\,pixels) have been excluded from the rest of the analysis. In particular, the excluded orders are: 8, 9, 10, 23, 24, 40, 41, 42, 43, 44, 45, 46, 47, 48, 49 (in the GIANO-B spectra, order 0 is the reddest and order 49 is the bluest).

\subsection{Telluric and stellar spectra removal}
At this stage of the analysis, the planet spectrum is overshadowed by the stellar and telluric spectra. However, the planet's orbital velocity has a non-zero radial component during transit ($v_{\rm p,\perp} \approx 10$\,km\,s$^{-1}$); consequently, while telluric and stellar lines are stationary or quasi-stationary (the stellar barycentric radial velocity changes by few m\,s$^{-1}$ during transit events) in wavelength, the planet spectrum experiences a detectable change in Doppler shift during the $\sim140$\,min of transit. This property can be used to disentangle the planetary signal from the stationary components that we have to remove.

In this work, in order to remove telluric and stellar spectra, a principal component analysis (PCA) was conducted, after having masked the deepest absorption lines. The idea behind the PCA technique, which has been successfully applied in the past by several authors (e.g. \citealt{dekok2013,damiano2019,piskorz2017,giacobbe2021, guilluy2022,carleo2022}), is to identify common trends in the spectra as a function of time (in this case represented by telluric and stellar lines in different spectral channels) and remove them. Before computing the PCA, we performed an optimal selection of the spectral orders, for each molecule and for each night, to discard the orders that do not contain enough signal (molecular lines) and/or are strongly contaminated by telluric and stellar lines, following the method explained by \citet{giacobbe2021}. The PCA was performed on a total of $K_i$ data matrices per night (where $K_i$ is the number of selected spectral orders for the $i$-th night). For each ($M$-rows; $N$-columns) data matrix ($M = 58-62$ images, $N = 2048$ pixels), the covariance matrix of the data was computed. Then, the covariance matrix was diagonalised by computing its eigenvectors (the Principal Components) ordered by their contribution to the variance of data (represented by the value of their associated eigenvalues). Following the procedure described in detail by \citet{giacobbe2021}, after having selected the appropriate number of principal components (between 9 and 23, depending on the quality of the night and the spectral order) that are supposed to mainly describe the telluric and stellar contaminations, these signals were reconstructed and removed from the data. The stages of the GIANO-B data reduction process are described with an example in Appendix~\ref{app_stage}, while the details of the spectral orders' selection and PCA procedures can be found in Appendix~\ref{app_os_pca}.

\subsection{Planet signal extraction via cross-correlation}
\label{pl_extr}
As the faint planetary signal is very dispersed by the high-resolution spectrograph, the single spectral lines are below the noise value ($\rm{S/N}_{\rm line} \lesssim 1$) of the residual signal obtained after the telluric removal. However, there are thousands of planetary lines observed at the same time in the large wavelength range of GIANO-B whose signal can be co-added, resulting in a boost in terms of S/N for the planet signal proportional to the square root of the number of lines observed, $N_{\rm lines}$: $\rm{S/N}\propto\sqrt{N_{\rm lines}}$ \citep{birkby2018}. The information contained in such a large number of lines can be combined by cross-correlating the residual data with template transmission spectra of the planet's atmosphere.

In particular, different models of the atmospheric transmission spectrum of HAT-P-11\,b have been simulated with PetitRADTRANS \citep{molliere2019}. All the simulated models assumed an isothermal atmosphere at the equilibrium temperature of the exoplanet at the transit epoch, assuming a heating re-distribution efficiency of $\epsilon=1$ ($T_{\rm {eq}}=699$\,K), as described in Sect.~\ref{albedo}, and are computed between $10$ bar and $10^{-8}$ bar in pressure. The models assume constant-with-altitude abundance (volume mixing ratio) profiles. The different models, one for each molecule that we wanted to test the presence of, assumed fixed values of the volume mixing ratios ($VMR$) of molecular hydrogen $VMR_{\ch{H2}}=0.855$ and helium $VMR_{\ch{He}}=0.145$ and assumed a $VMR_{\rm molecule} = 10^{-3}$ for the molecule to test (single-species models). Although these values do not match any specific chemical scenarios, this was the simpler framework we could adopt to probe the presence of a particular molecule. We investigated the presence of eight molecular species, very common in the atmosphere of hot giant planets: water vapour
(\ch{H2O}), methane (\ch{CH4}), ammonia (\ch{NH3}), acetylene (\ch{C2H2}), hydrogen cyanide (\ch{HCN}), carbon monoxide (\ch{CO}), carbon dioxide (\ch{CO2}), and hydrogen sulfide (\ch{H2S}).

To probe the presence of a particular molecule, after having performed the PCA, we computed the cross-correlation function (CCF) between the data and the template spectrum associated with the molecule. The CCF was evaluated shifting the model in wavelength on a fixed grid of RV lags ($\Delta RV=c\cdot\frac{\Delta\lambda}{\lambda}$) from $-270$\,km\,s$^{-1}$ to $+270$\,km\,s$^{-1}$, in steps of $0.1$\,km\,s$^{-1}$. The numeric computation was performed using the \textsc{C\_CORRELATE Pxy(L) IDL} function\footnote{\url{https://www.nv5geospatialsoftware.com/docs/C_CORRELATE.html}}, with null lag (L = 0), since the RV lags were applied to the wavelength array associated to the model and then the model was interpolated (via spline interpolation) on the same wavelength array of data before computing the CCF (i.e. with lag = 0). For every night and exposure, the CCFs calculated for each selected spectral order are co-added to obtain a single CCF for each exposure of each night.

Thanks to the high-resolution spectroscopy technique, it is possible to measure the Doppler shift of the spectral lines due to the planet's orbital motion. In this way, in order to be sure that a particular spectral feature is produced by a molecule in the atmosphere of an exoplanet, the signal should have a Doppler shift that `follows' the planetary movement and therefore we should observe that the peak of the CCF moves in wavelength as time passes according to the planetary motion-induced Doppler shift.
\begin{figure}
  \resizebox{\hsize}{!}{\includegraphics{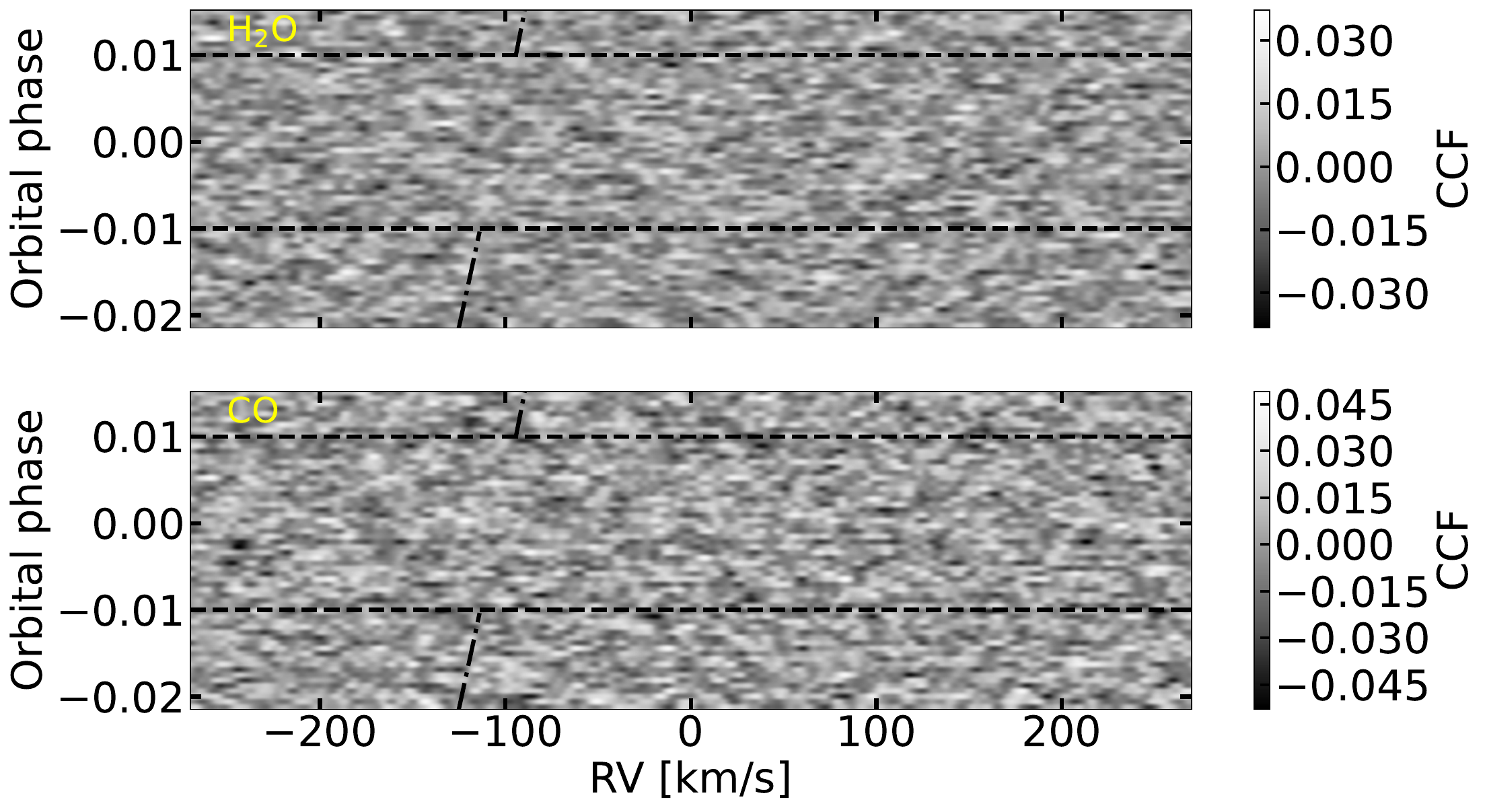}}
\caption{Examples of CCF values as a function of planetary orbital phase computed with data from the second observing night (18 June 2020) and model containing only \ch{H2O} (top panel) or \ch{CO} (bottom panel) lines. The horizontal dashed lines represent the transit ingress and egress while the dash-dotted line represents the expected CCF peak trail due to the planetary motion as measured in the observer rest frame. The expected CCF peak trail in transit is not represented for clarity.
As it can be seen, due to the faintness of the signal, the CCF peak trail is not visible by eye. This kind of plots serves as a visual check of any remaining telluric and stellar residuals, in this case showing no residuals and signifying that these are adequately corrected by the PCA.}
  \label{ccftrailreal}
\end{figure}
We assumed that the measured Doppler shift of the planetary spectral lines ($V_{\rm RV}=c\cdot \frac{\Delta\lambda}{\lambda}$) is made of three-velocity components:
\begin{equation}
V_{\rm RV}= V_{\rm p} + V_{\rm sys} - V_{\rm bary}\,,
\end{equation}
where $V_{\rm bary}$ is the velocity induced by Earth's motion around the barycentre of the Solar System (barycentric velocity), $V_{\rm sys}$ is the centre of mass velocity of the star-planet system with respect to the Earth (systemic velocity), and $V_{\rm p}$ is the planet RV. The time-dependent contribution $V_{\rm p}$ can be expressed as a function of two of the planetary orbital parameters (i.e. the eccentricity $e$ and the argument of periastron $\omega_{\rm p}$), the RV semi-amplitude $K_{\rm p}$ and the true orbital anomaly $\nu(t)$:
\begin{equation}
    V_{\rm p} (t) = K_{\rm p} \cdot[\cos{(\nu(t)+\omega_{\rm p})}+e\cdot \cos({\omega_{\rm p}})] \, .
\label{rvp}
\end{equation}
The radial velocity semi-amplitude $K_{\rm p}$ can be expressed as a function of the orbital eccentricity $e$, orbital inclination $i$, orbital period $P_{\rm orb}$, mass of the planet $M_{\rm p}$ and mass of the star $M_{\rm \star}$:
\begin{equation}
K_{\rm p}= \frac{M_{\rm p}\sin{i}}{(M_{\rm p}+M_{\rm \star})^{\frac{2}{3}}}\cdot\left(\frac{2\pi G}{P_{\rm orb}}\right)^\frac{1}{3}\cdot\frac{1}{\sqrt{1-e^2}} \, .
\label{kp_eq}
\end{equation}
For our analysis, it is convenient to re-express $K_{\rm p}$ isolating the term containing the eccentricity and grouping all the others in the constant $\Tilde{K}_{\rm p}$:
\begin{equation}
K_{\rm p}=\Tilde{K}_{\rm p}\cdot\frac{1}{\sqrt{1-e^2}} \, .
\label{kp_eq2}
\end{equation}

In Fig.~\ref{ccftrailreal} we report, as an example, a plot of the CCF values as a function of the planetary orbital phase computed between the models of \ch{H2O} and \ch{CO} and the data of the second observing night (18 June 2020). As it can be seen, the CCF peak trail is not visible by eye, since also at this stage of the analysis the planetary signal is too faint. This kind of plot helps in checking the telluric (stellar) spectrum-removal procedure since any strong residual signal due to a non-optimal telluric (stellar) subtraction would produce a spurious vertical trail of CCF peaks at a radial-velocity $RV = 0$\,km\,s$^{-1}$ ($RV = V_{\rm sys}$, neglecting the few m\,s$^{-1}$ motion induced by the planet onto the star) when data are cross-correlated with the \ch{H2O} (\ch{CO}) templates. A benefit of the eccentric orbit of HAT-P-11\,b is that the planetary radial velocity during the transit always remains strictly negative ($\sim-30$\,km\,s$^{-1}$ at the transit midpoint). This, combined with the high $V_{\rm sys}$, shifts the planetary signal at $\sim100$\,km\,s$^{-1}$ far from the signal of the telluric lines (that is at $RV=0$\,km\,s$^{-1}$), further reducing spurious contaminations due to the Earth's atmosphere.

If in the atmosphere of an exoplanet a molecule is present, its spectral signal has a null Doppler shift measured in the exoplanet rest frame ($V_{\rm rest}=0$\,km\,s$^{-1}$), in the absence of a Doppler shift induced by atmospheric dynamics. It follows that, after having subtracted the barycentric, systemic and planetary RV from the CCF trails (that is, after having moved to the exoplanet rest frame), all the CCF peaks should align at $V_{\rm rest}=0$\,km\,s$^{-1}$.

By subtracting different orbital solutions (in this work we explored a range of $K_{\rm p}$ values) from the CCF peaks trail, a different alignment of the CCF peaks is obtained. By co-adding the CCF values in phase (only considering the in-transit phases) into a single CCF for each trial $K_{\rm p}$, the planetary signal as a function of the rest-frame velocity $V_{\rm rest}$ and $K_{\rm p}$ is maximised and it is possible to build the so-called $K_{\rm p}-V_{\rm rest}$ map, in which, in the case of the detection of a molecule, a strong peak of the signal at the expected $K_{\rm p}$ and $V_{\rm rest}=0$\,km\,s$^{-1}$ is observed.
We took advantage of the high sampling of the CCF (larger than the GIANO-B pixel scale of $2.7$\,km\,s$^{-1}$) for a precise shift of the CCF trails into the planetary rest frame. However in order to avoid the use of correlated data points in our analysis, we binned the CCF values in radial velocity using a bin width of $2.7$\,km\,s$^{-1}$. We took the median of the $27$ CCF values in each radial velocity bin as the value of the CCF associated with each bin, before co-adding the CCF values in phase for each trial $K_{\rm p}$.

In this work, we explored a range of $K_{\rm p}$ values by varying the value of $\Tilde{K_{\rm p}}$ between $0$\,km\,s$^{-1}$ and $200$\,km\,s$^{-1}$ in steps of $2.7$\,km\,s$^{-1}$, having fixed the eccentricity and the other orbital parameters to the values reported in Table~\ref{tab1}. This means that the corresponding explored range of $K_{\rm p}$ is [$0;207$]\,km\,s$^{-1}$ (this is the $K_{\rm p}$ range reported in the $K_{\rm p}-V_{\rm rest}$ maps in the next section and it is computed from $\Tilde{K}_{\rm p}$ using equation~\ref{kp_eq2} with the value of $e$ from Table~\ref{tab1}). From Table~\ref{tab1}, the expected value of $K_{\rm p}$ is $\hat{K}_{\rm p}=123.4\pm9.9$\,km\,s$^{-1}$ and, consequently, the expected value of $\Tilde{K}_{\rm p}$ is $\hat{\Tilde{K}}_{\rm p}=119.3\pm9.5$\,km\,s$^{-1}$. However, the exploration of a large parameter space offers a strong diagnostic on all sources of noise and allows us to verify that no other spurious signal produces a significant detection near the planet's rest frame position.

\subsection{Results}

\label{results}
\begin{figure*}[!h]
\centering
   \includegraphics[width=17cm]{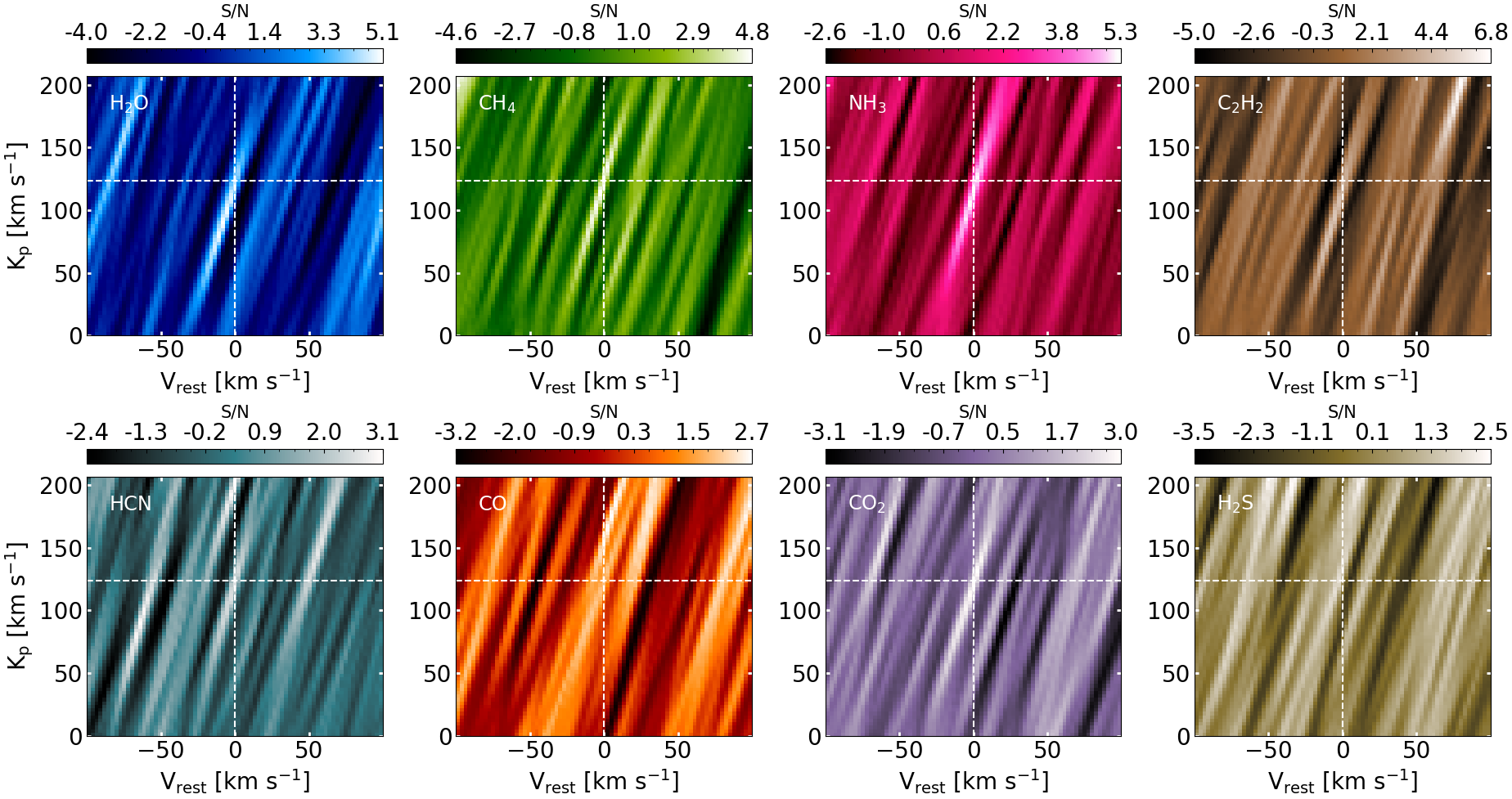}
     \caption{Signal-to-noise ratio $K_{\rm p}-V_{\rm rest}$ maps for the probed chemical species: \ch{H2O}, \ch{CH4}, \ch{NH3}, \ch{C2H2}, \ch{HCN}, \ch{CO}, \ch{CO2}, and \ch{H2S}. Each $K_{\rm p}-V_{\rm rest}$ map shows the S/N of the cross-correlation of the GIANO-B spectra (4 transits combined) with isothermal atmospheric models, as a function of the planet’s RV semi-amplitude ($K_{\rm p}$) and the planet’s rest-frame velocity ($V_{\rm rest}$). The S/N is computed by dividing the peak value of the cross-correlation function at each $K_{\rm p}$ by the standard deviation of the noise far from the peak, as described in the text. Negative S/N values correspond to anti-correlation. The vertical and horizontal white dashed lines correspond to $V_{\rm rest}=0$\,km\,s$^{-1}$ and the expected $K_{\rm p}$ value ($\hat{K}_{\rm p}$), respectively.}
     \label{snrplot}
\end{figure*}
\begin{figure*}
\centering
   \includegraphics[width=17cm]{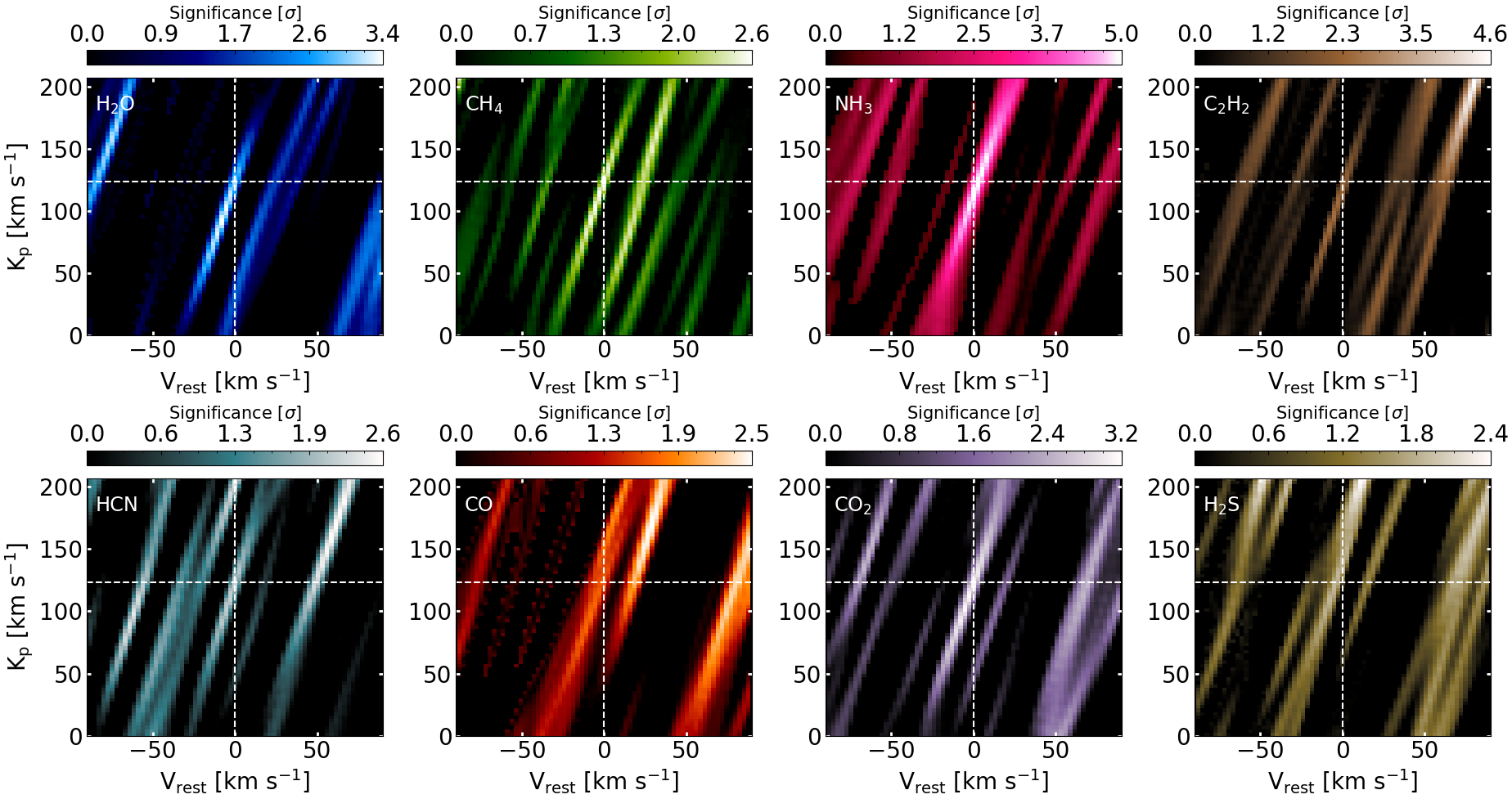}
     \caption{Significance $K_{\rm p}-V_{\rm rest}$ maps for the tested chemical species: \ch{H2O}, \ch{CH4}, \ch{NH3}, \ch{C2H2}, \ch{HCN}, \ch{CO}, \ch{CO2}, and \ch{H2S}. Each $K_{\rm p}-V_{\rm rest}$ map shows the significance of the cross-correlation signal of the GIANO-B spectra (4 transits combined) with isothermal atmospheric models, as a function of the planet’s RV semi-amplitude ($K_{\rm p}$) and the planet’s rest-frame velocity ($V_{\rm rest}$). The significance is computed with a Welch \textit{t-test} on two samples of cross-correlation values, that is, far from and near to the planet's RV respectively (see the text for more details). The vertical and horizontal white dashed lines correspond to $V_{\rm rest}=0$\,km\,s$^{-1}$ and the expected $K_{\rm p}$ value ($\hat{K}_{\rm p}$), respectively.}
     \label{map}
\end{figure*}

We first co-added the data of all the 4 observing nights and built the signal-to-noise ratio (S/N) $K_{\rm p}-V_{\rm rest}$ maps for the different probed molecules, obtained by cross-correlating data with models and dividing the result by the standard deviation of the noise far from the peak ($|V_{\rm rest}|>25$\,km\,s$^{-1}$), to search for signals following the expected planetary RV (potential detections). Then we computed the significance of each potential detection by performing a Welch \emph{t-test} \citep{welch1947} on two samples of CCF values: the former far ($|V_{\rm rest}|>25$\,km\,s$^{-1}$) from the planet's rest-frame velocity (``out-of-trail'') and the latter near to it ($|V_{\rm rest}|<3$\,km\,s$^{-1}$, ''in-trail''). The test rejects the null hypothesis ($H_0$) that the two samples have the same mean (and therefore that the CCF signal produced by the planet is only a statistical fluctuation of the background signal value) at a certain significance level that we adopted as the significance ($\sigma$) of our detections. Our significance calculations are based on the hypothesis of uncorrelated noise, which has been shown to be a valid approximation in previous works \citep{brogi2018,guilluy2019}. In order to build the significance $K_{\rm p}-V_{\rm rest}$ maps, the Welch \emph{t-test} was performed on CCF `in-trail' and `out-of-trail' distributions centred at the different $V_{\rm rest}$ explored, for each trial $K_{\rm p}$.

In Fig.~\ref{snrplot}, we report the S/N $K_{\rm p}-V_{\rm rest}$ maps for the different tested molecules, obtained by cross-correlating data with the different models and following the procedure explained in Sect. ~\ref{pl_extr}. As it can be seen, we obtain a signal around the expected planetary position in the $K_{\rm p}-V_{\rm rest}$ maps with S/N $>3$ for 4 molecular species: \ch{H2O} (S/N $=5.1$), \ch{CH4} (S/N $=4.8$), \ch{NH3} (S/N $=5.3$), and \ch{CO2} (S/N $=3.0$).

In Fig.~\ref{map}, we report the significance $K_{\rm p}-V_{\rm rest}$ maps for the different tested molecules, obtained by performing the Welch \emph{t-test} on in-trail and out-of-trail CCF distributions. As it can be seen from the maps, the peak of the signal for the four potentially detected species has a significance of $3.4\ \sigma$ (\ch{H2O}), $2.6\ \sigma$ (\ch{CH4}), $5.0\ \sigma$ (\ch{NH3}), and $3.2\ \sigma$ (\ch{CO2}). For the other four probed species, we measure no significant chemical signature at the planetary RV, and therefore we consider them as non-detections and focus our attention and following analyses on the signals of \ch{H2O}, \ch{CH4}, \ch{NH3}, and \ch{CO2} only. The distributions of CCF values in-trail and out-of-trail used for computing the significance of the detections via the Welch \emph{t-test} are reported in Fig.~\ref{ttest_histo4} for all the probed species.\\
It is interesting to notice the orientation of the signals in the $K_{\rm p}-V_{\rm rest}$ maps, which is typically vertical for atmospheric transmission studies (e.g. ~\citealt{giacobbe2021}) and sloped for atmospheric emission studies (e.g. ~\citealt{line2021}). This is due to the eccentric orbit of HAT-P-11\,b that makes the planetary RV to be not symmetrical around $0$\,km\,s$^{-1}$ during the transit (it is always strictly negative), as it happens for planets on circular orbit observed in emission (i.e. at orbital phases different from 0 or 0.5).

\begin{figure}
  \resizebox{\hsize}{!}{\includegraphics{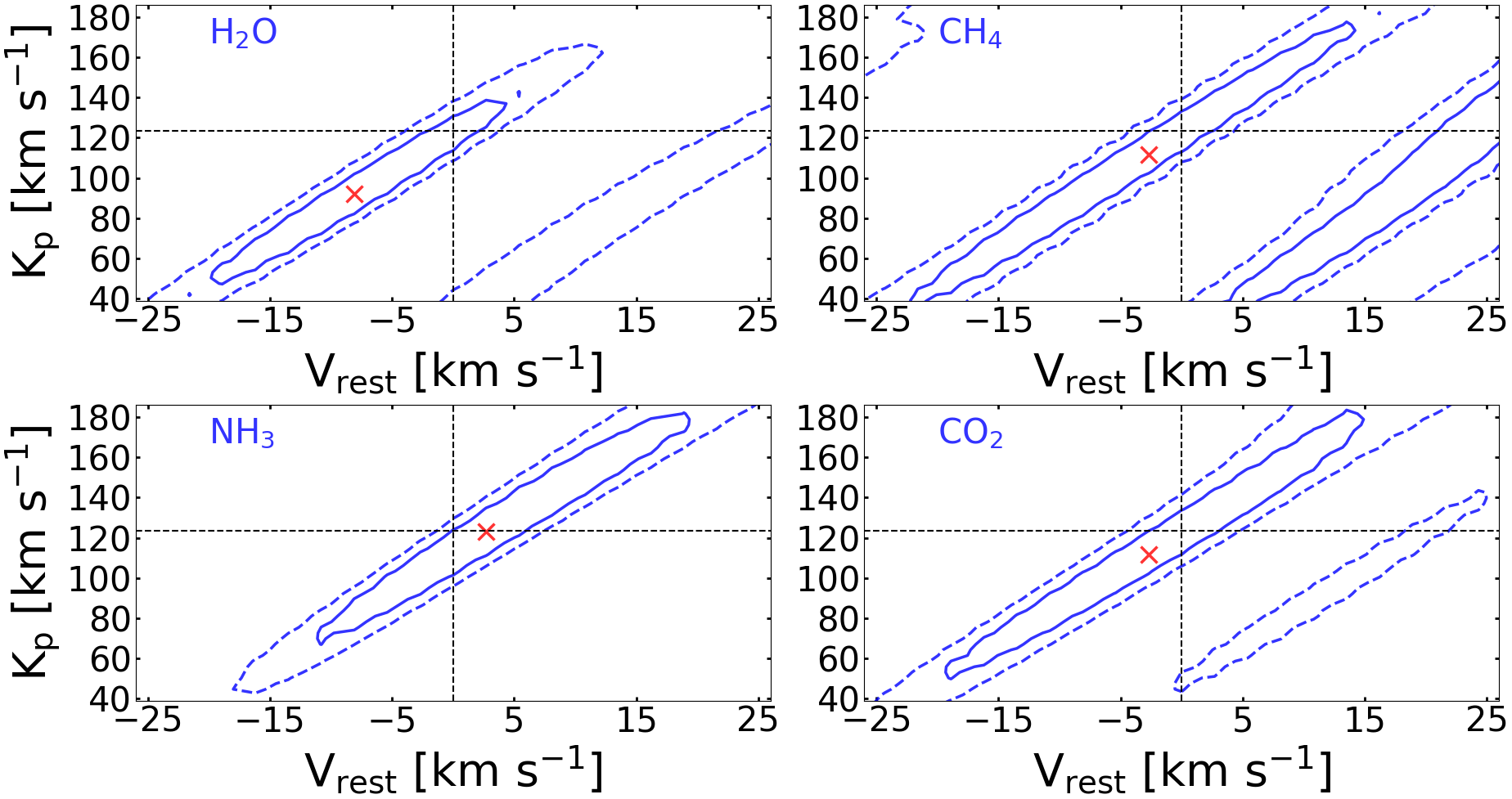}}
  \caption{Contour plots of the detection significance of the four selected chemical species. The solid (dashed) lines represent the $1\, \sigma$ ($2\, \sigma$) interval around the peak value of the significance (marked with a red cross). These intervals are computed as described in the text. The point of the $K_{\rm p}-V_{\rm rest}$ map in which the 2 black dashed lines (the horizontal one corresponds to the expected $K_{\rm p}$ value, while the vertical one corresponds to $V_{\rm rest} = 0$\,km\,s$^{-1}$) cross each other represents the expected detection significance peak position in the case in which the detected signal has a planetary origin. As it can be the significance peak position is compatible with the planetary origin hypothesis at better than $1\ \sigma$ for all the 4 species.}
  \label{contour}
\end{figure}

Finally, for the four selected species, we computed how much the Doppler signature of the signals is in accordance with the expected planetary RV. To do this, we built contour plots of the detection significance, defining the $1\,\sigma$ and $2\,\sigma$ significance intervals as the regions of the $K_{\rm p}-V_{\rm rest}$ maps where the significance drops by $1\, \sigma$ and $2\,\sigma$, respectively, with respect to the significance peak, and looked at where the expected position of the atmospheric signal in the $K_{\rm p}-V_{\rm rest}$ maps was with respect to those intervals. In Fig.~\ref{contour}, we report the contour plots of the detection significance for the four chemical species. As it can be seen, the signals we measure have a Doppler signature compatible with the planetary one at $<1\,\sigma$, for all four molecules.

Before claiming any detection, we checked the reliability of our results by performing a further test, described in the following. For each of the four chemical species, we combined the matrices of the CCF as a function of the orbital phase (see, e.g. the one in Fig.~\ref{ccftrailreal}) of the four observing nights in a single CCF matrix with the rows sorted in crescent orbital phase. Then, we randomly shuffled the CCF order in phase (this corresponds to shuffling the sequence of observed spectra in time, including the out-of-transit ones) 250 times. For each shuffle we built both the S/N and the significance $K_{\rm p}-V_{\rm rest}$ maps in a restricted interval of $K_{\rm p} =$ [$89.4; 156.5$]\,km\,s$^{-1}$ (corresponding to $\hat{K}_{\rm p}\pm34$\,km\,s$^{-1}$) and $V_{\rm rest}=$ [$-10;10$]\,km\,s$^{-1}$, in order to test the presence of spurious signals that have not a planetary origin but can produce significant features around the expected signal position in the maps due to some peculiar time-correlated noise. We repeated this test for the four selected chemical species and for each species we studied the distributions of the 43\,750 values of S/N and 43\,750 values of significance obtained in the chosen $K_{\rm p}-V_{\rm rest}$ interval combining the 250 permutations.

The results of this test are reported in Table~\ref{test_res}. As it can be seen, for all the 4 molecular species, 95\% of the test yields signals with S/N $<3$ and significance $\leq2.5\ \sigma$ in the selected $K_{\rm p}-V_{\rm rest}$ interval, while 99.73\% of the test yields S/N $\leq4.6$ and significance $\leq4.2\ \sigma$, in the same interval.

\begin{table}[!h]
\caption{Signal-to-noise ratio and significance from the \emph{t-test} ($\sigma$\ \textit{t-test}) values delimiting the 95\% and 99.73\% intervals of the distributions obtained by performing the reliability test described in the text for the four selected molecular species.} 
\label{test_res}
\centering
\begin{tabular}{c c c c c}
\hline
\hline
    Molecule & S/N & S/N & $\sigma$\ \textit{t-test} & $\sigma$\ \textit{t-test}\\
     & (95\%) & (99.73\%) & (95\%) & (99.73\%)\\
\hline\\ [-8pt]
    \ch{H2O} & $[-2.5;2.9]$ & $[-4.1;4.6]$ & $\leq2.5$ & $\leq4.2$\\
    \ch{CH4} & $[-2.8;2.6]$ & $[-4.0;4.1]$ & $\leq2.4$ & $\leq4.1$\\
    \ch{NH3} & $[-2.8;2.6]$ & $[-4.6;4.2]$ & $\leq2.3$ & $\leq3.7$\\
    \ch{CO2} & $[-2.5;2.7]$ & $[-3.7;4.5]$ & $\leq2.2$ & $\leq3.9$\\
\hline   
\end{tabular}
\end{table}

In Fig.~\ref{shuffle_snr} and Fig.~\ref{shuffle_tmap}, we report the distributions of S/N and significance values obtained in this test, respectively, for the four selected chemical species.

After having performed this test, for each of the four molecular species we computed the probability (\emph{p-values}) of randomly drawing from the distributions reported in Fig.~\ref{shuffle_snr} and Fig.~\ref{shuffle_tmap}, the S/N and significance values that we measure. To compute the \emph{p-values} for the S/N (significance) value, we summed the occurrences of S/N (significance) greater than the S/N (significance) measured and divided them by 43\,750. We report the \emph{p-values} in Table~\ref{sig}. As it can be seen, for all the four molecular species we obtain a \emph{p-value} $\leq2.5\%$ for the S/N and $\leq3.8\%$ for the significance, so there is less than $5\%$ of probability that these signals are due time-correlated noise.

In Table~\ref{sig}, we summarise the significance level of the signal of the four molecular species of interest obtained with different statistical methods. From these results, we conclude that we have a statistically robust detection of \ch{NH3}, which is the most significant signal ($5\ \sigma$) and the one with the lowest $p-values$ ($< 0.04\%$). We consider the \ch{H2O} as a detection, even though it is less statistically robust than the \ch{NH3} one, since it has already been detected at low-resolution multiple times in the atmosphere of HAT-P-11\,b (see the literature cited in Sec.~\ref{introduction}) and therefore we can be more confident about the reliability of the signal we measure.
Since the \emph{t-test} significance for the signal of \ch{CH4} is $<3\ \sigma$, the significance $K_{\rm p}-V_{\rm rest}$ map presents a second comparable peak at $V_{\rm rest} \approx 20$\,km\,s$^{-1}$, and the associated \emph{p-value} is the highest among the four chemical species, even if the signal has a S/N$=4.8$ and a Doppler signature that is compatible with the planetary one at $<1\ \sigma$, the detection of this molecular species remains tentative as it is not sufficiently robust from a statistical point of view. Even though the signal from \ch{CO2} has both the S/N and significance $\geq 3\ \sigma$ and the RV trail is compatible with the planetary one at $<1\ \sigma$, it has the lowest S/N and the highest S/N \emph{p-value} among the four selected species. For these reasons, we consider the detection of \ch{CO2} as tentative too. As it can be seen, for what concerns these two tentative detections, even though we obtain signals at the expected planetary RV, they are not sufficiently statistically robust and we suggest conducting further studies to unambiguously assess the presence of \ch{CH4} and \ch{CO2} in the atmosphere of HAT-P-11\,b.

To further assess the robustness of our analysis for what it concerns the impact of the number of principal components removed by the PCA (changing with the spectral order and night) on the final results, we also repeated the whole analysis removing a fixed number of principal components with the PCA for all the selected spectral orders per-molecule, an for all the observing nights. We performed this test twice, the first time we removed 9 principal components (the minimum number of principal components removed in our work among the different spectral orders and nights), the second time we removed 23 components (the maximum number of principal components removed in our work among the different spectral orders and nights). We obtain that none of these 2 extreme conditions changes our interpretation of which chemical species we detect (i.e. \ch{H2O} and \ch{NH3}), and which chemical species we tentatively detect and need further investigations (i.e. \ch{CH4} and \ch{CO2}), even if their S/N and significance slightly change (at less than $1\ \sigma$ level), as expected.

Finally, to assess the robustness of the final results (in particular the detection of \ch{H2O} and \ch{NH3}) in relation to the impact of the spectral orders' selection procedure on the CCF analysis, we performed an additional test. In this test, we repeated the analysis refining the orders' selection procedure to assess the presence of possible spurious signals near the expected planetary radial velocity that could lead to possible false positive detections. The description of this test and the results are reported in Appendix~\ref{test_ord}. We do not observe relevant changes in the results, further confirming the conclusion we reached with the main analysis.

In summary, in this work, we report the detection of two molecular species (i.e. \ch{H2O} and \ch{NH3}) in the atmosphere of HAT-P-11\,b and the tentative detection of two others (i.e. \ch{CH4} and \ch{CO2}), whose presence has to be assessed by further studies.

\begin{table*}[!h]
\caption{Significance of the detections calculated with different statistical methods.} 
\label{sig}  
\centering
\begin{tabular}{c c c c c c c}
\hline
\hline
    Molecule & S/N & Welch \emph{t-test} & Planetary RV & \emph{p-value} & \emph{p-value} & Status\tablefootmark{a}\\
     & & Significance & Compatibility & S/N & $\sigma$ \emph{t-test}\\
\hline\\ [-8pt]
    \ch{H2O} & $5.1$ & $3.4\ \sigma$ & $<1\ \sigma$ & $0.046\%$ & $1.3\%$ & \emph{D}\\
    \ch{CH4} & $4.8$ & $2.6\ \sigma$ & $<1\ \sigma$ & $0.023\%$ & $3.8\%$ & \emph{TD}\\
    \ch{NH3} & $5.3$ & $5.0\ \sigma$ & $<1\ \sigma$ & $0.039\%$ & $0.011\%$ & \emph{D}\\
    \ch{CO2} & $3.0$ & $3.2\ \sigma$ & $<1\ \sigma$ & $2.5\%$ & $1.3\%$ & \emph{TD}\\
\hline                                            
\end{tabular}
\tablefoot{
    \tablefoottext{a}{\emph{D} stands for `detected' and \emph{TD} stands for `tentatively detected'.}
    }
\end{table*}

\subsection{Discussion}
\label{discussion}
The significance of the detections reported in this work is lower than that of the species detected in other hot and warm Jupiters' atmospheres, which even reached the $10\,\sigma$ level (e.g. \citealt{giacobbe2021}).
This is mainly due to the lower atmospheric signal level of warm Neptune-size exoplanets (the atmospheric signal level is $\approx2H_{\rm s}R_{\rm p}/R^2_\star$, where $H_{\rm s}$ is the atmospheric scale height, and therefore it is $\sim55$\,ppm\footnote{For comparison, HD\,209458\,b investigated by ~\citet{giacobbe2021} has a $2H_{\rm s}R_{\rm p}/R^2_\star\sim164$\,ppm.} for HAT-P-11\,b) and underlines the difficulty in probing the atmospheric features of this class of exoplanets.

\subsubsection{Detected species in the HAT-P-11\,b planetary context}
Our detection of water vapour in the atmosphere of HAT-P-11\,b is in accordance with the results of \citet{fraine2014}, \citet{tsiaras2018}, \citet{chachan2019}, and \citet{cubillos2022}. The signal from the methane, which we measure at the expected planetary radial velocity, supports the results obtained by \citet{chachan2019} and \citet{cubillos2022}, who suggested the presence of \ch{CH4} from the analysis of HST data however, as told in the previous section, the significance of the signal is very low and therefore we cannot confirm the presence of this molecular species in the atmosphere of HAT-P-11\,b and further studies are required. 

The hot and warm Neptunes' atmospheric chemistry and dynamics are different from the ones that act in the hot and warm Jupiters' atmospheres, mainly because of the smaller radius and mass and the possible higher metallicity of the former~\citep{moses2013}. The atmospheric composition of individual exoplanets also depends on their effective temperature, formation history, atmospheric evolution, orbital parameters, and irradiation environment, so it is difficult to predict their exact atmospheric properties. However, for what concerns the chemical composition, some general trends with temperature, metallicity and C/O ratio can be found, as shown in the works of \citet{moses2013}, who particularly focused on hot Neptunes, \citet{madhusudhan2012} and \citet{fonte2023}. 

Even though with this work we are not able to constrain HAT-P-11\,b atmospheric physical and chemical properties (such as the elemental abundances), we can still make some general considerations about our detections. In particular, under thermochemical equilibrium, the presence of hydrocarbons, like \ch{HCN} and \ch{C2H2}, is particularly favoured in carbon-rich environments (C/O $\gtrsim1$) at high temperatures ($T\gtrsim1000$\,K). This is in line with our non-detections of these two chemical species, given the relatively low-temperature atmosphere ($T_{\rm eq}\sim700$\,K) of the target we analysed. \citet{chachan2019} measured an atmospheric C/O ratio close to unity (C/O$=0.97^{+0.59}_{-0.46}$). A C/O value close to unity together with a low-temperature environment favours the formation of ammonia, which is the molecule that we detected with the highest significance and the second nitrogen-bearing species we probed. In addition, at temperatures lower than $T \lesssim 1300$\,K, the formation of \ch{H2O} and \ch{CH4} is favoured, in accordance with our detection of \ch{H2O} and tentative detection of \ch{CH4}, considering the HAT-P-11\,b equilibrium temperature.

Another interesting result of our work is the tentative detection of \ch{CO2} in the atmosphere of HAT-P-11\,b. Carbon dioxide is an important indicator of the metal enrichment of the atmosphere of exoplanets and therefore it can give important information about the formation processes of the primary atmospheres of gas giants. Indeed several models for warm gas giant atmospheres predict that \ch{CO2} is one of the molecules most strongly enhanced with increasing atmospheric metallicity, becoming detectable for metallicities greater than $\approx10$ times that of the Sun \citep{lodders2002,moses2013}. Until now, only \citet{jwst2023} have firmly detected the presence of \ch{CO2} in the atmosphere of an exoplanet (the warm Saturn WASP-39\,b) at low spectral resolution, while at high spectral resolution \citet{carleo2022} tentatively detected it in the atmosphere of the warm Jupiter WASP-80\,b. Therefore, our tentative detection of \ch{CO2} represents the first hint of the presence of this molecule in the atmosphere of a warm Neptune and, if confirmed, could point towards a super-solar metallicity for the atmosphere of HAT-P-11\,b. However, as underlined before, further studies are needed and only a statistical comparison between different atmospheric chemical-physical models (for example through atmospheric retrievals), could robustly determine the atmospheric chemical-physical characteristics of this target.

For the case of HAT-P-11\,b, the estimated metallicity is $Z<4.6\ Z_{\odot}$ at $2\, \sigma$ level ~\citep{chachan2019} with $Z_{\odot}$ the solar metallicity, that is, $Z<2.3\ Z_{\star}$ referred to the host-star metallicity $Z_{\star} = 2.0\pm0.2\, Z_{\odot}$ ~\citep{bakos2010}, while ~\citet{welbanks2019} estimated a substellar \ch{H2O}/\ch{H} metallicity. At this level of metallicity, \ch{CO2} should be scarce (see \citealt{moses2013}, Fig.~5), but if we consider the $3\, \sigma$ confidence level upper limit on metallicity ($Z<86\ Z_{\odot}=43\ Z_{\star}$) by ~\citet{chachan2019}, the \ch{CO2} abundance could increase up to the point of being detectable with our kind of analysis.

It is worth noting that the cross-correlation analysis is more sensitive to molecular lines' position in wavelength rather than their depths with respect to the continuum. This means that a more significant detection for a particular chemical species does not necessarily imply that species is more abundant with respect to the other ones because it could arise for example from a denser forest of lines and consequently from a stronger correlation peak. In addition, our non-detections do not necessarily imply the absence of such chemical species, for which further investigations are needed.
Finally, for what concerns the non-detections, the $K_{\rm p}-V_{\rm rest}$ significance maps show spurious signals that are not related to the planetary signal. They are typically aliases generated by the autocorrelation function of the template that can also be seen in the maps of the detected species far from the planetary RV.

\begin{figure*}
\centering
\includegraphics[width=0.95\textwidth]{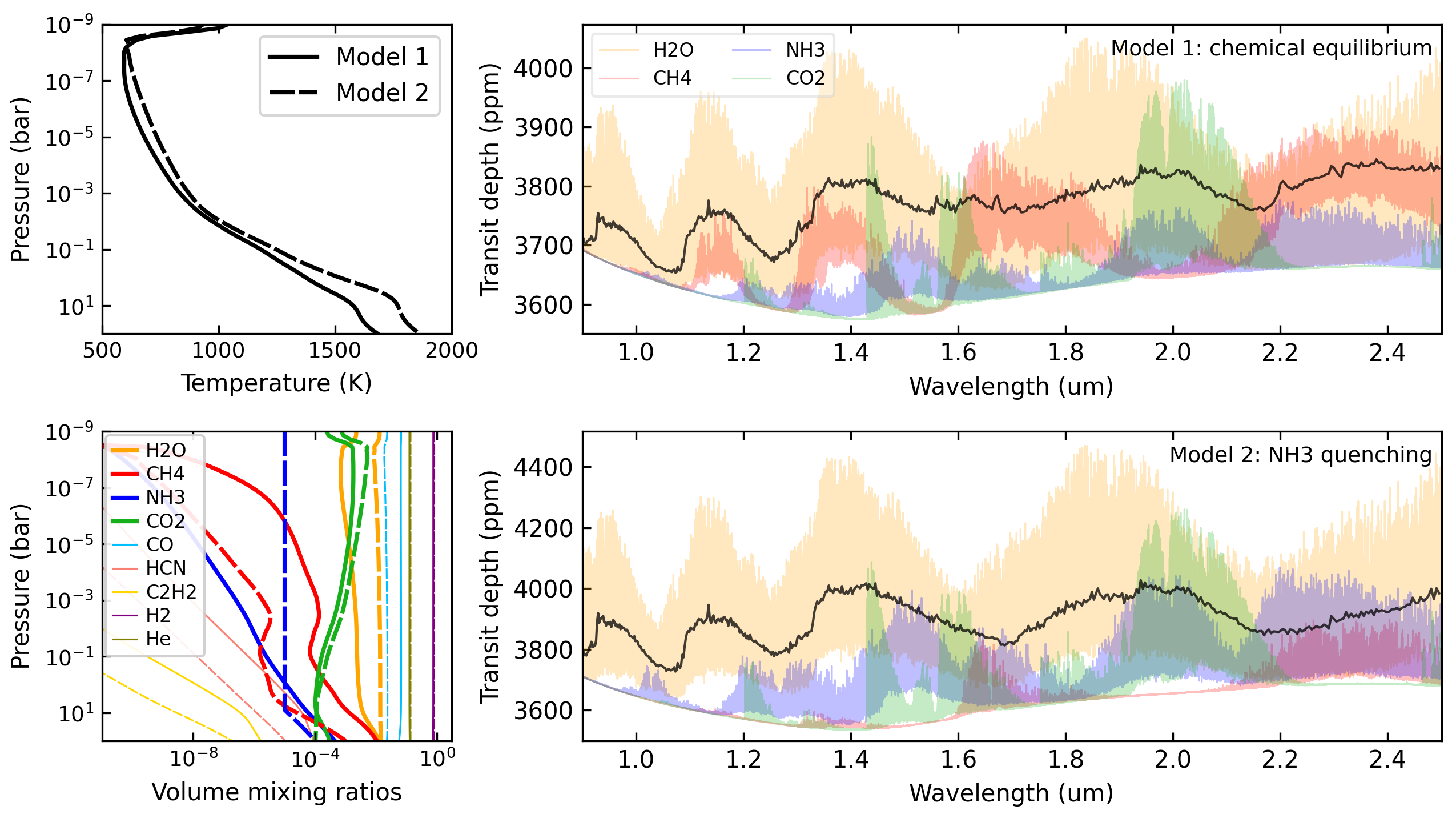}
\caption{Most favourable atmospheric chemical models. Top-left panel: HAT-P-11\,b atmospheric temperature-pressure profile corresponding to the most favourable chemical-equilibrium model that shows spectral features from all the detected species, including the tentatively detected ones (Model 1: [M/H]=2.0, C/O=0.9, N/O=0.85, $\beta_{\rm irr}$=0.5; solid line) and to a grid model modified including \ch{NH3} vertical quenching at the 10 bar level (Model 2: [M/H]=1.7, C/O=0.59, N/O=0.14, $\beta_{\rm irr}$=0.5; dashed lines). Bottom-left panel: volume mixing ratios of the dominant chemical species (see legend) as a function of the atmospheric pressure corresponding to Model 1 (solid lines) and to Model 2 (dashed line).
Right panels: theoretical transmission spectra of HAT-P-11\,b for atmospheres in thermochemical equilibrium (top-right panel) and with \ch{NH3} quenching (bottom-right panel). The coloured curves show the contribution to the synthetic spectra of the four species detected by GIANO-B, including the tentatively detected ones (see legend). The black curve shows a lower-resolution ($R = 500$) spectrum combining the absorption from all atmospheric species in the model.}
\label{fig:chemistry}
\end{figure*}

\subsubsection{Atmospheric chemical modelling}
\label{sec:chemistry}
In order to better characterise the atmospheric properties of HAT-P-11\,b, we tried to find the chemical scenarios that are most compatible with the detection of \ch{H2O} and \ch{NH3}, and possibly \ch{CH4} and \ch{CO2} (while also considering a non-detection of the other species) in the atmosphere of HAT-P-11\,b, by exploring with a grid of theoretical models the possible radiative state and elemental compositions. To this end, we employed the \textsc{Pyrat Bay} \citep{CubillosBlecic2021mnrasPyratBay} modelling framework to compute transmission spectra from an atmosphere in radiative and thermochemical equilibrium.  
The code iterates over a two-stream radiative-transfer calculation until the atmosphere converges towards a stable radiative equilibrium solution at each layer \citep{HengEtal2014apjsTwoStreamRT, MalikEtal2017ajHELIOS}.
The end products of this approach are the temperature and composition profiles of an atmosphere in radiative equilibrium, from which we can produce transmission-spectrum models to determine which species have observable spectral features.

The inputs of the models are the known system parameters, a stellar spectrum \citep{Castelli&Kurucz2003IAUS}, and the atmospheric elemental composition.  We explored a range of plausible scenarios by varying the elemental composition over a range of metallicities (from [M/H] = -1.0 to 2.0), C/O ratios (from 0.1 to 1.5), N/O ratios from (0.14 to 0.85), and over a range of heat radiation regimes by varying the $\beta_{\rm irr} = (1-A_{\rm b})/f$ parameter (from $\beta_{\rm irr} = 0.5$ to $1.0$), where $A_{\rm b}$ is the Bond albedo and $f$ is the day-night heat redistribution efficiency.

The atmospheric model spans a pressure
range from 100 to $10^{-9}$~bar, and a wavelength grid
ranging from 0.3 to 30 {$\mu$m} sampled at a resolving power of $R=15\,000$, sufficient to contain the bulk of the stellar and planetary fluxes.  The chemical network includes 45 neutral and ionic species that are the main carriers of H, He, C, N, O, Na, Si, S, K, Ti, V, and Fe. The opacity sources include line-list data for 
\ch{CO}, \ch{CO2}, and \ch{CH4} from HITEMP \citep{RothmanEtal2010jqsrtHITEMP,LiEtal2015apjsCOlineList, HargreavesEtal2020apjsHitempCH4}, and \ch{H2O}, \ch{HCN}, \ch{NH3}, and \ch{C2H2} from ExoMol
\citep{PolyanskyEtal2018mnrasPOKAZATELexomolH2O,ChubbEtal2020mnrasC2H2acetyExomol,YurchenkoEtal2011mnrasNH3opacities,HarrisEtal2006mnrasHCNlineList, HarrisEtal2008mnrasExomolHCN,ColesEtal2019mnrasNH3coyuteExomol}.
We preprocessed these large data sets with the \textsc{repack} package \citep{Cubillos2017apjRepack} to extract the dominant line transitions. Additionally, we included Na and K opacities
\citep{BurrowsEtal2000apjBDspectra}; \ch{H}, \ch{H2}, and \ch{He} Rayleigh opacities
 \citep{Kurucz1970saorsAtlas}; \ch{H2}-\ch{H2} and \ch{H2}-\ch{He} collision-induced
absorption \citep{BorysowEtal1988apjH2HeRT, BorysowFrommhold1989apjH2HeOvertones, BorysowEtal1989apjH2HeRVRT, BorysowEtal2001jqsrtH2H2highT, Borysow2002jqsrtH2H2lowT, JorgensenEtal2000aaCIAH2He}, and H$^{-}$ free-free and bound-free opacity \citep{John1988aaHydrogenIonOpacity}.
Once we obtained a grid of radiative-equilibrium atmospheric models, we post-produced transmission spectra at the GIANO-B spectral resolution.

In general, our grid of HAT-P-11\,b models indicates that the strong \ch{H2O} absorption bands dominate the transmission spectra at most wavelengths. The scenarios that are more consistent with the detected species required super-solar metallicity, since this enhances the \ch{CO2}/\ch{H2O} abundance ratio, which in turn makes \ch{CO2} detectable. Similarly, the lower-heat models improve the detection of both \ch{NH3} and \ch{CH4}, since both of these species are more abundant than at higher temperatures.  However, \ch{NH3} is never the dominant nitrogen bearer at these pressure and temperature conditions.  Consequently, only for the highest N/O ratios tested (N/O=0.85) the \ch{NH3} absorption lines impact the transmission spectra (Fig.~\ref{fig:chemistry}), this is well above the solar N/O ratio of 0.14 \citep{AsplundEtal2021aaTheSun}. Alternatively, one can invoke disequilibrium-chemistry processes to enhance the abundance of \ch{NH3} with respect to the other species \citep{Moses2014rsptaChemicalKinetics}.
Specifically, vertical quenching can transport \ch{NH3} from deeper layers where it is significantly more abundant (Fig.~\ref{fig:chemistry}, bottom-left panel). Combining these constraints with future observations of HAT-P-11\,b (e.g. JWST GO proposals 2950 and 4150) has the potential to place stronger constraints on the planet's atmospheric composition and further infer which physical processes shape its atmospheric properties.

The right-hand panels in Fig.~\ref{fig:chemistry} also highlight the complementary potential of combining low- and high-resolution observations for atmospheric characterisation.
For a molecule to be detectable, its absorption lines must be at least at the same level as the dominant species in the atmosphere; in this case, they overlap with the \ch{H2O} lines, as shown by the coloured curves on the right panels of Fig.~\ref{fig:chemistry}.
While this can be the case for several species at high resolution, instead, at low resolution (black curves) many of these individual line features are washed out and blend into the absorption of the dominant species. Therefore, with sufficient S/N, high-resolution observations can pick up molecular features that may go undetected at low resolution.

\subsubsection{How the possible presence of planet c could have influenced the formation and the atmospheric composition of planet b}
The study of the atmospheric chemical composition of close-in sub-Neptunes and Neptunes, such as HAT-P-11\,b, can give important constraints about giant-planet formation in the pebble-accretion scenario. In particular, in this scenario, the formation of these kinds of planets in the inner part of the protoplanetary disc depends on the flux of water ice-rich pebbles drifting inward from the outer regions of the protoplanetary disc ~\citep{bitsch2022}. At the water ice line, the ice on the pebbles evaporates enriching the inner part of the disc with water vapour. Substructures and cavities in the protoplanetary disc, for example, due to the formation of a Jupiter-size exoplanet, create pressure bumps that might block the inward migration of pebbles ~\citep{Banzatti2020}. In this way the presence of a rapidly growing external Jupiter-size companion and its position with respect to the water ice line could influence the chemical composition of the inner protoplanetary disc and therefore of the close-in Neptune: if the giant outer companion forms outside the water ice line it could block the water-rich pebbles before they cross the water ice line and therefore it prevents them from enriching the inner disc with water; if the giant outer companion forms inside the water ice line, the inward migrating pebbles might be blocked after they cross the water ice line and their ice evaporates enriching the inner disc with water. In the first case the close-in Neptune would form in a ``dry'' environment, in the second case it would form in a ``wet'' environment accreting a water-rich (up to a few per cent) atmosphere ~\citep{bitsch2022}. A third scenario is also possible: the close-in Neptune forms beyond the water ice line by the accretion of a large amount of water ice before migrating inwards. In this case, the Neptunian planet would contain up to 50\% of its total mass in water ice (`very wet' Neptune) and the formation time and location of the giant companion would not matter. Also carbon-bearing species can be blocked by growing giant planets, reducing the metal enrichment of the inner disc and therefore the atmosphere of the close-in Neptune. Of course, it is important to underline that protoplanetary discs evolve over time by cooling down, shifting the ice lines inwards over time and therefore both the time and the position of the forming planets inside the protoplanetary disc play an important role in this scenario (e.g. ~\citealt{eistrup2016,eistrup2022a,eistrup2022b}).

In addition, a late accretion of planetesimals or planet-planet scattering could further influence the atmospheric composition of the Neptunian planet.
For example, the high eccentricity of the orbit of HAT-P-11\,b could be explained by the interaction with a possible external companion (if any) through planet-planet scattering migration. During the migration, HAT-P-11\,b might have been enriched in heavy elements mainly through the accretion of planetesimals present in the protoplanetary disc instead of pebbles (blocked by the external companion). In the orbital region populated by HAT-P-11\,b, planetesimals are more effective in delivering \ch{O} than \ch{N} to the accreting planet ~\citep{turrini2021}. As their accretion acts to lower the high N/O ratio of giant planets formed by pebble accretion, this scenario favours the second atmospheric model we considered in Sect.~\ref{sec:chemistry}. As it can be seen, the presence of an external companion could have played a crucial role in the HAT-P-11\,b formation, influencing its atmospheric composition, as also highlighted by ~\citet{chatziastros2023}. Of course, the investigation of the presence of an outer giant companion and its characterisation, as well as the retrieval of the elemental abundance and the elemental ratios (e.g. C/O, N/O) of the chemical species present in the atmosphere of HAT-P-11\,b, are crucial to probe the formation history of the planetary system and to enlarge our knowledge about the planetary formation process.

\subsubsection{Planetary radial-velocity semi-amplitude shift investigation}
As can be seen from the $K_{\rm p}-V_{\rm rest}$ maps and from the contour plots, even though the signals of all the molecules for which we obtain a detection (including the tentative ones) cross the expected $K_{\rm p}$ and $V_{\rm rest}$ values, they span a large range of $K_{\rm p}$ and $V_{\rm rest}$ values, of the order of $\approx 100$\,km\,s$^{-1}$ and $\approx30$\,km\,s$^{-1}$, respectively. The possible presence of winds and the planetary rotation, combined with other complex atmospheric dynamics effects involving global circulation\footnote{See, for example, the results of the global circulation models that \citet{lewis2010} obtained for the eccentric warm Neptune GJ\,436\,b, whose physical and orbital characteristics are similar to those of HAT-P-11\,b.}, introduces distortions of molecular line profiles during the transit event that produce shifts of the detection signal peak in $K_{\rm p}$ and $V_{\rm rest}$ (e.g. ~\citealt{wardenier2021,kesseli2022,pelletier2023}). In particular, for the case of HAT-P-11\,b, \citet{allart2018} found evidence for a high-altitude wind flowing from the day-side to the night-side of the planet at a velocity of $v_{\rm wind}\approx3$\,km\,s$^{-1}$. For what concerns the planetary rotation speed, due to its eccentric orbit, the most probable rotation configuration for HAT-P-11\,b is the pseudo-synchronous one \citep{hut1981}, in which the planet rotation is synchronous at periastron, implying a rotation period $P_{\rm rot}$ that is shorter than the orbital one by about 29\% (for an eccentricity of $e = 0.2577$). In the case of HAT-P-11\,b, this rotation period would correspond to an equatorial rotation velocity of $v_{\rm eq}\approx1$\,km\,s$^{-1}$. As it can be seen, these kinds of effects have typical velocity scales of a few km\,s\,$^{-1}$, which is about one order of magnitude smaller than our precision. This implies that we are not sensitive to them. With future more precise measurements, it will be possible to investigate the complex atmospheric dynamics of this warm Neptune.

Even though our results do not allow us to probe the presence of atmospheric dynamics effects, we observe (see Fig.~\ref{contour}) a systematic shift of the signal peak at $K_{\rm p}$ values smaller than the expected one (with the exception of the signal of \ch{NH3} that is at the expected $K_{\rm p}$). In order to quantify this shift, we made a weighted average of the $K_{\rm p}$ values corresponding to the peak of significance for each of the detected molecules (including the tentatively detected one), assuming as the extremes of the $1\,\sigma$ uncertainty intervals the $K_{\rm p}$ values where the significance drops by $1\,\sigma$ from the peak (after having marginalised over the $V_{\rm rest}$ values).
The averaged value of the measured $K_{\rm p}$ is: $K_{\rm p_{meas}}=107\pm29$\,km\,s$^{-1}$, which is compatible at $0.54\,\sigma$ with the expected value reported in Table~\ref{tab1}.

We note that the large uncertainty in $K_{\rm p}$, due to the relatively small change in planet RV during transit (a few tens of km\,s$^{-1}$), makes the systematically observed downward shift in $K_{\rm p}$ of the order of $\approx10$\,km\,s$^{-1}$ not significant on a statistical basis. However, we decided to investigate the possible presence of a systematic effect by studying how the uncertainties on $e$ and $\omega_{\rm p}$ affect the retrieved value of $K_{\rm p}$ and therefore the position of the peak of significance in the $K_{\rm p}-V_{\rm rest}$ maps. These two parameters are the two main sources of uncertainties in determining the planetary orbital configuration and the atmospheric Doppler signature \citep{montalto2011}.

We generated $N=500\,000$ couples of values of orbital eccentricity and planetary argument of periastron ($e$; $\omega_{\rm p}$), extracted randomly from two Gaussian distribution (one for $e$ and one for $\omega_{\rm p}$) with mean equal to the values of these two parameters that we used in this work ($e=0.2577$; $\omega_{\rm p}=192.0$\,deg) and standard deviation equal to the $1\,\sigma$ uncertainties on these values. For each ($e$; $\omega_{\rm p}$) couple we computed the planetary RV curve during the transit combining equation~\ref{rvp} and ~\ref{kp_eq2}:
\begin{equation}
RV= \Tilde{K}_{\rm p} \cdot[\cos{(\nu(t)+\omega_{\rm p})}+e\cdot \cos({\omega_{\rm p}})]\cdot\frac{1}{\sqrt{1-e^2}}
\label{kp_eq3}
\end{equation}
fixing the $\Tilde{K}_{\rm p}$ value to the expected one ($\hat{\Tilde{K}}_{\rm p}=119.3$\,km\,s$^{-1}$). In this way, we simulate different possible orbital configurations according to our limited knowledge of the ($e$; $\omega_{\rm p}$) parameters. Subsequently, we fitted the $\Tilde{K}_{\rm p}$ value for each of the $N$ RV curves keeping fixed the ($e$; $\omega_{\rm p}$) values to the ones we used in this work. 

\begin{figure}
  \resizebox{\hsize}{!}{\includegraphics{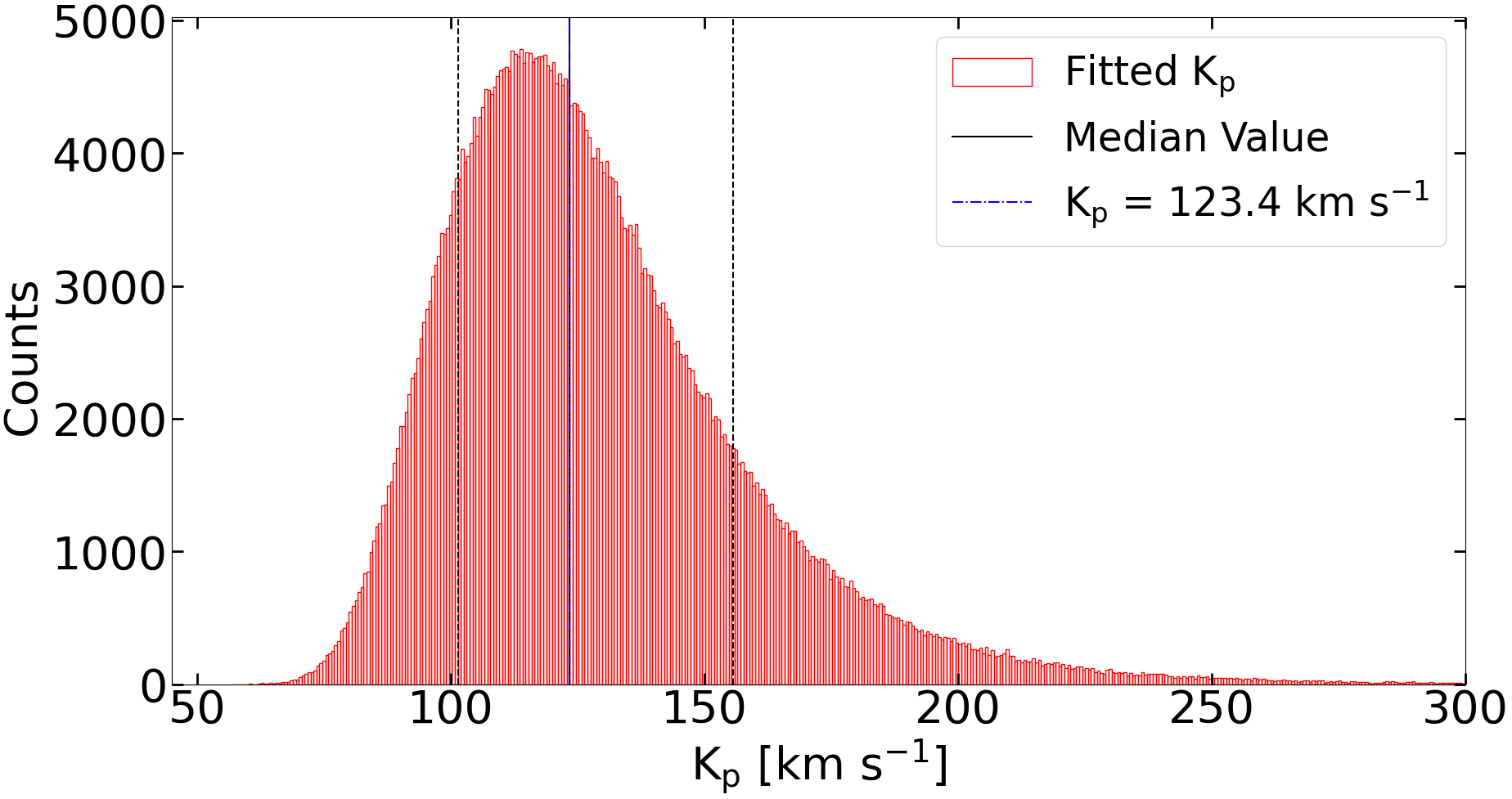}}
  \caption{Distribution of $K_{\rm p}$ values fitted on randomly generated planetary RV curves with properties described in the text. The vertical black solid line represents the median value of the distribution, while the 2 vertical black dashed lines represent the 2 quantile values delimiting the $68\%$ ($1\,\sigma$) interval of $K_{\rm p}$ around the median. The vertical dash-dotted blue line represents the expected $K_{\rm p}$ value: $\hat{K}_{\rm p}=123.4$\,km\,s$^{-1}$. The median and the expected values are coincident and therefore they are not visibly distinguishable.}
  \label{kpplot}
\end{figure}

Following this procedure, we obtained a distribution of retrieved $\Tilde{K}_{\rm p}$ values that we transformed into a distribution of $K_{\rm p}$ values (using the $e$ value reported in Table~\ref{tab1}) in order to compare them with the $K_{\rm p}$ value obtained from the significance maps. This distribution is reported in Fig.~\ref{kpplot}. With this procedure we simulated the effect of the choice of ($e$; $\omega_{\rm p}$) used in this work on the $K_{\rm p}$ value estimation from the peak of the CCF signal, taking into account our limited knowledge of the orbital solution.

The RV semi-amplitude value is strictly related to the slope of the planetary RV curve. In our data analysis, the slope of the planetary RV curve is not directly fitted but, since it influences the alignment of the CCF trails, it changes the position of the peak of significance in the $K_{\rm p}-V_{\rm rest}$ maps and therefore the value of $K_{\rm p}$ we can estimate from the maps.

The first result worth noting is that the $K_{\rm p}$ distribution is asymmetric with a long tail extending towards $K_{\rm p} > 150$\,km\,s$^{-1}$ values. In terms of the properties of the distribution, its median value coincides with the expected value and by computing the $1\, \sigma$ interval of this distribution, the estimated value of $K_{\rm p}$ can be obtained: $\overline{K}_{\rm p} = 123^{+32}_{-22}$\,km\,s$^{-1}$. This is consistent with the expected value but, as it can be seen, the distribution has not a peak at the expected $K_{\rm p}$ value, instead, the modal value is: $K_{\rm p_{\rm mode}} = 114$\,km\,s$^{-1}$. This means that the most probable measured $K_{\rm p}$ value is $\approx10$\,km\,s$^{-1}$ smaller than the expected one, which is consistent with our measured $K_{\rm p_{\rm meas}}$ value. From this investigation, two important considerations can be made: 1) even if the values of eccentricity and argument of periastron are very precise ($\approx 1$\%), the $K_{\rm p}$ value that can be extracted from our kind of data analysis cannot be known at better than $\approx25$\%, due to the relatively small change in planet RV during transit; 2) this investigation suggests the presence of a systematic effect that we do not fully understand that makes more probable to observe a shift of the signal of the order of $\approx10$\,km\,s\,$^{-1}$ towards $K_{\rm p}$ values lower than the expected one, at least in the case of the analysed target, related to the uncertainties about the orbital solution adopted. This has to be taken into consideration for future atmospheric dynamics studies, since in that case precise but also accurate measurements are needed.

\section{Conclusion}
\label{conclusion}
In this work, we revisited the HAT-P-11 planetary system. Using {\it Kepler} and HIRES at Keck archival data, we refined the orbital and physical parameters of HAT-P-11\,b. We further showed that the long-term RV signal with a semi-amplitude of 30\,m\,s$^{-1}$ and periodicity of $\sim 9-10$~yr is more likely due to the stellar activity cycle than to the presence of the planet HAT-P-11\,c \citep{yee2018}. Nonetheless, the Hipparcos--Gaia difference in proper-motion anomaly suggests that an outer-bound companion might still exist, though with $\rm S/N\lesssim 5$. The continuation of RV monitoring and, even more importantly, a combined analysis of RV and future DR4 Gaia astrometric data, will 
help to characterise this possible companion.

This review of the HAT-P-11 planetary system allowed us to perform a consistent analysis of the atmosphere of planet b at high spectral resolution.
Moreover, HAT-P-11\,b represents a remarkable target as it can provide a better understanding of the atmospheres of warm Neptunes, which remain to be explored in detail.

In particular, we probed the presence of eight chemical species in the atmosphere of HAT-P-11\,b by cross-correlating template atmospheric models with data taken with the NIR GIANO-B high-resolution spectrograph at the $3.58$\,m TNG telescope during four planetary transits. We detect the presence of two molecular species, \ch{H2O} and \ch{NH3}, with a S/N of $5.1$ and $5.3$, and a significance of $3.4\,\sigma$ and $5.0\,\sigma$, respectively. The signals from these molecules have a Doppler signature compatible with the planetary one at $<1\, \sigma$. We also tentatively detect the presence of \ch{CH4}, whose signal has a S/N of $4.8$ but a significance level of $2.6\,\sigma$, and of \ch{CO2}, with a S/N of $3.0$ and a significance level of $3.2\,\sigma$. Further studies are necessary to confirm the presence of \ch{CH4} and \ch{CO2} in the atmosphere of HAT-P-11\,b. These results constitute the first simultaneous observation of multiple molecular species in the atmosphere of a warm Neptune-type planet.

Our results are in accordance with the previous water vapour detections made by \citet{fraine2014}, \citet{tsiaras2018}, \citet{chachan2019}, and \citet{cubillos2022} and with the suggestion of the presence of methane by \citet{chachan2019} and \citet{cubillos2022}. We hereby enlarge the number of chemical species known to be present in the atmosphere of this exoplanet, which allows us to start exploring plausible physical conditions for the atmosphere of HAT-P-11\,b. Our models suggest two scenarios that are more in accordance with the observations: the first model describes an atmosphere in chemical equilibrium with supersolar metallicity and enhanced C/O and N/O ratios relative to solar values; the second model describes an atmosphere with disequilibrium chemistry (i.e. \ch{NH3} vertical quenching), lower metallicity, and C/O and N/O ratios close to solar values.

We note that the significance peak of the detected species (including the tentatively detected ones) is shifted by $\approx10$\,km\,s$^{-1}$ towards $K_{\rm p}$ values smaller than the expected one ($\hat{K}_{\rm p}=123.4$\,km\,s$^{-1}$). Indeed the average value of the planetary RV semi-amplitude estimated from our significance maps is $K_{\rm p_{\rm meas}}=107\pm29$\,km\,s$^{-1}$, which is compatible at $0.54\ \sigma$ with the expected one due to the large uncertainty. We show how a small error ($\approx 1$\%) on the eccentricity and argument of periastron parameters translates into a large uncertainty ($\approx25$\%) on the retrieved $K_{\rm p}$ parameter due to the small change in RV during the transit, and can introduce the observed systematic $K_{\rm p}$ shift.

The next step in our analysis will be to statistically retrieve the chemical physical properties of the atmosphere of HAT-P-11\,b (e.g. the elemental abundances and the temperature-pressure profile) in order to better characterise its atmosphere and to constrain its formation path. Finally, a future combination of high- and low-resolution observations will improve the characterisation of the atmosphere of this target and our knowledge about warm Neptunes.

\begin{acknowledgements}
This paper includes data collected by the {\it Kepler} mission and obtained from the MAST data archive at the Space Telescope Science Institute (STScI). Funding for the {\it Kepler} mission is provided by the NASA Science Mission Directorate. STScI is operated by the Association of Universities for Research in Astronomy, Inc., under NASA contract NAS 5–26555. 
This work has made use of data from the European Space Agency (ESA) mission {\it Gaia} (\url{https://www.cosmos.esa.int/gaia}), processed by the {\it Gaia} Data Processing and Analysis Consortium (DPAC,
\url{https://www.cosmos.esa.int/web/gaia/dpac/consortium}). Funding for the DPAC has been provided by national institutions, in particular, the institutions participating in the {\it Gaia} Multilateral Agreement.
The authors acknowledge the financial contribution from PRIN INAF 2019, and from the European Union - Next Generation EU RRF M4C2 1.1 PRIN MUR 2022 project 2022CERJ49 (ESPLORA) and project 2022J4H55R.
A.\,S. acknowledges support from the Italian Space Agency (ASI) under contract 2018-24-HH.0 ``The Italian participation to the Gaia Data Processing and Analysis Consortium (DPAC)'' in collaboration with the Italian National Institute of Astrophysics.
The following internet-based resources were used in research for this paper: the ESO Digitized Sky Survey; the NASA Astrophysics Data System; the SIMBAD database operated at CDS, Strasbourg, France; and the ar$\chi$iv scientific paper preprint service operated by the Cornell University.
A.\,M. acknowledges partial support from the ASI-INAF agreement n.2018-16-HH.0 (THE StellaR PAth project).
P.\,E.\,C. is funded by the Austrian Science Fund (FWF) Erwin Schroedinger Fellowship, program J4595-N.
\end{acknowledgements}

\bibliographystyle{aa}
\bibliography{bibl.bib}

\begin{appendix}
\section{Stages of the GIANO-B data reduction process}
\label{app_stage}
\begin{figure*}
\centering
\resizebox{\hsize}{!}{\includegraphics{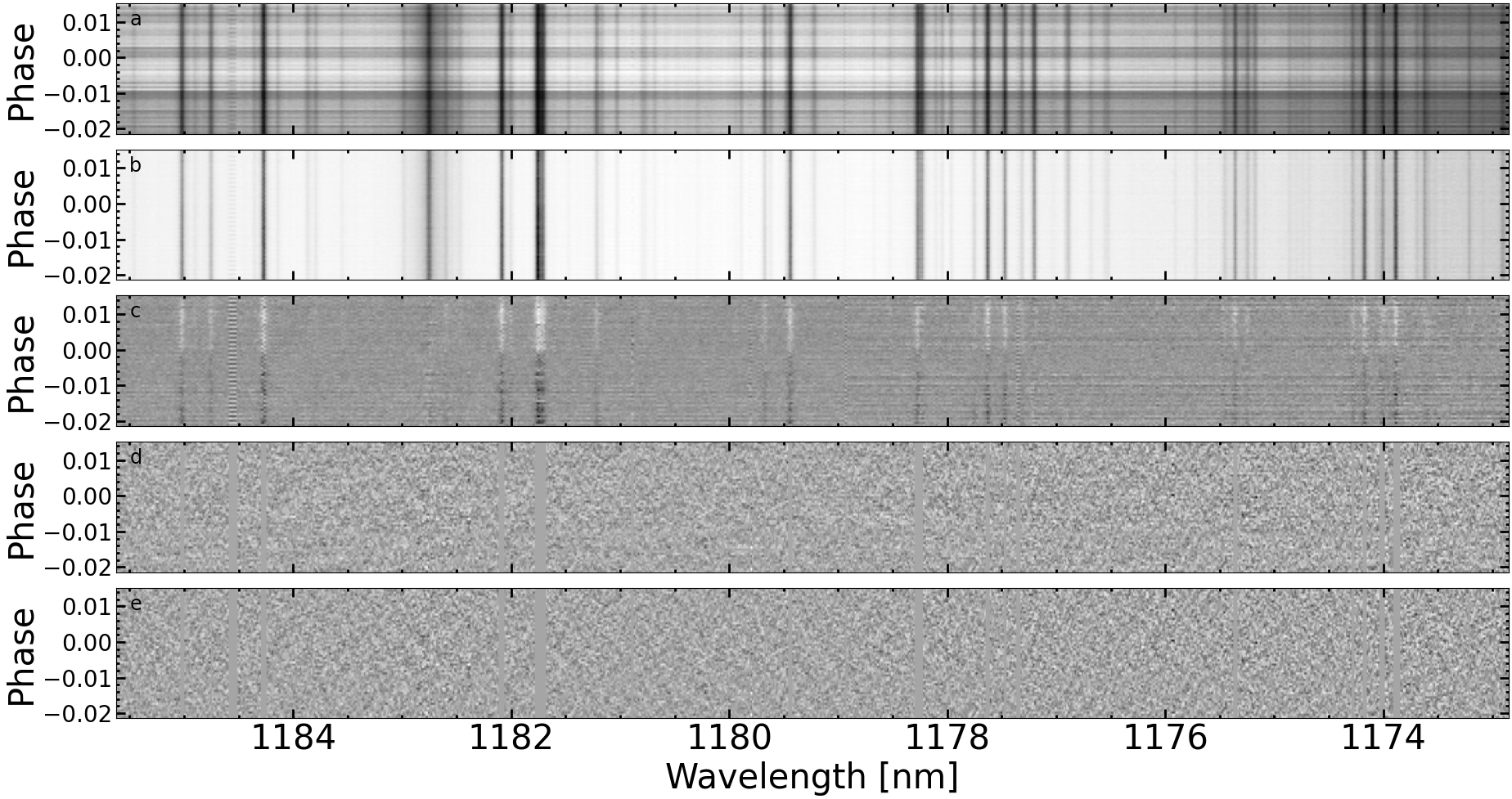}}
     \caption{Example of the stages of the GIANO-B data reduction process applied to real GIANO-B spectra over a short wavelength interval (\textit{x}-axis). The planetary orbital phase is reported on the \textit{y}-axis . From top to bottom panel: (a) the extracted spectra; (b) residual spectra after having normalised each spectrum (each row) by its median value to correct baseline flux differences between the spectra that are due, for example, to variable transparency of the atmosphere, imperfect telescope pointing, or instability of the stellar point spread function; (c) residuals after each spectral channel (each column) had its mean subtracted;
     (d) residuals after having masked the strongest or saturated telluric lines and having removed the telluric and stellar spectra with the PCA; as it can be seen also other spurious effects such as the alternating pattern visible at certain wavelengths likely related to the instrument A-B nodding are effectively removed with PCA; (e) residuals after having applied a high-pass filter to each row of the residual matrix (this step mitigates any residual correlation between different spectral channels).}
     \label{pca}
\end{figure*}

Figure~\ref{pca} shows an example of the stages of our data reduction process applied to real GIANO-B spectra, over a short wavelength interval, from the extracted spectra to the residuals after the PCA application.\\

\section{Distribution of the cross-correlation function values }
\begin{figure*}[!h]
\centering
   \includegraphics[width=17cm]{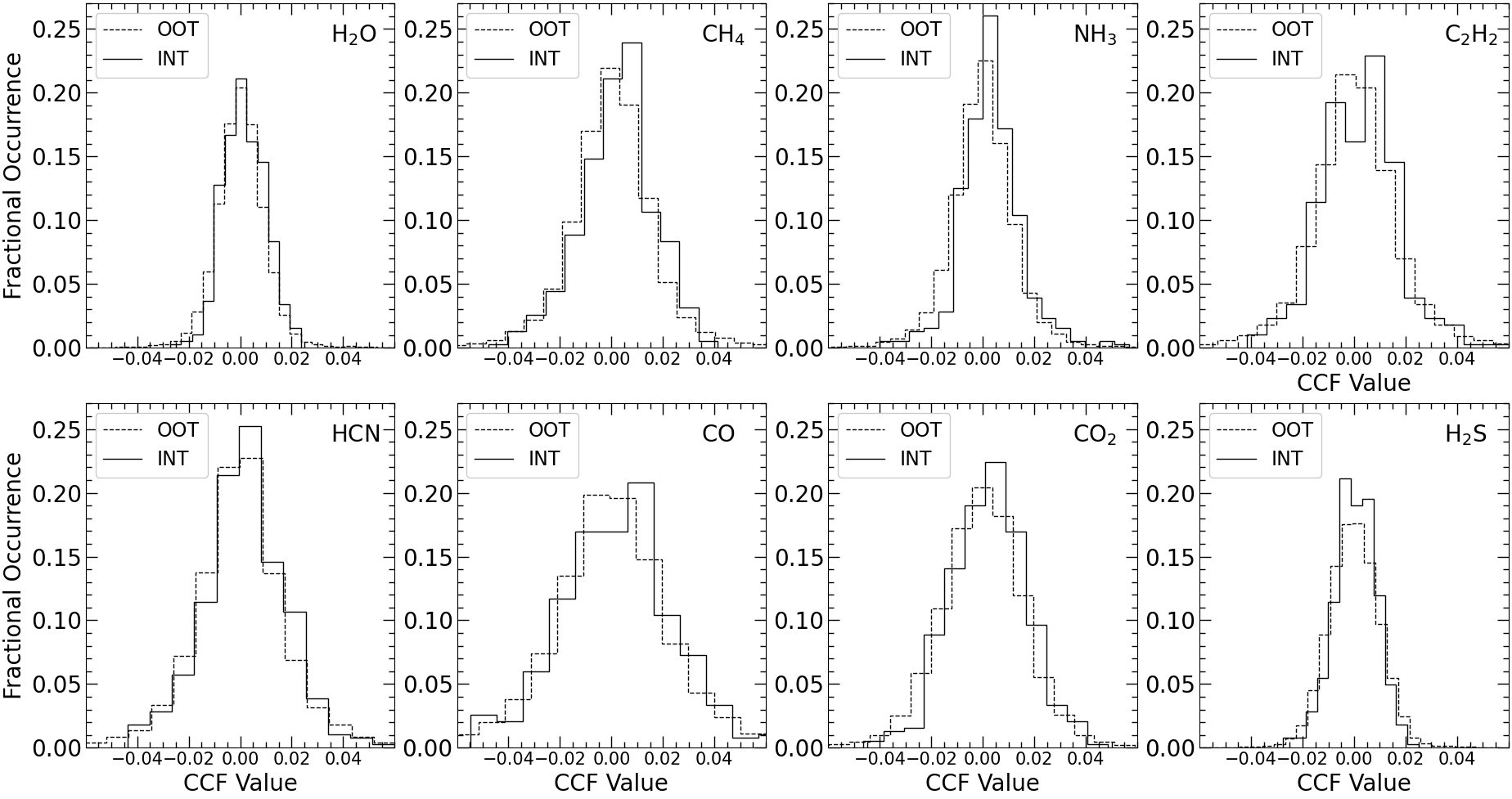}
     \caption{Distribution of CCF values far ($|V_{\rm rest}|>25$\,km\,s$^{-1}$) from the planet's rest-frame velocity ("out-of-trail", OOT, dashed lines) and near to it ($|V_{\rm rest}|<3$\,km\,s$^{-1}$, "in-trail", INT, solid lines) for the probed chemical species.}
     \label{ttest_histo4}
\end{figure*}

Figure~\ref{ttest_histo4} shows the distribution of CCF values far ($|V_{\rm rest}|>25$\,km\,s$^{-1}$) from the planet's rest-frame velocity (``out-of-trail'') and near to it ($|V_{\rm rest}|<3$\,km\,s$^{-1}$, ``in-trail'') for the 3 detected molecular species. The Welch \textit{t-test} used to estimate the significance of the detections was performed on these distributions.

\section{Test on the reliability of the detections}
\begin{figure*}
\centering
   \includegraphics[width=17cm]{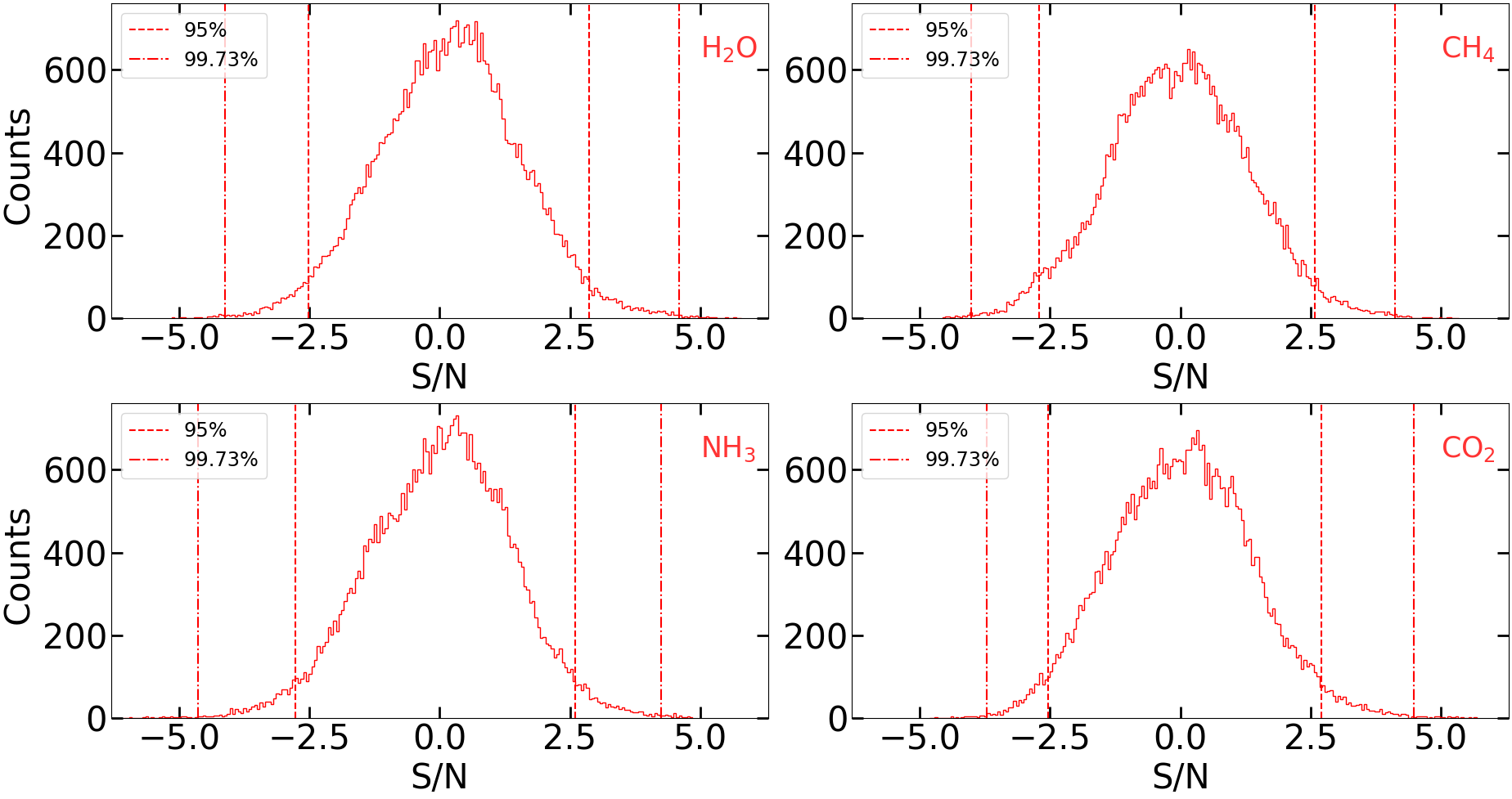}
     \caption{Distribution of S/N obtained in the selected $K_{\rm p}-V_{\rm rest}$ interval by shuffling 250 times the time order of the observed spectra, for each of the four selected chemical species. The vertical dashed (dash-dotted) lines represent the borders of the intervals enclosing the 95\% (99.73\%) of the S/N values, corresponding to the $2.5\%-97.5\%$ ($0.135\%-99.865\%$) quantiles of the distributions. These intervals are reported in Table~\ref{test_res}.}
     \label{shuffle_snr}
\end{figure*}

\begin{figure*}
\centering
   \includegraphics[width=17cm]{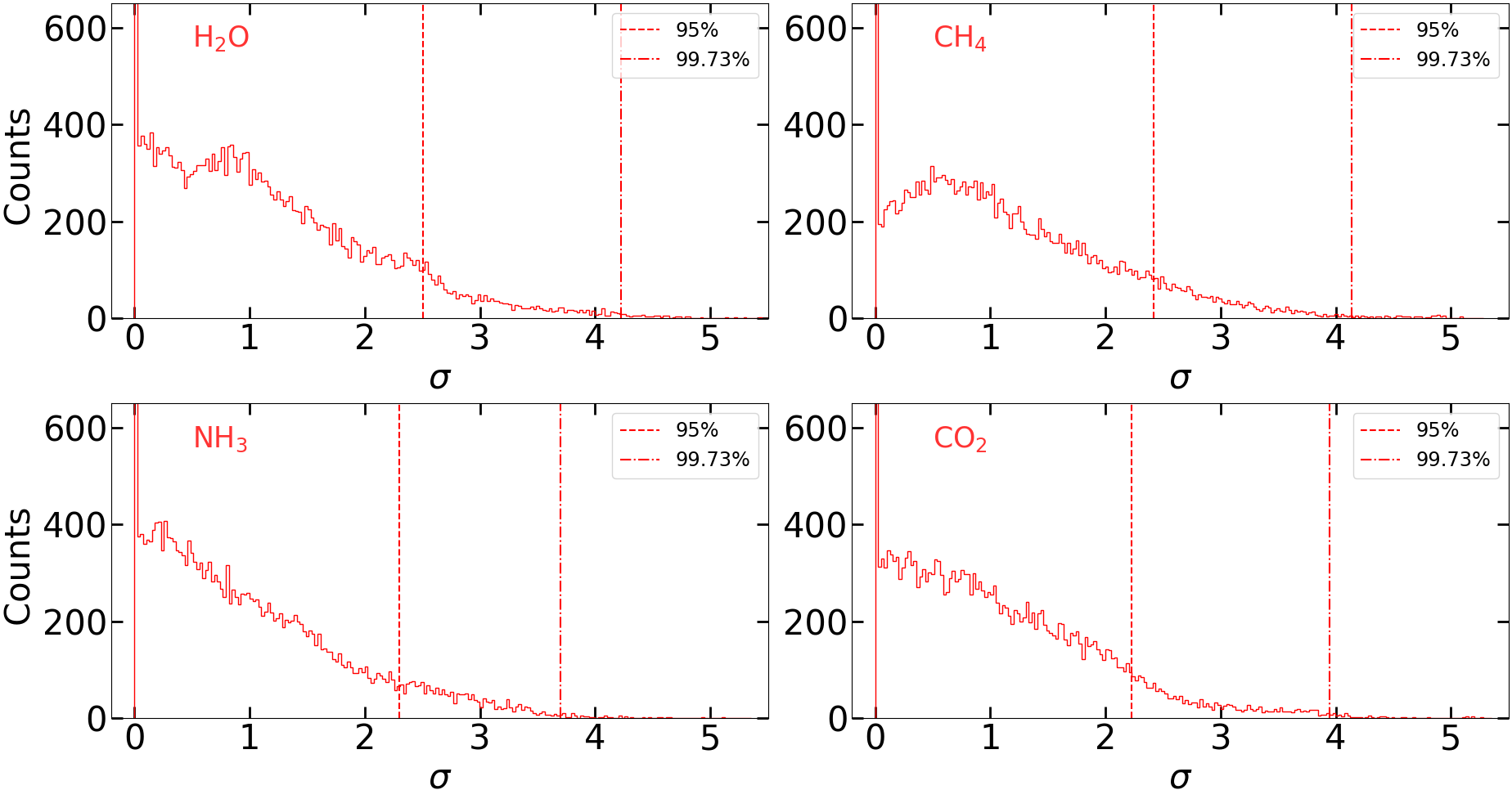}
     \caption{Distribution of significance (from the \emph{t-test}) obtained in the selected $K_{\rm p}-V_{\rm rest}$ interval by shuffling 250 times the time order of the observed spectra, for each of the four selected chemical species. The vertical dashed (dash-dotted) lines represent the borders of the intervals enclosing the 95\% (99.73\%) of the significance values, corresponding to the $95\%$ ($99.73\%$) quantile of the distributions. These intervals are reported in Table~\ref{test_res}. Since the 50\% of the significance values ($\sim20\,000$ values) are $<0.2\ \sigma$ for all the 4 chemical species, the plots of the distributions are limited on the \textit{y}-axis in the interval [0; 650] for clarity.}
     \label{shuffle_tmap}
\end{figure*}

In order to check the reliability of the detections, we combined the matrices of the CCF as a function of the orbital phase of the four observing nights in a single CCF matrix with the rows sorted in the crescent orbital phase. We then randomly shuffled the CCF order in phase 250 times. For each shuffle, we built both the S/N and \emph{t-test} significance $K_{\rm p}-V_{\rm rest}$ maps in an interval of $K_{\rm p} =$ [$89.4; 156.5$]\,km\,s$^{-1}$ ($\hat{K}_{\rm p}\pm34$\,km\,s$^{-1}$) and $V_{\rm rest}=$ [$-10;10$]\,km\,s$^{-1}$, in order to test the presence of spurious signals due to correlated noise that can produce significant features around the expected signal position in the maps. We repeated this test for all four detected chemical species for a total of 250 combinations, 43\,750 values of S/N, and 43\,750 values of significance for each species. In Fig.~\ref{shuffle_snr} and Fig.~\ref{shuffle_tmap} we report the distributions of S/N and significance values obtained in the selected $K_{\rm p}-V_{\rm rest}$ interval, respectively, for the 4 detected species.

\section{Orders' selection and PCA procedure}
\label{app_os_pca}
In this section we briefly describe the orders' selection procedure and the PCA procedure we applied to data.

We followed ~\citet{giacobbe2021} for the spectral orders' selection procedure, per molecule and per night. In particular, we injected single-species model planetary spectra at planetary radial velocity into the observations just before the PCA procedure. We then searched for these artificial signals and measured their significance order by order, by using the same procedure as for the molecule detection. The injected models were computed as those used for cross-correlation, except that they were amplified to be detectable at $> 3\,\sigma$ potentially in each order. Therefore an order was selected when the most significant signal was recovered within $\pm3$\,km\,s$^{-1}$ from $V_{\rm rest}=0$\,km\,s$^{-1}$ and $\pm30$\,km\,s$^{-1}$ from the expected $K_{\rm p}$, with significance greater than $3\,\sigma$. We underline that the injected signal is not used to optimise the number of PCA components via its recovery maximisation, but it is used only with the aim of selecting the orders where molecular absorption for a single species is likely to occur and/or the contamination by telluric residuals and other systematic effects (e.g. the wavelength-dependent efficiency of GIANO-B) is minimised.

Once we selected the spectral orders, we applied the PCA procedure. This procedure was applied to data of each spectral order selected for a particular molecular species, for each night. The aim of the PCA procedure is to find common spectral trends in the different wavelength channels (i.e. telluric and stellar lines) as a function of time and remove them. It consists of computing the principal components of the data matrix of each spectral order. In particular, each spectral order is an (M-rows; N-columns) data matrix ($M=58-62$ spectra, $N = 2048$ pixels) and the principal components correspond to the eigenvectors of its covariance matrix. For each eigenvector, the associated eigenvalue denotes the contribution of that component to the data variance.

First, we normalised each spectrum to its median value, to correct baseline flux variations. Then, we masked the spectral channels with stronger or saturated telluric lines. To do this, we subtracted the time-averaged spectrum to all the spectra (i.e. we subtracted to each spectral channel its mean value), computed the standard deviation of data in each spectral channel and its median value ($\sigma_{\rm m}$), and then we masked the spectral channels that had a standard deviation $\sigma_{\lambda} > 1.5\cdot\sigma_{\rm m}$. After subtracting the time-averaged spectrum from all the spectra and masking highly contaminated spectral channels, we reduced data to a null-mean variable by subtracting to each spectrum its mean value computed on the different spectral channels. Then, the principal components were computed using the \textsc{PCOMP} IDL routine\footnote{\url{https://www.nv5geospatialsoftware.com/docs/PCOMP.html}}. The output eigenvectors were ordered in decreasing contribution to the variance and we selected the minimum number of components that together contributed up to 70\% to the data-variance. Once the number of principal components was selected, the matrix that should mainly describe the telluric and stellar contaminations was built via a linear combination of the principal components and removed from the original data matrix, obtaining an ideal residual telluric- and stellar-free matrix. Finally, we applied a high-pass filter to each row of the residual matrix to remove any possible residual correlation between different spectral channels. The goodness of the telluric (or stellar) lines removal was evaluated by a visual inspection of the residual matrix of each spectral order and of the cross-correlation values as a function of planetary orbital phase computed between data and the \ch{H2O} (or \ch{CO}) model (see, e.g. Fig.~\ref{ccftrailreal}).

In Table~\ref{pca_comp} we report the selected spectral orders and the number of principal components removed by PCA for each selected spectral order, for the different molecular species and nights.

\begin{table*}[!h]
\caption{Selected orders (Sel. Ord.) and number of principal components removed by PCA (PCA Comp.) for each selected order, for the different molecular species and nights.} 
\label{pca_comp} 
\centering
\begin{tabular}{l c c c c}
\hline
\hline\\ [-2pt]
     & 7 July 2019 & 18 June 2020 & 19 September 2020 & 13 June 2023\\[4pt]
\hline\\[2pt]
\multicolumn{1}{l}{\textbf{\ch{H2O}}} \\[2pt]
    Sel. Ord. & $[3,5,6,7,11,12,16,$ & $[21,22,26,32,33,34,35,36]$ & $[6,12,15,17,26,34]$ & $[2,6,13,16,17,18,19,$\\
     & $21,26,27,32,36,37]$ & & & $20,21,26,28,34,36,37]$ \\
    PCA Comp. & $[20,17,9,12,17,19,20,$ & $[9,9,16,17,17,16,10,9]$ & $[9,21,23,22,14,14]$ & $[22,9,21,21,20,20,16,$\\
     & $14,14,15,14,9,10]$ & & & $9,9,9,9,9,9,9]$\\[10pt]
\multicolumn{1}{l}{\textbf{\ch{CH4}}} \\[2pt]
    Sel. Ord. & $[4,13,18,20]$ & $[1,3,7,12,15]$ & $[3]$ & $[15,38]$\\
    PCA Comp. &  $[19,20,20,15]$ & $[20,20,9,22,23]$ & $[21]$ & $[21,9]$\\[10pt]
\multicolumn{1}{l}{\textbf{\ch{NH3}}} \\[2pt]
    Sel. Ord. & $[3,5,6,11,14,15,17,$ & $[2,3,19,21]$ & $[6,7]$ & $[2,5,7,11,12,13,19,$\\
     & $18,26,35]$ & & & $20,31]$\\
    PCA Comp. &  $[20,17,9,17,21,20,19,$ & $[21,20,20,9]$ & $[9,9]$ & $[22,12,9,9,20,21,16,$\\
     & $20,14,11]$ & & & $9,9]$\\[10pt]
\multicolumn{1}{l}{\textbf{\ch{C2H2}}} \\[2pt]
    Sel. Ord. & $[13,25]$ & $[0,2,3,18,28]$ & $[36]$ & $[3,4,17,19]$\\
    PCA Comp. &  $[20,9]$ & $[23,21,20,23,21]$ & $[9]$ & $[20,21,20,16]$\\[10pt]
\multicolumn{1}{l}{\textbf{\ch{HCN}}} \\[2pt]
    Sel. Ord. & $[15]$ & $[2,17]$ & $[4,16,20]$ & $[13,15,16,17,19,20,22]$\\
    PCA Comp. &  $[20]$ & $[21,22]$ & $[21,22,12]$ & $[21,21,21,20,16,9,9]$\\[10pt]
\multicolumn{1}{l}{\textbf{\ch{CO}}} \\[2pt]
    Sel. Ord. & $[1]$ & $[14,15]$ & $[16,17]$ & $[1]$\\
    PCA Comp. &  $[21]$ & $[23,23]$ & $[22,22]$ & $[21]$\\[10pt]
\multicolumn{1}{l}{\textbf{\ch{CO2}}} \\[2pt]
    Sel. Ord. & $[16,17,21]$ & $[14,20]$ & $[6,30]$ & $[6,7,17]$\\
    PCA Comp. &  $[9,19,14]$ & $[23,13]$ & $[9,20]$ & $[9,9,20]$\\[10pt]
\multicolumn{1}{l}{\textbf{\ch{H2S}}} \\[2pt]
    Sel. Ord. & $[14,15,16,17,18,19]$ & $[0,4,6,7,14,15,16,$ & $[14,15,16,17,18,19]$ & $[14,15,16,17,18,19]$\\
     & & $26,27]$ & & \\
    PCA Comp. &  $[21,20,20,19,20,18,]$ & $[23,21,9,9,23,23,22,$ & $[22,23,22,22,23,21]$ & $[22,21,21,20,20,16]$\\
     & & $16,20]$ & & \\
\hline                                            
\end{tabular}

\end{table*}

\section{Test on the reliability of the orders' selection procedure}
\label{test_ord}
The spectral orders' selection procedure described in the previous section is important to select the orders where molecular absorption for a single species is likely to occur considering different systematic effects (e.g. night-by-night variations of the observing conditions, telluric contamination, efficiency of wavelength calibration). To probe the possible presence of spurious positive signals near the expected planetary radial velocity that could lead to possible false positive molecular signals through the orders' selection procedure we adopted, we performed the following test.\\

\begin{figure*}
\centering
   \includegraphics[width=15cm]{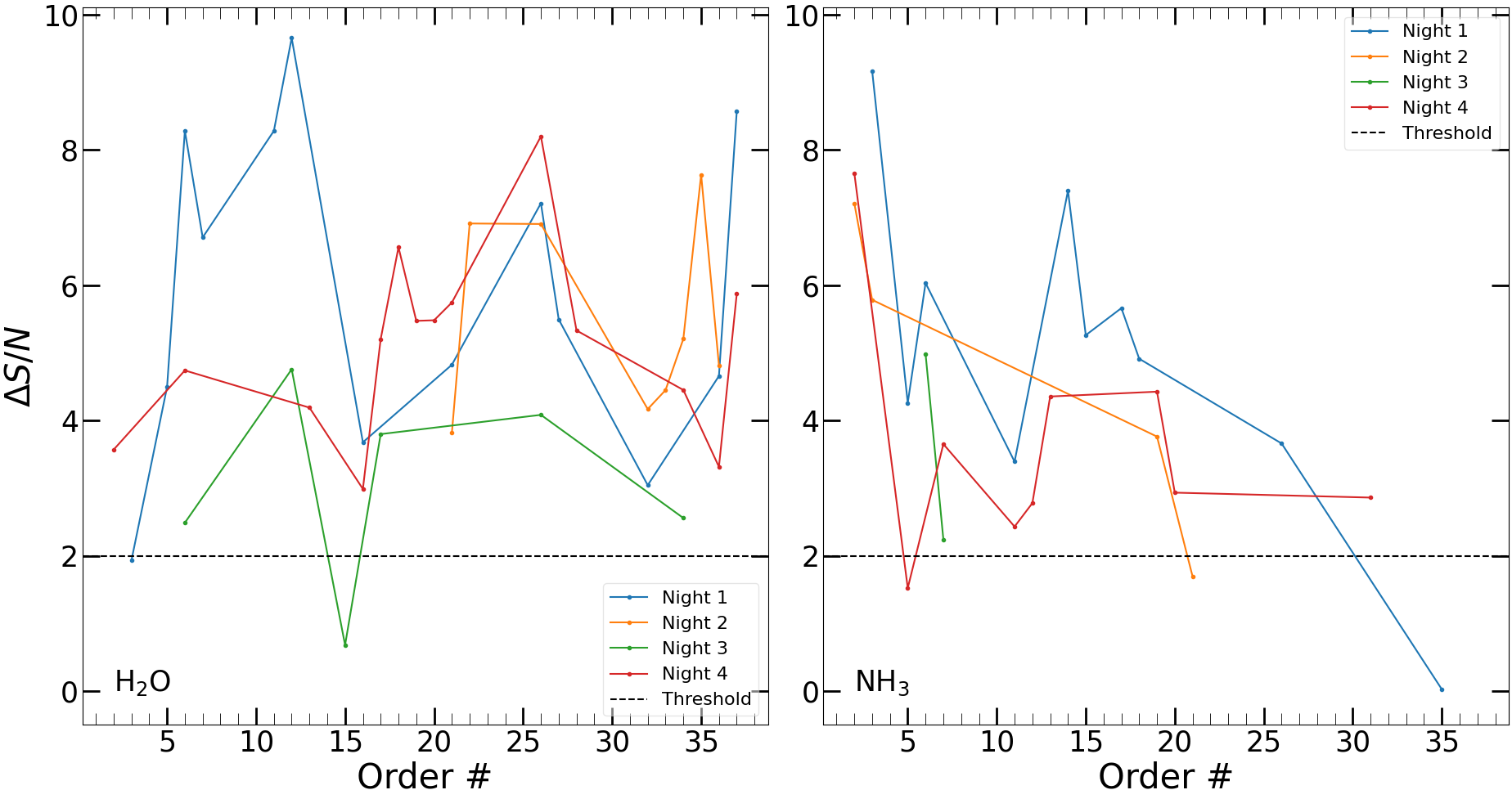}
     \caption{Change in S/N ($\Delta S/N$) due to the injection of the model of \ch{H2O} (left) and \ch{NH3} (right) into the data, as a function of the selected spectral orders in the different nights. The black horizontal dashed line represents the $2\ \sigma$ threshold.}
  \label{deltasnr}
\end{figure*}

For each of the two detected molecules (i.e. \ch{H2O} and \ch{NH3}) and for each night, we started from the spectral orders' lists selected with the procedure described in Appendix~\ref{app_os_pca} (and reported in Table~\ref{pca_comp}) and refined the selection in the following way:\\
- step 1) for each of the selected spectral order, we applied the PCA and computed the signal-to-noise ratio (S/N) $K_{\rm p}-V_{\rm rest}$ maps by cross-correlating the spectra of that order with the single-species model for the tested molecule;\\
- step 2) for each of the selected spectral order, we injected the single-species model for the tested molecule into the data at the planetary radial velocity and then applied the PCA and computed the signal-to-noise ratio (S/N) $K_{\rm p}-V_{\rm rest}$ maps by cross-correlating them with the model itself;\\
- step 3) finally, for each spectral order, we computed the change ($\Delta S/N$) in the S/N of the CCF peak at the planetary radial velocity obtained in the two previous steps (i.e. data alone and data+injected model) and selected the orders that showed a $\Delta S/N \geq 2\ \sigma$.\\
With this kind of selection, we selected the orders that not only allowed to retrieve the injected signal but also showed a significant increase of the S/N of the CCF peak when the model itself was injected into the data, further minimising the possible presence of spurious systematics in the final dataset that could lead, for example, to false positive signals. In Fig.~\ref{deltasnr} we report the $\Delta S/N$ for the different selected orders and nights. As it can be for most of the selected spectral orders the increase of S/N is $> 2\ \sigma$.

Given the $2\ \sigma$ threshold we discarded the following orders from the original lists:\\
- for \ch{H2O}: order 3 (night 1), order 15 (night 3);\\
- for \ch{NH3}: order 35 (night 1), order 21 (night 2), order 5 (night 4).

\begin{figure*}
\centering
   \includegraphics[width=14cm]{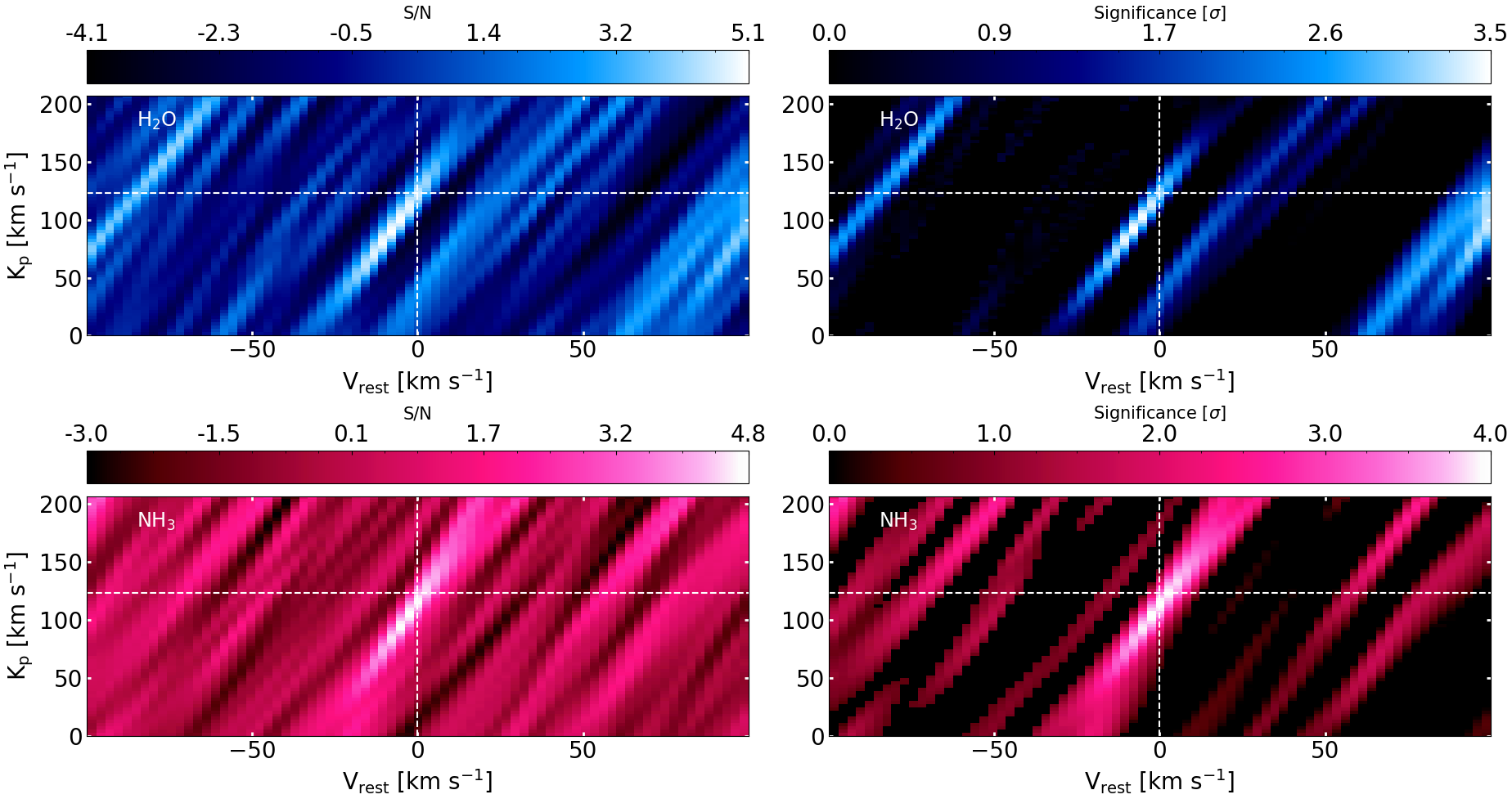}
     \caption{Signal-to-noise ratio (left panels) and Welch \emph{t-test} significance (right panels) $K_{\rm p}-V_{\rm rest}$ maps for the 2 tested chemical species (top for \ch{H2O}, bottom for \ch{NH3}) obtained by combining data from the four observing nights excluding spectral orders with $\Delta S/N<2\ \sigma$. The point where the 2 white dashed lines cross each other is the expected position of a signal with planetary origin.}
  \label{mappe_test1}
\end{figure*}

We then repeated the telluric removal procedure and the CCF analysis using the updated selected orders' lists and built the $K_{\rm p}-V_{\rm rest}$ maps (Fig.~\ref{mappe_test1}), obtained by combining the signal of the four nights.
As it can be seen, we confirm the detection of both the molecules obtained in the main analysis: for \ch{H2O} we obtain a CCF peak at the expected planetary radial velocity with a S/N$= 5.1$ and a \emph{t-test} significance of $3.5\ \sigma$; for \ch{NH3} we obtain a CCF peak at the expected planetary radial velocity with a S/N$= 4.8$ and a \emph{t-test} significance of $4.0\ \sigma$. As it can be seen, removing the orders that showed a $\Delta S/N<2\ \sigma$, the result of this test does not lead to significative differences ($\leq 1\ \sigma$) with the conclusion reached in our work, giving more robustness to our analysis.

\section{Log of the HAT-P-11\,b GIANO-B observations}
In Table~\ref{logobs} we report the log of the GIANO-B observations of the four transits of HAT-P-11\,b.

\begin{table*}[!h]
\caption{Log of the GIANO-B observations of the four transits of HAT-P-11\,b.\tablefootmark{a}} 
\label{logobs}  
\centering
\begin{tabular}{c c c c c c}
\hline
\hline
    Night & $AM_{\rm min}$ to $AM_{\rm max}$ & $N_{\rm obs}$ & $t_{\rm exp}$ & S/N$_{\rm avg}$ & S/N$_{\rm min}$ to S/N$_{\rm max}$ \\
\hline
    7 July 2019 & $1.06-1.17$ & $60$ & $200$\,s & $58$ & $12-115$ \\
    18 June 2020& $1.06-1.23$ & $60$ & $200$\,s & $59$ & $9-110$ \\
    19 September 2020 & $1.06-1.24$ & $58$ & $200$\,s & $52$ & $2-100$ \\
    13 June 2023& $1.09-1.39$ & $62$ & $200$\,s & $48$ & $2-101$ \\
\hline                                            
\end{tabular}

\tablefoot{
    \tablefoottext{a}{From left to right we report: the date at the start of the observing night; the range of airmass $AM_{\rm min}$ to $AM_{\rm max}$ of the individual observing night during the planetary transit; the number of observed spectra $N_{\rm obs}$; the exposure time per spectrum $t_{\rm exp}$; the signal-to-noise ratio averaged across the whole spectral range S/N$_{\rm avg}$; the range of signal-to-noise ratios S/N$_{\rm min}$ to S/N$_{\rm max}$ in the individual spectral orders.}
}
\end{table*}
\end{appendix}

\end{document}